\documentclass{sig-alternate-mccown}

\usepackage{array}

\usepackage{url}

\def\more-auths{%
\end{tabular}
\begin{tabular}{c}}

\begin{document}
%
\CopyrightYear{2005} 

\title{Reconstructing Websites for the Lazy Webmaster}

%
%

\numberofauthors{1}
%

\author{
%
\alignauthor Frank McCown \; Joan A. Smith \; Michael L. Nelson \; Johan Bollen\\
       \affaddr{Old Dominion University}\\
       \affaddr{Computer Science Department}\\
       \affaddr{Norfolk, Virginia, USA  23529}\\
       \email{\{fmccown, jsmit, mln, jbollen\}@cs.odu.edu}
}

%
%


\date{12 November 2005}
\maketitle
\begin{abstract}
Backup or preservation of websites is often not considered until
after a catastrophic event has occurred.  In the face of complete
website loss, ``lazy'' webmasters or concerned third parties may be
able to recover some of their website from the Internet Archive.
Other pages may also be salvaged from commercial search engine
caches.  We introduce the concept of ``lazy preservation''- digital
preservation performed as a result of the normal operations of the
Web infrastructure (search engines and caches).  We present Warrick,
a tool to automate the process of website reconstruction from the
Internet Archive, Google, MSN and Yahoo.  Using Warrick, we have
reconstructed 24 websites of varying sizes and composition to
demonstrate the feasibility and limitations of website
reconstruction from the public Web infrastructure.  To measure
Warrick's window of opportunity, we have profiled the time required
for new Web resources to enter and leave search engine caches.
\end{abstract}




\section{Introduction}

\begin{quote}
\textit{``My old web hosting company lost my site in its entirety
(duh!) when a hard drive died on them.  Needless to say that I was
peeved, but I do notice that it is available to browse on the
wayback machine... Does anyone have any ideas if I can download my
full site?''} - A request for help at archive.org \cite{ross:web}
\end{quote}

Sometimes websites are lost due to negligence, sometimes to
laziness, and sometimes because the resources to backup a website
are just not available.  Even when backups are performed, they may
not have been performed correctly.  When a website is lost due to
some calamity, many webmasters will turn to the Internet Archive
(IA) ``Wayback Machine'' for help. The IA performs permanent
archiving of all types of Web resources when crawling the Web.
Although the IA can sometimes help reconstruct a website, it is
strictly a best-effort approach that performs sporadic, incomplete
and slow crawls of the Web (the IA repository is at least 6 months
out-of-date \cite{ia:faq}). Missing content can also be found from
search engines (SEs) like Google, MSN and Yahoo that scour the Web
looking for content to index and cache. Unfortunately the SEs do not
keep web pages long after they have gone missing (404), and they do
not preserve canonical copies of all the web resources they cache.
We will refer to the IA holdings and the SE caches collectively as
\emph{web repositories}.

We have built Warrick\footnote{Warrick is named after a fictional
forensic scientist with a penchant for gambling.}, a command-line
tool that reconstructs websites by recursively crawling the contents
of 4 web repositories (IA, Google, MSN and Yahoo). We used Warrick
to reconstruct 24 websites of various sizes and subject matter to
measure how well websites can be reconstructed from the 4 web
repositories. We measured the time it takes for the SEs to crawl and
cache web pages that we have created on .com and .edu websites. In
June 2005, we created synthetic web collections consisting of HTML,
PDF and images. For 90 days we systematically removed web pages and
measured how long they remained cached by the SEs.


\section{Background and Related Work}

Prior work has focused on 1) web archiving as a means of digital
preservation and 2) improving the ability of SEs to index the Web.
But no research has been performed that uses the byproduct of
commercial SE activity for archiving the Web.

In regards to archiving websites, organizations like the Internet
Archive and national libraries are currently engaged in archiving
the external (or client's) view of selected websites
\cite{day:collecting} and improving that process by building better
web crawlers and tools \cite{marill:tools}.  Systems have been
developed to ensure long-term access to Web content within
repositories and digital libraries \cite{reich:lockss}.

Numerous systems have been built to archive individual websites and
web pages. InfoMonitor archives the server-side components (e.g.,
CGI scripts and datafiles) and filesystem of a web server
\cite{cooper:infomonitor}. It requires an administrator to configure
the system and a separate server with adequate disk space to hold
the archives. Other systems like TTApache \cite{dyreson:managing}
and iPROXY \cite{chungwwa:webarchiving} archive requested pages from
a web server but not the server-side components.  TTApache is an
Apache module which archives different versions of web resources as
they are requested from a web server. Users can view archived
content through specially formatted URLs. iPROXY is similar to
TTApache except that it uses a proxy server and archives requested
resources for the client from any number of web servers. A similar
approach using a proxy server with a content management system for
storing and accessing Web resources was proposed in
\cite{feise:approach}. Commercial systems like Furl
(\url{http://furl.net/}) and Spurl.net (\url{http://spurl.net/})
also allow users to archive selected web resources that they deem
important.

A great deal of research has focused on improving the ability of SEs
to crawl and index content.  Work in this area focuses on issues
related to crawling performance
\cite{broder:efficient,cho:parallel,hafri:high}, choosing what web
pages to crawl
\cite{cho:incremental_crawler,cho:efficient_crawling,meczer:evaluating}
choosing when to re-crawl
\cite{cho:effective,coffman:optimal,edwards:adaptive}, and how to
crawl the deep web \cite{ntoulas:downloading,raghavan:crawling}.
Research has been performed showing how to find duplicate Web
content
\cite{bharat:mirror,cho:finding_replicated,shivakumar:finding_near_replicas}
and how to measure differences between text documents
\cite{broder:syntactic,shivakumar:scam}. Work related to measuring
observed web page change rates and their effects on SEs have also
been performed
\cite{cho:incremental_crawler,fetterly:large-scale,ntoulas:whats}.
A body of work proposes software that runs on web servers to
increase web crawling efficiency
\cite{brandman:crawler-friendly,gupta:internet,nelson:modoai,thati:crawlets}.


Estimates of SE coverage of the indexable Web have been performed
most recently in \cite{gulli:indexable}, but no measurement of SE
cache sizes or types of files stored in the SE caches has been
performed.  We are also unaware of any research that documents the
crawling and caching behavior of commercial SEs.


\section{Web Repositories}
\label{web_repositories}

\begin{table}
\centering \caption{Web repository-supported data types}
\begin{small}
\begin{tabular}{|l|c|c|c|c|c}
\hline
Type        &   G & Y & M & IA \\
\hline
HTML &  C   & C   & C   & C\\
Plain text  & M   & M   & M   & C\\
GIF, PNG, JPG  & M  & M  & $\sim$R  & C\\
JavaScript  & M  & & M  & C\\
MS Excel &  M  & $\sim$S & M & C\\
MS PowerPoint & M & M & M & C\\
MS Word  & M   & M   & M   & C\\
PDF    & M &  M & M & C \\
PostScript & M   & $\sim$S  &    & C\\
\hline
\end{tabular}
\end{small}
\begin{scriptsize}
\begin{tabular}{l}
\\
C = Canonical version is stored \\
M = Modified version is stored (image thumbnails or \\
\; \; \; \; HTML conversions) \\
$\sim$R = Stored but not retrievable with direct URL\\
$\sim$S = Indexed but stored version is not accessible \\
\end{tabular} \label{tbl:mimetypes}
\end{scriptsize}
\end{table}


To limit the implementation complexity, we have focused on what we
consider to be the 4 most popular web repositories.  It has been
shown that Google, MSN and Yahoo index significantly different
portions of the Web and have an intersection of less than 45\%
\cite{gulli:indexable}. Adding additional web repositories like
ask.com, gigablast.com, incywincy.com and any other web repository
that allows direct URL retrieval would likely increase Warrick's
ability to reconstruct websites.

Although SEs often publish index size estimates, it is difficult to
estimate the number of resources in each SE cache.  An HTML web page
may consist of numerous web resources (e.g., images, applets, etc.)
that may not be counted in the estimates, and not all indexed
resources are stored in the SE caches. Google, MSN and Yahoo will
not cache an HTML page if it contains a NOARCHIVE meta-tag
\cite{google:remove,msn:help,yahoo:help}, and the http Cache-control
directives `no-cache' and `no-store' may also prevent caching of
resources \cite{berghel:responsible_web_caching}.

Only IA stores web resources indefinitely. The SEs have proprietary
cache replacement and removal policies which can only be inferred
from observed behavior.  All four web repositories perform sporadic
and incomplete crawls of websites making their aggregate performance
important for website reconstruction.

Table \ref{tbl:mimetypes} shows the most popular types of resources
held by the four web repositories.  This table is based on our
observations when reconstructing websites with a variety of content.
IA keeps a canonical version of all web resources with only small
changes to HTML documents (some hyperlinks may get changed and extra
HTML is placed in the foot of the document). When adding an HTML
file to their cache, the SEs typically add extra HTML in the header.
The extra HTML can be removed to produce the canonical version for
Google and MSN and near canonical version for Yahoo (they convert
some characters like `\&nbsp;' into binary encodings).

\begin{figure*}
\begin{center}
\scalebox{0.8}{\includegraphics{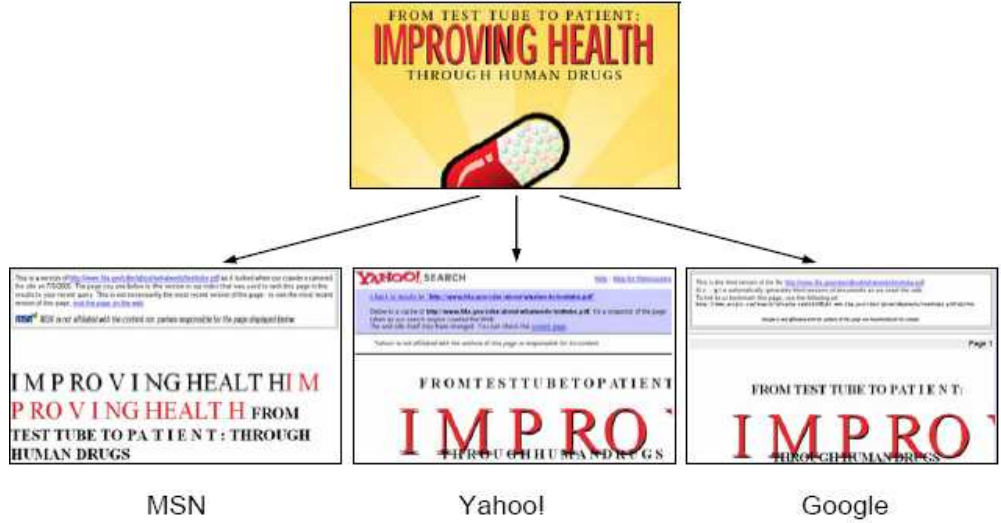}}
\caption{Original PDF and the HTML cached versions}
\label{fig:cachedpdfs}
\end{center}
\end{figure*}

When adding PDF, PostScript and Microsoft Office (Word, Excel,
PowerPoint) resources to their cache, the SEs create HTML versions
of the resources which are stripped of all images.  In most cases it
is not possible to recreate the canonical version of the document
from the HTML version. Figure \ref{fig:cachedpdfs} shows a PDF as it
was cached by Google, Yahoo and MSN. Although ``IMPROVING'' looks
like an image in two of the caches, it is text displayed in HTML
using a style sheet.

The SEs have separate search interfaces for their images.  They keep
only a thumbnail version of the images they cache due to copyright
law \cite{olsen:court}. MSN uses Picsearch for their image crawling;
unfortunately Picsearch and MSN do not support direct URL queries
for accessing these images, so they cannot be used for
reconstructing a website.
\section{Web Crawling and Caching}
\label{web_crawling_experiment}

\subsection{Lifetime of a Web Resource}

In order for a website to be reconstructed, it needs to have been
crawled and cached by at least one search engine and/or the Internet
Archive.  There are some methods that web masters can use to make
their websites crawler-friendly, and they can submit their URLs
to each of the SEs and IA in the hopes of being crawled and indexed
sometime in the near future.  Although there are mechanisms in place
to stop crawlers from indexing and caching pages from a website
(e.g., robots.txt), there is no mechanism to tell crawlers when to
start crawling or caching a website. SEs operate like a black box
whose external behavior can only be observed.

\begin{figure}
\begin{center}
\scalebox{0.48}{\includegraphics{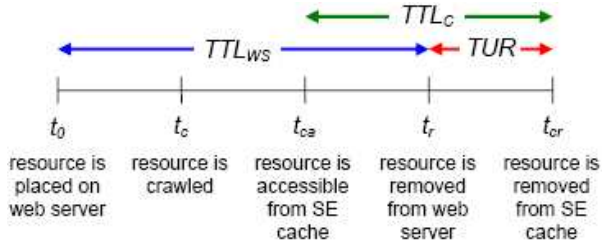}} \caption{Timeline of
SE resource acquisition and release} \label{fig:timeline}
\end{center}
\end{figure}

Figure \ref{fig:timeline} illustrates the life span of a web
resource from when it is first made available on a web server to
when when it is finally purged from a SE cache.  A web resource's
time-to-live (TTL) on the web server is defined as the number of
days until the resource is removed from the server:
\begin{equation}
    \mbox{\emph{TTL}}_{ws} = t_{r} - t_0
\end{equation}

If a SE crawls and caches a resource, it will typically remain
cached until the SE performs another crawl and discovers the
resource is no longer available on the web server.  The resource's
TTL in the SE cached is defined as:
\begin{equation}
    \mbox{\emph{TTL}}_c = t_{cr} - t_{ca}
\end{equation}
and the time until the resource is removed from the cache, the
time-until-removal (TUR), is defined as:
\begin{equation}
    \mbox{\emph{TUR}} = t_{cr} - t_r
\end{equation}

The $\mbox{\emph{TTL}}_{ws}$ and $\mbox{\emph{TTL}}_c$ values of a
resource may not necessarily overlap.  A SE that is trying to
maximize the freshness of its index will try to minimize the
difference between $\mbox{\emph{TTL}}_{ws}$ and
$\mbox{\emph{TTL}}_c$. A SE that is slow in updating its index,
perhaps because it purchases crawling data from a third party, may
experience late caching where $t_r < t_{ca}$.  It is also possible
to have negative values for \emph{TUR} if for some reason a resource
is removed from the cache before it has been removed from the web
server.

A resource is considered `vulnerable' until it is cached. A
\textbf{vulnerable resource} has an undefined $t_{ca}$ value. A
\textbf{recoverable resource} is defined as a resource where $t_c <
t_r$ and $\mbox{\emph{TTL}}_c > 0$.  A recoverable resource can only
be recovered during the $\mbox{\emph{TTL}}_c$ period with a
probability of $P_r$ (the observed number of days that a resource is
retrievable from the cache divided by $\mbox{\emph{TTL}}_c$). We
would like a resource to have a minimal $t_{ca}$ value to reduce its
vulnerability. SEs may also share this goal if they want to index
new content as quickly as possible.  We would also like a resource
to experience large values of \emph{TUR} so it can be recovered for
many days after its disappearance. SEs on the other hand may want to
minimize \emph{TUR} in order to purge missing content from their
index.

\begin{table*}
\centering \caption{Caching of HTML resources from 4 web
collections}
\begin{small}
\begin{tabular}{|l|m{0.5cm}|m{0.5cm}|m{0.5cm}|m{0.5cm}|m{0.5cm}|m{0.5cm}|l|l|l|l|l|l|l|l|l|l|} \hline
Web         &   \multicolumn{3}{c|}{\% URLs crawled} &   \multicolumn{3}{c|}{\% URLs cached} & \multicolumn{3}{c|}{$t_{ca}$} & \multicolumn{3}{c|}{$TTL_c$ / $P_r$} & \multicolumn{3}{c|}{\emph{TUR}}  \\
collection  & G & M & Y                           & G & M* & Y                          & G & M & Y                     & G & M & Y                    & G & M & Y \\
\hline
fmccown  & 91 & 41  & 56  & 91 & 16 & 36 & 13 & 65 & 47 & 90 / 0.78 & 20 / 0.87 & 35 / 0.57 & 51 & 9 & 24  \\
jsmit    & 92 & 31  & 92  & 92 & 14 & 65 & 12 & 66 & 47 & 86 / 0.82 & 20 / 0.91 & 36 / 0.55 & 47 & 7 & 25 \\
mln      & 94 & 33  & 84  & 94 & 14 & 49 & 10 & 65 & 54 & 87 / 0.83 & 21 / 0.90 & 24 / 0.46 & 47 & 8 & 19 \\
owenbrau & 18 & 0   & 0   & 20 & 0 & 0 & 103 & N/A & N/A & 40 / 0.98 & N/A & N/A & 61 & N/A & N/A \\
\hline
Ave      & 74 & 26  & 58  & 74 & 11 & 37 & 35   & 66 &  50 & 76 / 0.86 & 20 / 0.89  &  32 / 0.53  &  51  &  8 & 23 \\
\hline
\end{tabular}
\label{tbl:cached_collections}
\end{small}
\begin{scriptsize}
\begin{tabular}{l}
* Due to a query error, the MSN results could be higher (up to the
percentage crawled).\\
\end{tabular}
\end{scriptsize}
\end{table*}

\subsection{Web Collection Design}

We created 4 synthetic web collections with the same number of HTML,
PDF and image resources. The web collections were deployed in June
2005 at 4 different websites (1 .com and 3 .edu websites).  The .com
website (owenbrau.com) was new and had never been crawled before.
The 3 .edu websites fmccown, jsmit and mln (all subsites of
www.cs.odu.edu) had existed for over 1 year and had been previously
crawled by multiple SEs. In order for the web collections to be
found by the SEs, we placed links to the root of each web collection
from the .edu websites, and we submitted owenbrau's base URL to
Google, MSN and Yahoo 1 month prior to the experiment. For 90 days
we systematically removed resources from each collection. We
examined the server web logs to determine when resources were
crawled, and we queried Google, MSN and Yahoo daily to determine
when the resources were cached.

We organized each web collection into a series of \emph{update bins}
or directories which contained a number of HTML pages referencing
the same three inline images (GIF, JPG and PNG) and a number of PDF
files. An index.html file (with a single inline image) in the root
of the web collection pointed to each of the bins. An index.html
file in each bin pointed to the HTML pages and PDF files so a web
crawler could easily find all the resources. All these files were
static and did not change throughout the 90 day period except the
index.html files in each bin which were modified when links to
deleted web pages were removed.

The number of resources in the web collections were determined by
the number of update bins $B$, the last day that resources were
deleted from the collection $T$ (the \emph{terminal day}), and the
bin $I$ which contained 3 images per HTML page. Update bins were
numbered from 1 to $B$, and resources within each bin $b$ were
numbered from 1 to $\lfloor{T / b}\rfloor$. Resources were deleted
from the web server according to their bin number. Every \emph{n}
days we would deleted one HTML page (and associated images for pages
in bin $I$) and one PDF file from bin \emph{n}. For example,
resources in bin 1 were deleted daily, resources in bin 2 were
deleted every other day, etc. We also removed the links to the
deleted HTML and PDF files from bin \emph{n}'s index.html file.

At any given day $d$ during the experiment (where $d = 0$ is the
starting day and $d \le T$), the total number of resources in the
web collection is defined as:
\begin{equation}
Total_c(d) = 2 + \sum_{i=1}^{B} Total_b(i,d)
\end{equation}

\begin{figure}
\begin{center}
\scalebox{0.3}{\includegraphics{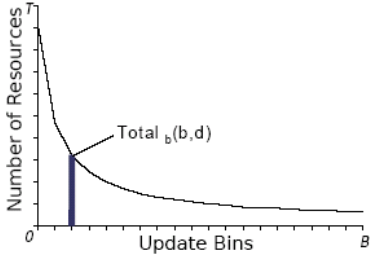}}
\caption{Number of resources in web collection}
\label{fig:web_collection_graph}
\end{center}
\end{figure}

The total number of HTML, PDF and image files in bin $b$ on any day
$d$ is defined as:
\begin{equation}
Total_b(b,d) = \mbox{\emph{HTML}}(b,d) + \mbox{\emph{PDF}}(b,d) +
\mbox{\emph{IMG}}(b,d)
\end{equation}
The total number of resources in each update bin deceases with the
bin's periodicity as show in Figure \ref{fig:web_collection_graph}.
The number of HTML, PDF and image files in each bin $b$ on any day
$d$ is defined as:
\begin{equation}
\mbox{\emph{HTML}}(b,d) = \lfloor{T / b}\rfloor - \lfloor{d /
b}\rfloor + 1
\end{equation}
\begin{equation}
\mbox{\emph{PDF}}(b,d) = \lfloor{T / b}\rfloor - \lfloor{d /
b}\rfloor
\end{equation}
\begin{equation}
\mbox{\emph{IMG}}(b,d) =
\begin{cases}
3(\mbox{\emph{HTML}}(b,d) - 1) & \mbox{if } b = I \\
0 & \mbox{if } \mbox{\emph{HTML}}(b,d) = 1 \\
3 & \mbox{otherwise}
\end{cases}
\end{equation}


In each of our web collections we created 30 update bins ($B=30$)
that completely decayed by day 90 ($T=90$), and we chose bin 2
($I=2$) to have the supplemental images. So the total number of
files in each collection on day 0 was $Total(0) = 954$. We limited
the web collections to less than 1000 resources in order to limit
the number of daily queries to the SEs. We created a fewer number of
images than HTML and PDF pages because we hypothesized that images
were not cached as frequently as other resources and the cost of
querying for images (number of queries issued per resource) was
higher than for HTML and PDF resources.

The \emph{TTL}$_{ws}$ for each resource in the web collection is
determined by its bin number $b$, page number $p$, and the web
collection terminal day $T$:
\begin{equation}
\mbox{\emph{TTL}}_{ws} = b (\lfloor{T / b}\rfloor - p + 1)
\end{equation}
%

\begin{figure}
\begin{center}
\scalebox{0.46}{\includegraphics{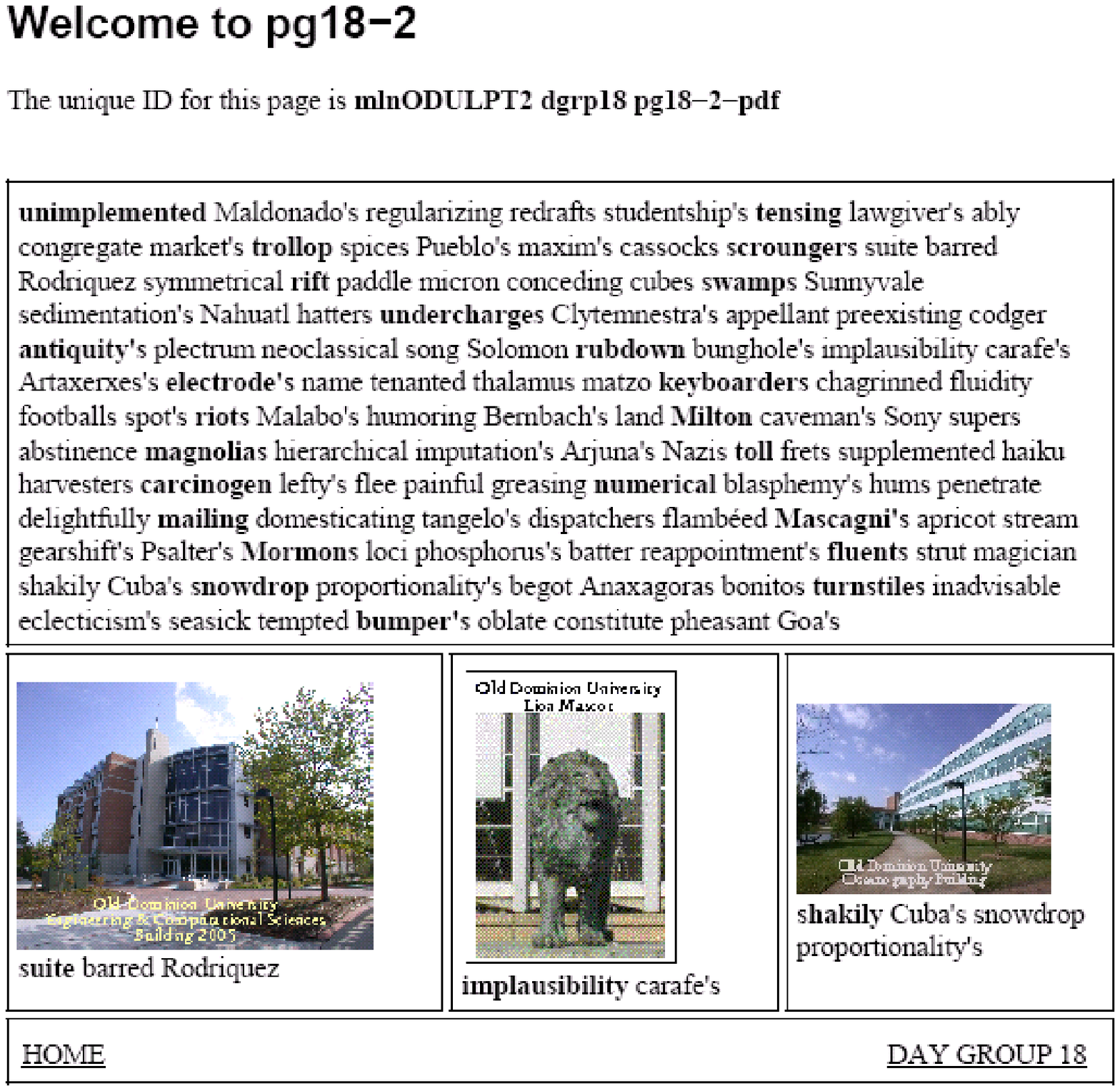}} \caption{Example PDF
page} \label{fig:example_page}
\end{center}
\end{figure}

An example PDF page from one of the web collections is shown in
Figure \ref{fig:example_page}. HTML pages look very similar.  Each
HTML and PDF page contain a unique identifier (UID) at the top of
each page that included 4 identifiers: the web collection (e.g.,
`mlnODULPT2' means the `mln' collection), bin number (e.g., `dgrp18'
means bin 18), page number and resource type (e.g., `pg18-2-pdf'
means page number 2 from bin 18 and PDF resource). The UID contains
spaces to allow for more efficient querying of the SE caches.  The
text for each page was randomly generated from a standard
English dictionary so there would be no keyword skew (e.g., ``Bush
Iraq'') that might impact crawler behavior.  By using random words
we avoided creating duplicate pages that a SE may reject.  SEs could
use natural language processing to determine that the words of each
page were random and therefore might punish such pages by refusing
to index them, but the SE caching behavior we observed seems to
indicate this is not the case.

\subsection{Daily SE Queries}

In designing our daily SE queries, care was taken perform a minimal
number of daily queries to not overburden the SEs. We could have
queried the SEs using the URL for each resource, but this might have
led to our resources being cached prematurely; it is possible that
if a SE is queried for a URL it did not index that it would add the
URL to a list of URLs to be crawled at a later date. This is how IA
handles missing URLs that are used in users' queries.

In order to determine which HTML and PDF resources had been cached,
we used the UID for each resource to query each SE.  We queried for
the top 100 results of those items matching the resource type (`PDF'
or `HTML') using the first 2 parts of the UID (e.g., `mlnODULPT2
dgrp18') which uniquely identified the web collection and bin
number\footnote{MSN only allows limiting the results page to 50 and
does not support limiting by resource type. Therefore large bins
might require more than 1 query.}. We then parsed the single result
page looking for links to cached versions of the resources. Although
we could have queried for the complete UID, this would have
unnecessarily increased the load put on the SEs.  If there are $w$
web collections with $b$ bins and $t$ types of resources in each
bin, and there are $s$ SEs which need to be queried daily, then the
total number of daily queries $Q_b$ is:
\begin{equation}
Q_b = w \cdot b \cdot t \cdot s
\end{equation}

Querying the SE caches for images requires a different strategy than
the one used for querying for text-based resources.  We gave each
image a globally unique filename so that a query for the image's
filename would result in a found or not found result.  If there are
$w$ web collections with $i$ images and $s$ SEs need to be queried
daily, then the total number of daily queries $Q_i$ is:
\begin{equation}
Q_i = w \cdot i \cdot s
\end{equation}

\subsection{Crawling and Caching Observations}
Although the web server logs registered visits from a variety of
crawlers, we report only on crawls from Google, Inktomi (Yahoo) and
MSN.  Alexa Internet (who provides crawls to IA) only accessed our
collection once (induced through our use of the Alexa toolbar). A
separate IA robot accessed less than 1\% of the collections, likely
due to several submissions we made to their Wayback Machine's
advanced search interface early in the experiment.

We report only detailed measurements on HTML resources (PDF
resources were similar). Images were crawled and cached far less
frequently; Google and Picsearch (the MSN Images provider) were the
only ones to crawl a significant number of images.  The 3 .edu
collections had 29\% their images crawled, and owenbrau had 14\% of
its images crawled. Only 4 unique images appeared in Google Images,
all from the mln collection. Google likely used an image duplication
detection algorithm to prevent duplicate images from different URLs
from being cached. Only one image (from fmccown) appeared in MSN
Images.  None of the cached images fell out of cache during our
experiment.




Table \ref{tbl:cached_collections} summarizes the performance of
each SE to crawl and cache HTML resources from the 4 web
collections\footnote{Due to a technical mishap we lost days 41-55 of
crawling data for owebrau and parts of days 66-75 and 97 for the
.edu web collections. We also lost all cache data from all 4 web
collections for days 53, 54, 86 and 87.}. This table does not
include index.html resources which had an infinite $TTL_{ws}$. We
believe there was an error in the MSN query script which caused
fewer resources to be found in the MSN cache, but the percentage of
crawled URLs provides an upper bound on the number of cached
resources; this has little to no effect on the other measurements
reported.

The three SEs showed equal desire to crawl HTML and PDF resources.
Inktomi (Yahoo) crawled 2 times as many resources as MSN, and Google
crawled almost 3 times as many resources than MSN. Google was the
only SE to crawl and cache any resources from the new owenbrau
website.

From a website reconstruction perspective, Google out-performed MSN
and Yahoo in nearly every category.  Google cached the highest
percentage of HTML resources (76\%) and took only 12 days on average
to cache new resources from the .edu web collections. On average,
Google cached HTML resources for the longest period of time (76
days), consistently provided access to the cached resources (86\%),
and were the slowest to remove cached resources that were deleted
from the web server (51 days). Although Yahoo cached more HTML
resources and kept the resources cached for a longer period than
MSN, the probability of accessing a resource on any given day was
only 53\% compared to 89\% for MSN.

\begin{figure*}
\begin{center}
\scalebox{1.08}{\includegraphics[clip=false]
{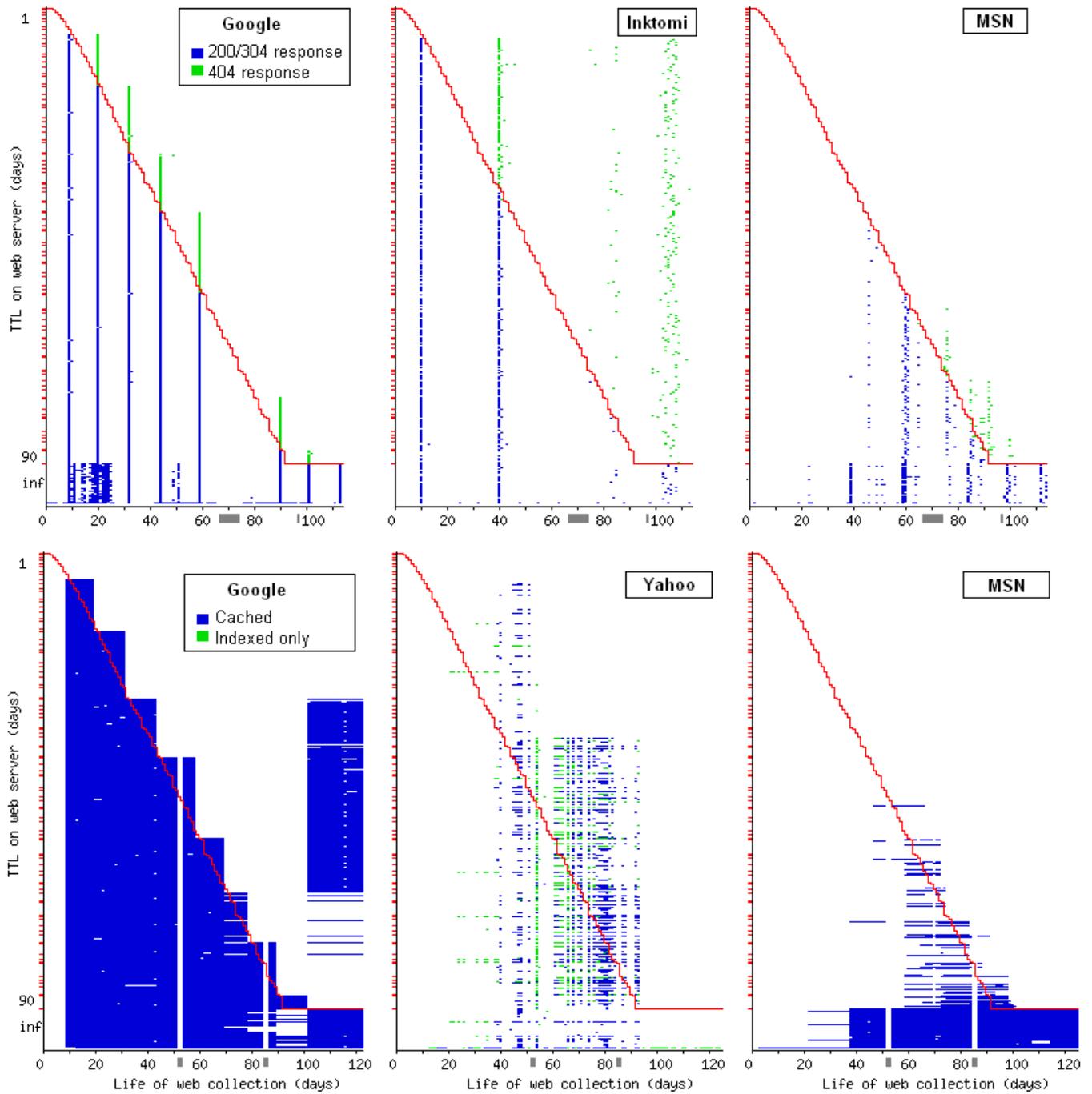}} \caption{Crawling (top) and
caching (bottom) of HTML resources from the mln web collection}
\label{fig:crawling_caching_mln_html}
\end{center}
\end{figure*}

%

Figure \ref{fig:crawling_caching_mln_html} provides an interesting
look at the crawling and caching behavior of Google, Yahoo and MSN.
These graphs illustrate the crawling and caching of HTML resources
from the mln collection; the other 2 .edu collections exhibited
similar behavior. The resources are sorted by
$\mbox{\emph{TTL}}_{ws}$ with the longest-living resources appearing
on the bottom.  The index.html files which were never removed from
the web collection have an infinite TTL (`inf'). The red line
indicates the decay of the web collection.  On the top row of Figure
\ref{fig:crawling_caching_mln_html}, blue dots indicate resources
that were crawled on a particular day. As resources were requested
that had been deleted, the web server responded with a 404 (not
found) code represented by green dots above the red line. The bottom
row of Figure \ref{fig:crawling_caching_mln_html} shows the cached
HTML resources (blue) resulting from the crawls. Some pages in Yahoo
were indexed but not cached (green).

Google was by far the most active of the crawlers and cached more
resources than the other two SEs.  Google was quick to purge
resources from their cache when a crawl revealed the resources were
no longer available on the web server.  On day 102 many previously purged resources
reappeared in the Google cache and stayed
for the remainder of our measurements.  The other 2 .edu
web collections recorded similar behavior where HTML resources
reappeared in the Google cache long after being removed.  Cached PDF
resources did not experience the same reappearance.

Yahoo performed sporadic caching of resources.  As shown in Figure
\ref{fig:crawling_caching_mln_html}, resources tended to fluctuate
in and out of the Yahoo cache.  There is a lag time of about 30 days
between Inktomi crawling a resource and the resource appearing in
the Yahoo cache, and many crawled resources never appear in the
Yahoo cache.  Although Inktomi crawled nearly every available HTML
resource on day 10, only half of those resources ever became
available in the Yahoo cache.

MSN was very slow to crawl the HTML and PDF resources in the update
bins. After day 40 they began to crawl some of the resources and
make them available in their cache.  Like Google, MSN was quick to
remove 404 resources from their cache.

\begin{figure*}
\begin{center}
\scalebox{0.7}{\includegraphics{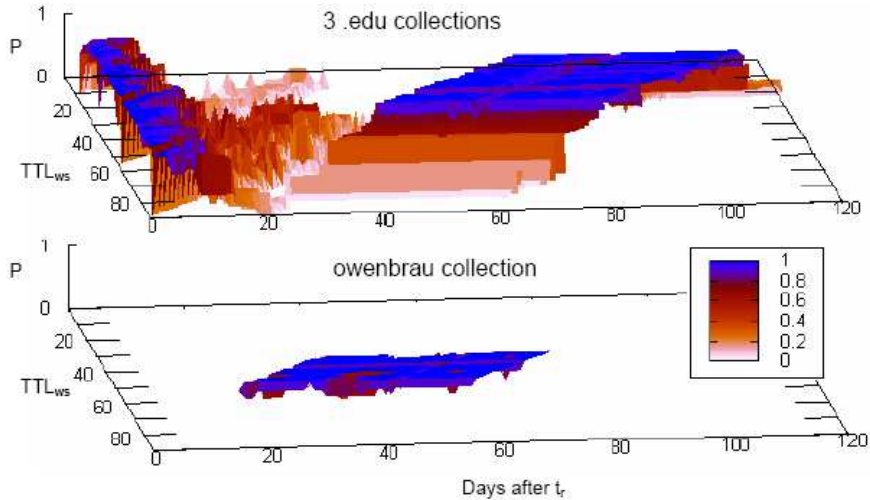}}
\caption{HTML resources in at least one SE cache}
\label{fig:surface_map}
\end{center}
\end{figure*}


Figure \ref{fig:surface_map} graphs the observed probability of
finding HTML resources from the web collections in at least one SE
cache after they have been removed from the web server ($t_r$). The
top graph is from the 3 .edu collections grouped together, and the
bottom graph is from owenbrau. On examination of the .edu web
collections, most HTML resources were purged from the SE caches 10
days after $t_r$ as illustrated by the ridge on the left. Some
resources continued to be available (mostly from Yahoo) which form
the hills in days 20-40. The large plateau on the center-right are
those HTML resources that re-appeared in Google's cache days after
being purged.  None of the HTML resources from owenbrau appeared in
cache until 20-30 days after they were removed, and they remained
cached for 40 days before being purged.

We have observed from our measurements that nearly all new HTML and
PDF resources that were placed on known websites were crawled and
cached by Google several days after they were discovered. Resources
on a new website were not cached for months. Yahoo and MSN were 4-5
times slower than Google to acquire new resources, and Yahoo incurs
a long transfer delay from Inktomi's crawls into their cache. We
have also observed that cached resources are often purged from all 3
caches as soon as a crawl reveals the resources are missing. But we
also observed many HTML resources reappearing in Google's cache
weeks after being removed.  We have not performed experiments on
enough websites to conclude that the SEs always perform as we have
witnessed, but our observations do suggest that websites can be
recovered more successfully if they are recovered quickly after
disappearing, and running Warrick over a period of a month or more
will likely increase recovery success.

\section{Warrick}
\label{warrick}

\subsection{Operation}
Warrick is able to reconstruct a website when given a URL pointing
to the host page where the site used to exist.  The web repositories
can be queried with a URL to produce the stored version of the
resource in their holdings. For example, we can download Google's
cached version of \url{http://foo.edu/page1.html} like so:
\url{http://search.google.com/search?q=cache:http://foo.edu/page1.html}.
If Google has not cached the page, an error page will be generated.
Otherwise the cached page can be stripped of any Google-added HTML,
and the page can be parsed for links to other resources from the
foo.edu domain.

\begin{table}
\caption{Number of queries required to obtain resources from the web
repositories}
\begin{center}
\begin{small}
\begin{tabular}
{|m{2.5cm}|m{1.5cm}|m{1.5cm}|} \hline
Web repository & Non-image resource & Image \; \; resource \\
\hline
Google & 1 & 2 \\
MSN & 2 & N/A \\
Yahoo & 2 & 2 \\
Internet Archive & $\ge$ 2 & $\ge$ 2 \\
\hline
\end{tabular}
\end{small}
\end{center}
\label{tbl:num_queries}
\end{table}

As shown in Table \ref{tbl:num_queries}, most repositories require 2
queries to obtain a resource because the URL of the stored resource
cannot be created automatically like Google's.  MSN, for example,
requires an initial query that will produce a web page with the
cached URL within it. The cached URL must be extracted from the
result page and accessed to find the cached resource. Yahoo and MSN
also provide APIs that require 2 queries for obtaining resources
without the necessity of page scraping. IA may require more than 2
queries because it will occasionally produce links to stored content
that is not retrievable, and additional queries must be performed to
check for older versions of the resource.

\begin{figure}
\begin{center}
\scalebox{0.9}{\includegraphics[clip=true,viewport=0 10 250
205]{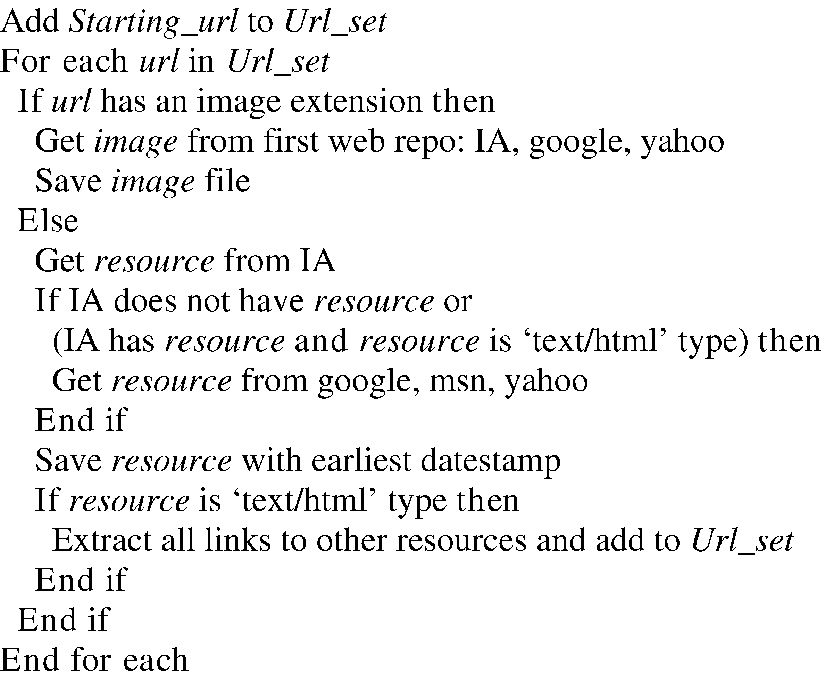}}  \caption{Pseudocode for recovering resources from
web repositories} \label{fig:pseudocode}
\end{center}
\end{figure}

Figure \ref{fig:pseudocode} shows the logic Warrick uses when
reconstructing a website. For each URL, the file extension (if
present) is examined to determine if the URL is an image (.png,
.gif, .jpg, etc.) or other resource type.  All three SEs use a
different method for retrieving images than for other resource
types.  IA has the same interface regardless of the file type.  We
would have better accuracy at determining if a given URL referenced
an image or not if we knew the URL's resource MIME type, but this
information is not available to us.

IA is the first web repository queried by Warrick because it keeps a
canonical version of all web resources.  When querying for an image
URL, if IA does not have the image then Google and Yahoo are queried
one at a time until one of them returns an image.  Google and Yahoo
do not publicize the cached date of their images, so it is not
possible to pick the most recently cached image.

If a non-image resource is being retrieved, again IA is queried
first.  If IA has the resource and the resource does not have a MIME
type of `text/html', then the SEs are not queried since they only
have canonical versions of HTML resources. We assume the user would
prefer to have the canonical copy of a resource rather than the HTML
version, but Warrick provides an option to select the most recent
version if that is preferred over the canonical version. If the
resource does have a `text/html' MIME type (or IA did not have a
copy), then all three SEs are queried. If more than one SE produces
a result, the cache dates are compared, and the most recent resource
is chosen.  MSN always provides a cache date for its contents.
Yahoo, unfortunately, does not provide a cache date for any of its
resources (it does sometimes provide a Last-Modified date for the
resource), and Google will only provide a cache date for HTML
resources.  Therefore if a non-HTML resource is returned from all 3
SEs (assuming IA did not have the resource), the MSN version will
always be the resource selected.

If the resource has a MIME type of `text/html', Warrick will look
for URLs to other resources by examining a variety of tags (e.g.,
$<$A$>$, $<$IMG$>$, $<$LINK$>$, $<$EMBED$>$, etc.).  The web
repositories are queried with each of these URLs until there are no
more URLs to be found. There are numerous options to limit how many
files are reconstructed and the types of files that should be
reconstructed.

All resources are stored to disk in a directory/file that matches
the reconstructed URL.  For example, the resource with URL
\url{http://foo.org/ dir1/ abc.html} would be saved as
\url{foo.org/dir1/abc.html}. HTML versions of PDFs and other
converted types are stored with their original filenames.  Warrick
provides an option to rename these types of files with an .html
extension so they can be viewed in a browser.  All links can be made
relative and point to the newly renamed files.

Warrick keeps a log of each reconstructed resource that includes the
MIME type (`MISSING' if the resource could not be found), the web
repository the resource was restored from and the date of the
resource as indicated by the web repository.
\subsection{Cost of Website Reconstruction}
Warrick relies on the services provided by the web repositories for
reconstructing websites.  Warrick ``respects'' the web repositories
in that it issues a limited number of queries per 24 hours and
delays before issuing repeated queries.  After exceeding a certain
number of queries from a particular IP address, some SEs may quit
responding to requests (we personally experienced this once).
Unfortunately there is no standard way for Warrick to
automatically find this limit. Instead we must rely on published
standards for whatever public API is released or rely on personal
experience.  In general, a SE will have a limited number of queries
$L$ that an agent may make on a SE in a time period of $P$. It may
also prefer that an agent wait $W$ seconds between requests. If the
number of queries performed in time $P$ exceeds $L$, the agent
should cease making queries until a time span of $P$ has elapsed. If
an agent follows these guidelines, the agent is said to ``respect''
the SE or web repository.

The SEs all provide public APIs that specify a limited number of
queries per 24 hours with no $W$ requirements.  Google allows 1000
queries, Yahoo allows 4000, and MSN allows 10,000. Although we
currently do not use the Google and MSN APIs, their API limits are
good indicators for what they deem reasonable use of their web-based
search interfaces.  IA does not publish an API with daily query
limits, so we chose a daily limit of 1000 queries.

Warrick self-imposes a $W$ requirement of 1-4 seconds (random)
between query rounds (1 round = queries to all web repositories). It
will sleep for 24 hours if the number of queries it performs on any
web repository exceeds the web repository's query limit and then
pick back up where it left off. Unfortunately files may start to
leave a SE cache before being able to complete the reconstruction of
a large website. A more urgent reconstruction effort might increase
the number of queries per day.

The collective cost incurred by the web repositories for
reconstructing a website is the total number of queries they must
respond to from Warrick. The query cost $C_q$ can be defined as the
total number of queries needed to pull $r$ non-image resources from
$n$ web repositories and $i$ image resources from $m$ web
repositories that house images:
\begin{equation}
C_q(r,i) = r  \sum_{j=1}^{n} Q_r(j) + i \sum_{j=1}^{m} Q_i(j)
\end{equation}

An upper bound for $Q_r$ is 7 if a non-image resource was found in
all 4 web repositories.  The lower bound is 2 (canonical version was
found in IA). An upper bound for $Q_i$ is 4 if an image resource was
not found until the last web repository.  The lower bound is 2
(image was found in the first web repository). The maximum number of
queries required to reconstruct a website with 50 HTML files, 50 PDF
files, and 50 images would be $C_q(100,50) = 900$ queries, and the
minimum number of queries would be 300.

\section{Warrick Evaluation}
\label{recon_websites}

\subsection{Reconstruction Measurements}
A website is a set of interlinked web pages that are made up of one
or more web resources (e.g., style sheets, JavaScript, images,
etc.), each identified by a URI \cite{lavoie:web_characterization}.
Each web page (which is itself a web resource) may link to other web
pages if it is an HTML resource, or it may be a self-contained
resource like a PDF, MS PowerPoint slide show or image.

\begin{figure}
\begin{center}
\scalebox{0.5}{\includegraphics{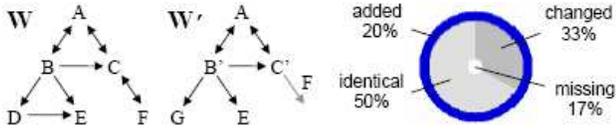}} \caption{Original
website (left), reconstructed website (center), and reconstruction
diagram (right)} \label{fig:website_graphs}
\end{center}
\end{figure}

We can construct a web graph $G = (V,E)$ for a website where each
resource $r_i$ is a node $v_i$, and there exists a directed edge
from $v_i$ to $v_j$ when there is a hyperlink or reference from
$r_i$ to $r_j$.
This graph may be
constructed for any website by downloading the host page (e.g.,
\url{http://foo.com/}) and looking for links or references to other
Web resources, a method employed by most web crawlers.  The left
side of Figure \ref{fig:website_graphs} shows a web graph
representing some website $W$ if we began to crawl it beginning at
A.

Suppose the website $W$ was lost and reconstructed forming the
website $W'$ represented in the center of Figure
\ref{fig:website_graphs}. For each resource $r_i$ in $W$ we may
examine its corresponding resource $r'_i$ in $W'$ that shares the
same URI and categorize $r'_i$ as:
\begin{enumerate}
  \item \textit{identical} -- $r'_i$ is byte-for-byte identical to $r_i$
  \item \textit{changed} -- $r'_i$ is not identical to $r_i$
  \item \textit{missing} -- $r'_i$ could not be found in any web
  repository and does not exist in $W'$
\end{enumerate}
We would categorize those resources in $W'$ that did not share a URI
with any resource in $W$ as:
\begin{enumerate}
\setcounter{enumi}{3}
  \item \textit{added} -- $r'_i$ was not a part of the current website
but was recovered due to a reference from $r'_j$
\end{enumerate}

Figure \ref{fig:website_graphs} shows that resources A, G and E were
reconstructed and are identical to their original versions.  An
older version of B was found (B') that pointed to G, a resource that
does not currently exist in $W$.  Since B' does not reference D, we
did not know to recover it.  It is possible that G is actually D
renamed, but we do not test for this.  An older version of C was
found, and although it still references F, F could not be found in
any web repository.

A measure of change between the original website $W$ and the
reconstructed website $W'$ can be described using the following
\textbf{difference vector}:
\begin{equation}\mbox{difference}(W,W') =
\left(
    \frac{R_{changed}}{|W|}, \frac{R_{missing}}{|W|},
    \frac{R_{added}}{|W'|}
\right)
\end{equation}
For Figure \ref{fig:website_graphs}, the difference vector is (2/6,
1/6, 1/5) = (0.333, 0.167, 0.2).

The following bounds apply to the difference vector:
\begin{itemize}
  \item (0, 0, 0) Best case: a complete reconstruction of a website.
  \item (1, 0, 0) Every resource of the website was found, but they are all
changed.
  \item (0, 1, 0) Worst case: no resources were found in any web repository.
  \item (0, 0, 1) Impossible case: It would not be possible to find only added
resources if we did not start with a single resource that was either
identical or changed.
\end{itemize}

The difference vector for a reconstructed website can be illustrated
as a \textbf{reconstruction diagram} as shown on the right side of
Figure \ref{fig:website_graphs}. The changed, identical and missing
resources form the core of the reconstructed website. The dark gray
portion of the core grows as the percentage of changed resource
increases.  The hole in the center of the core grows as the
percentage of missing resources increases. The added resources
appear as crust around the core.  This representation will be used
later in Table \ref{tbl:recons} when we report on the websites we
reconstructed in our experiments.

\subsection{Website Reconstruction Results}
\label{sec-results}
We chose 24 websites covering a variety of topics based on our
personal interests and random samples from the Open Directory
Project (dmoz.org).  Some of the websites we selected are actually
subsites, but we will use the term `website' when referring to all
of them. We chose 8 small websites (1-149 URIs), 8 medium websites
(150-499 URIs) and 8 large websites (500-2000 URIs).  We also chose
only websites that were crawler friendly (did not use JavaScript to
produce dynamic links, did not use Flash exclusively as the main
interface, etc.) and did not use the robots exclusion protocol
(robots.txt) to stop web crawlers from indexing their websites.

In August 2005 we downloaded the 24 websites using Wget. All files
needed to produce each web page were downloaded (HTML, style sheets,
external JavaScript files, images, etc.)\footnote{We used Wget 1.10
with the following options: -np, -p, -w 2, -r, -l 0, -e, -t 1.}. For
simplicity, we restricted the download to only those files that were
in and beneath the starting directory. For example, if we downloaded
\url{http://foo.edu/abc/}, only URLs matching
\url{http://foo.edu/abc/*} were downloaded.
Immediately after downloading the websites, we ran Warrick to
reconstruct all 24 websites.  For each website we reconstructed 5
different versions: 4 using each web repository separately, and 1
using all web repositories together.

The results of the aggregate website
reconstructions are shown in Table \ref{tbl:recons} sorted by
website size (number of files in the original website). The `PR'
column is Google's PageRank for the root page of each website at the
time of the experiments. The PageRank is the importance (0-10 with
10 being the most important) that Google assigns to a web page. MSN
and Yahoo do not publicly disclose their `importance' metric.  We
were unable to find any statistical correlation between percentage
of recovered files and PageRank or between recovered files and
website size.

For each website the total number of files in the original website
is shown along with the total number of files that were recovered
and the percentage.  The files are also totalled by MIME type. The
difference vector for the website accounts for recovered files that
were added.

The `Almost identical' column shows the percentage of text-based
resources (e.g., HTML, PDF, PostScript, Word, PowerPoint, Excel)
that were \emph{almost identical} to the originals.  The last column
shows the reconstruction figure for each website if these almost
identical resources are moved from the `Changed' category to
`Identical' category.  We counted the number of shared fixed-size
shingles to determine text document similarity.  Shingling (as
proposed by Broder et al. \cite{broder:syntactic}) is a popular
method for quantifying similarity of text documents when word-order
is important
\cite{bharat:mirror,fetterly:large-scale,ntoulas:whats}. We
considered any two documents to be almost identical if they shared
at least 75\% of their shingles of size 10.  Other document
similarity metrics that take word order into account could also have
been used \cite{shivakumar:scam}. We used open-source tools
\cite{ockerbloom:mediating} to convert non-HTML resources into text
before computing shingles.  We did not use any image similarity
metrics.

\begin{table*}
\centering \caption{Results of website reconstructions}
\begin{small}
\begin{tabular}{|m{4.3cm}|m{0.3cm}|m{1.3cm}|m{1.0cm}|m{1.0cm}|m{1.0cm}|m{2.6cm}|m{0.7cm}|m{0.8cm}|m{0.7cm}|} \hline
         &    & \multicolumn{4}{c|}{MIME type groupings (orig/recovered)} &   Difference vector       &       &                  & \\
Website & PR & Total & HTML & Images & Other             & (Changed, Missing, Added) & Recon diag & Almost identical & New recon diag \\
\hline
1. www.smoky.ccsd.k12.co.us & 4 & 63/27 43\% & 20/20   100\% & 23/5   22\% & 20/2   10\% & (0.111, 0.571, 0.000) & \scalebox{0.05}{\includegraphics{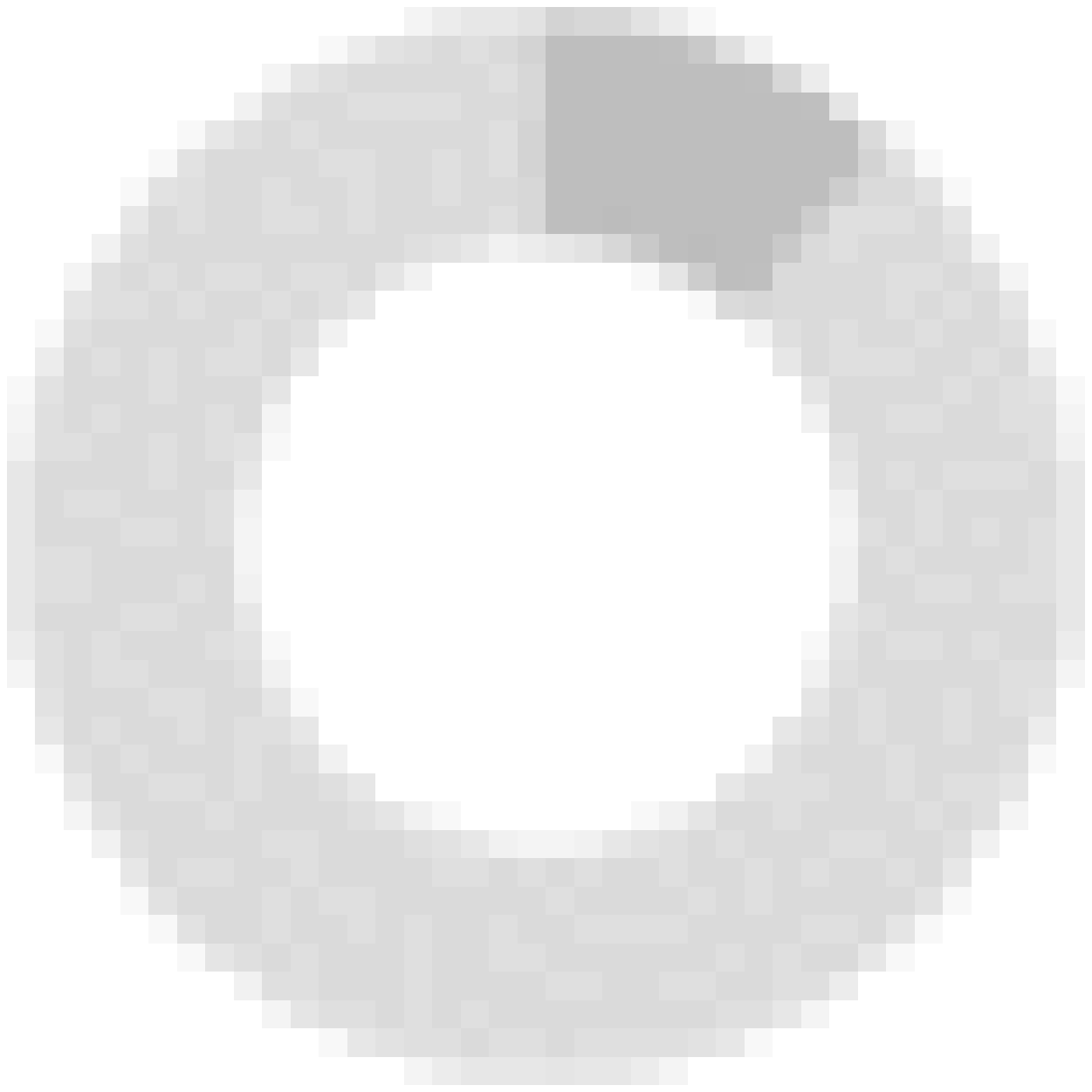}} & 100\% & \scalebox{0.05}{\includegraphics{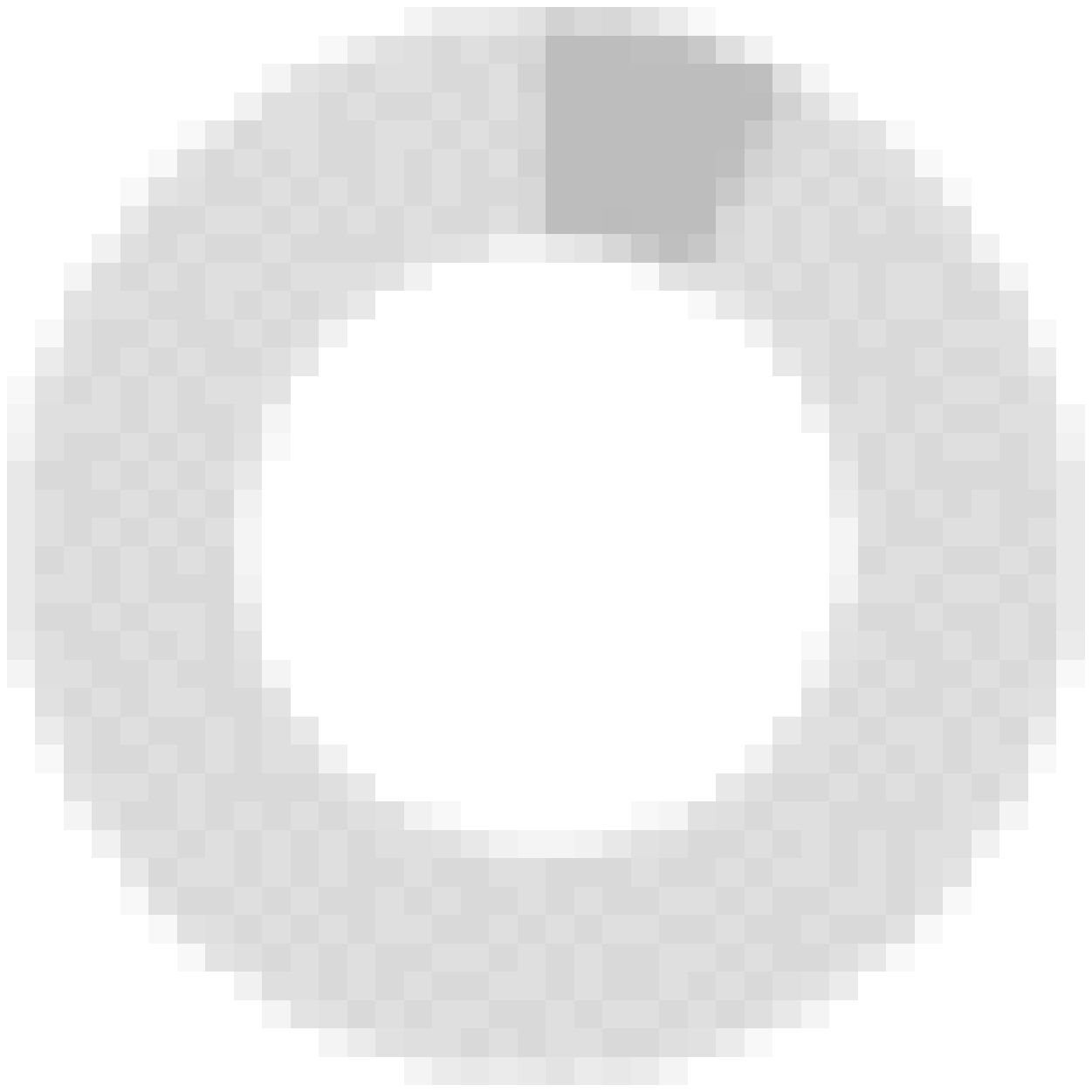}} \\
2. www.genesis427.com & 2 & 65/53 82\% & 10/10   100\% & 50/40   80\% & 5/3   60\% & (0.662, 0.185, 0.000) & \scalebox{0.05}{\includegraphics{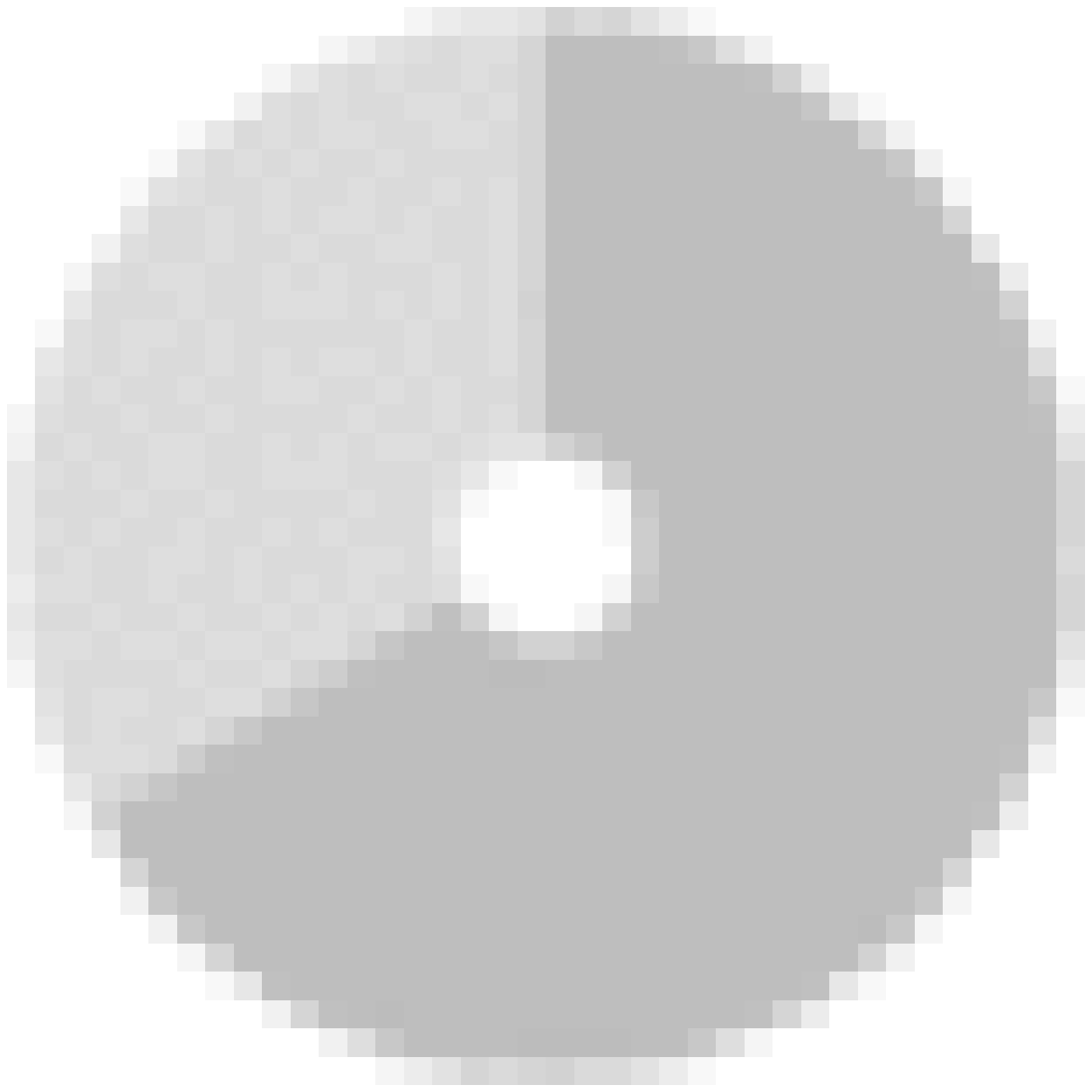}} & 33\% & \scalebox{0.05}{\includegraphics{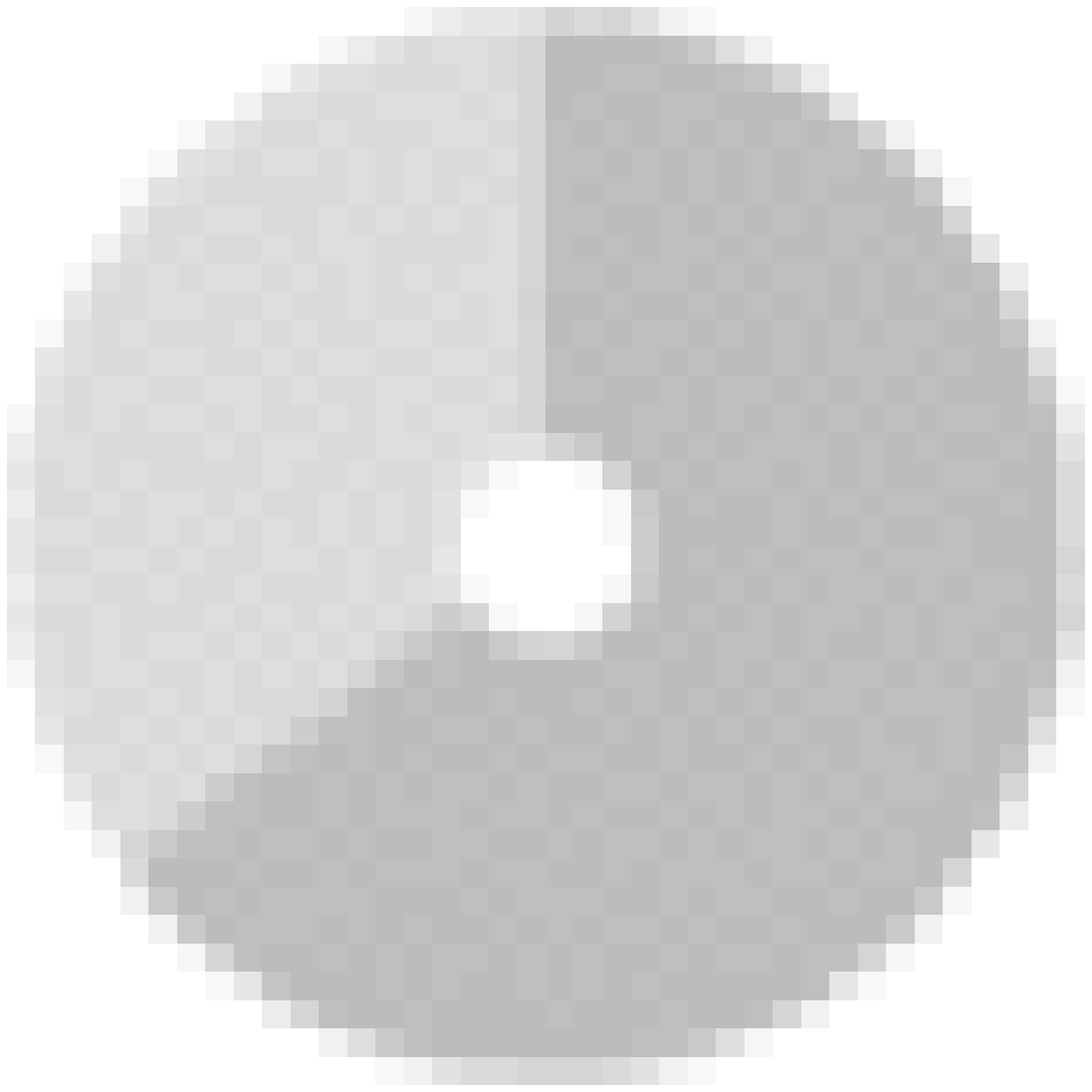}} \\
3. englewood.k12.co.us/schools/ clayton & 3 & 68/58 85\% & 32/29   91\% & 36/29   81\% & 0/0    & (0.426, 0.147, 0.000) & \scalebox{0.05}{\includegraphics{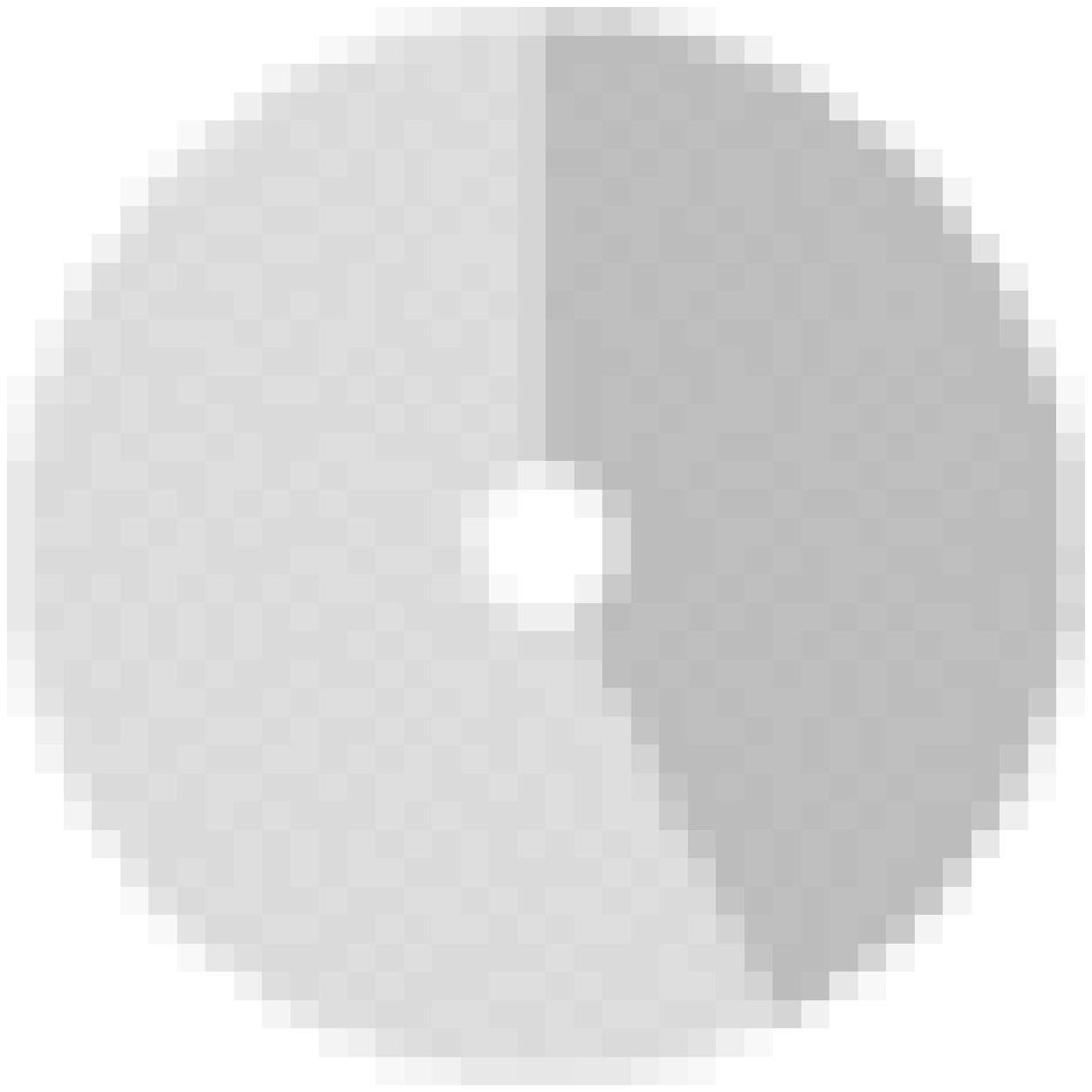}} & N/A & \scalebox{0.05}{\includegraphics{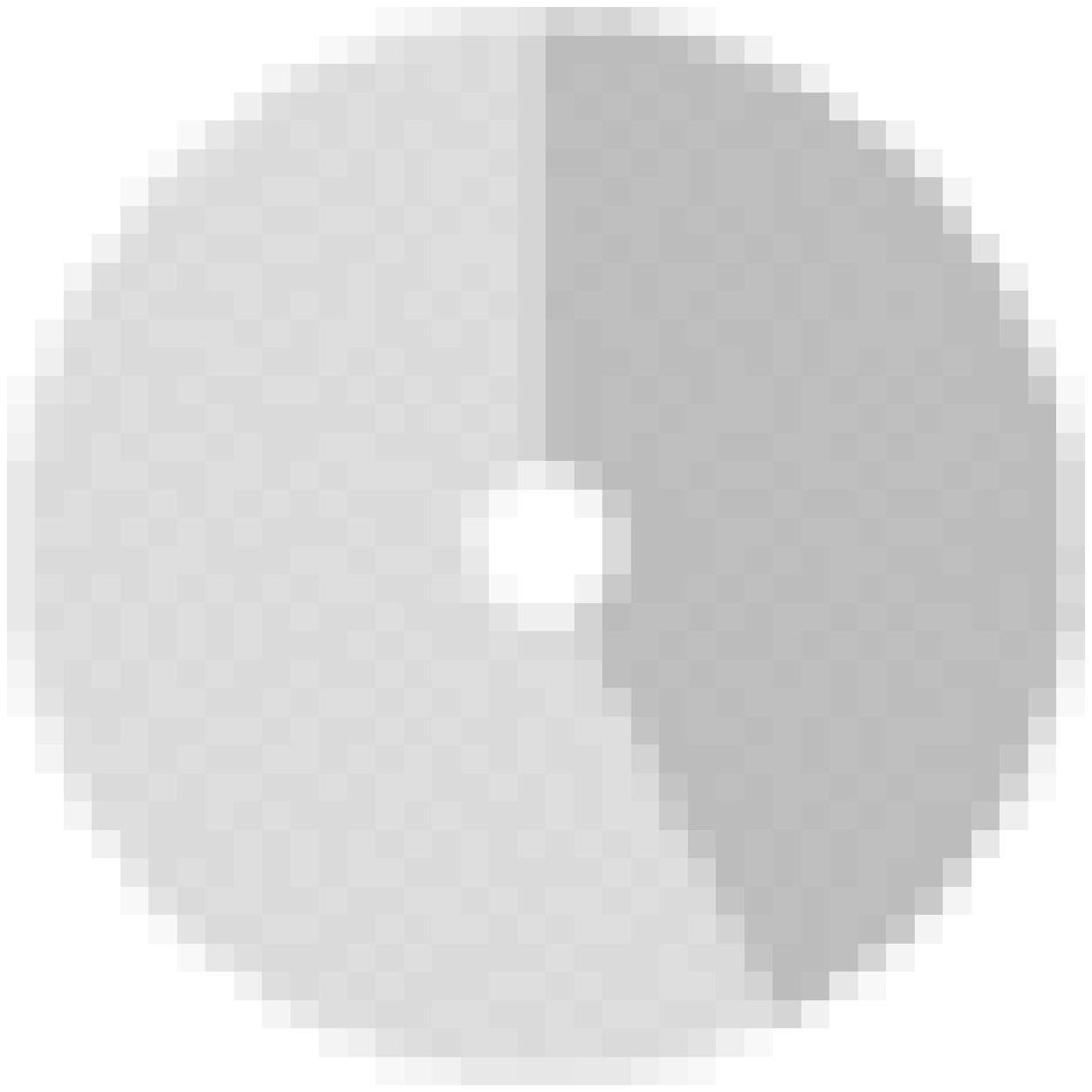}} \\
4. www.harding.edu/hr & 4 & 73/47 64\% & 19/19   100\% & 25/2   8\% & 29/26   90\% & (0.438, 0.356, 0.145) & \scalebox{0.05}{\includegraphics{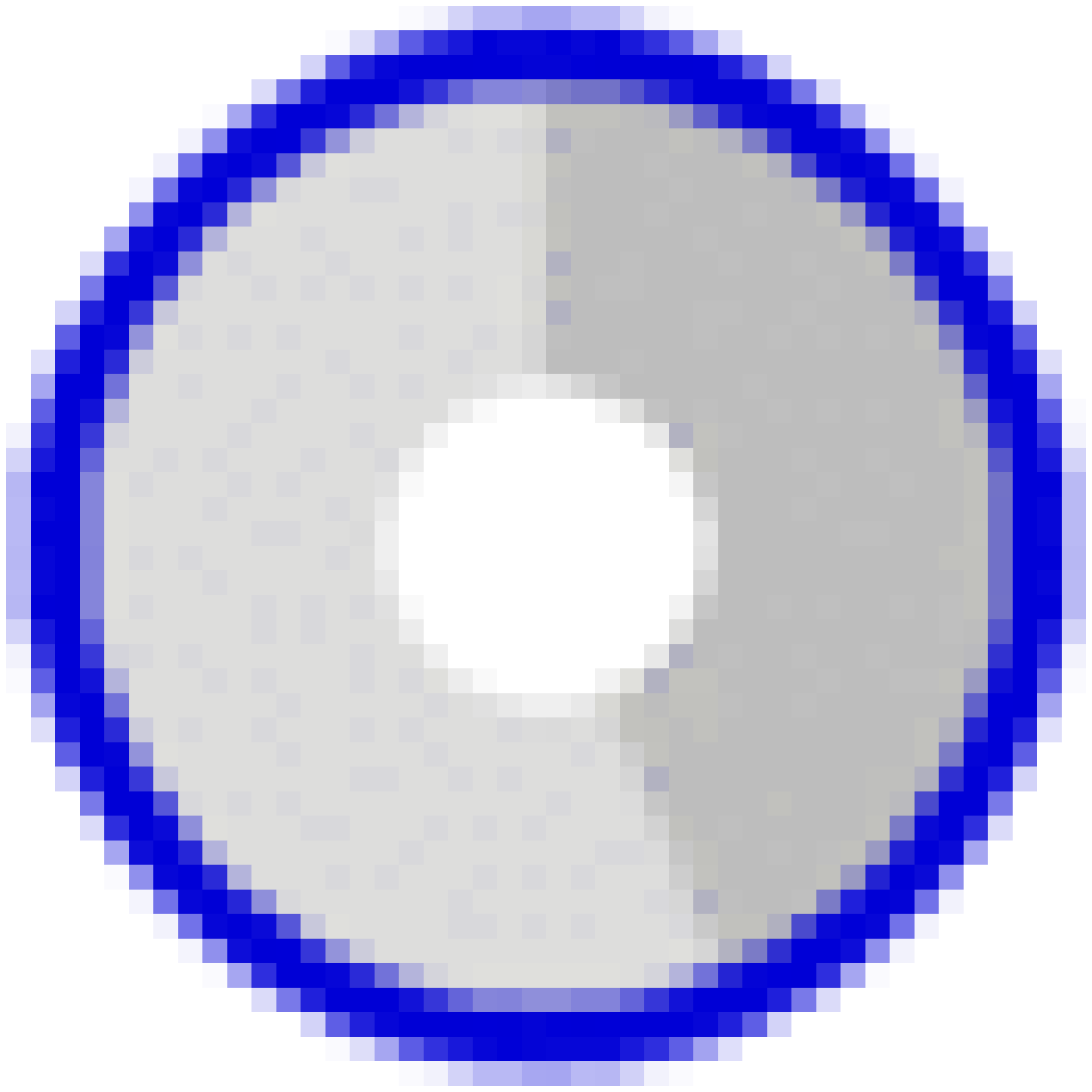}} & 83\% & \scalebox{0.05}{\includegraphics{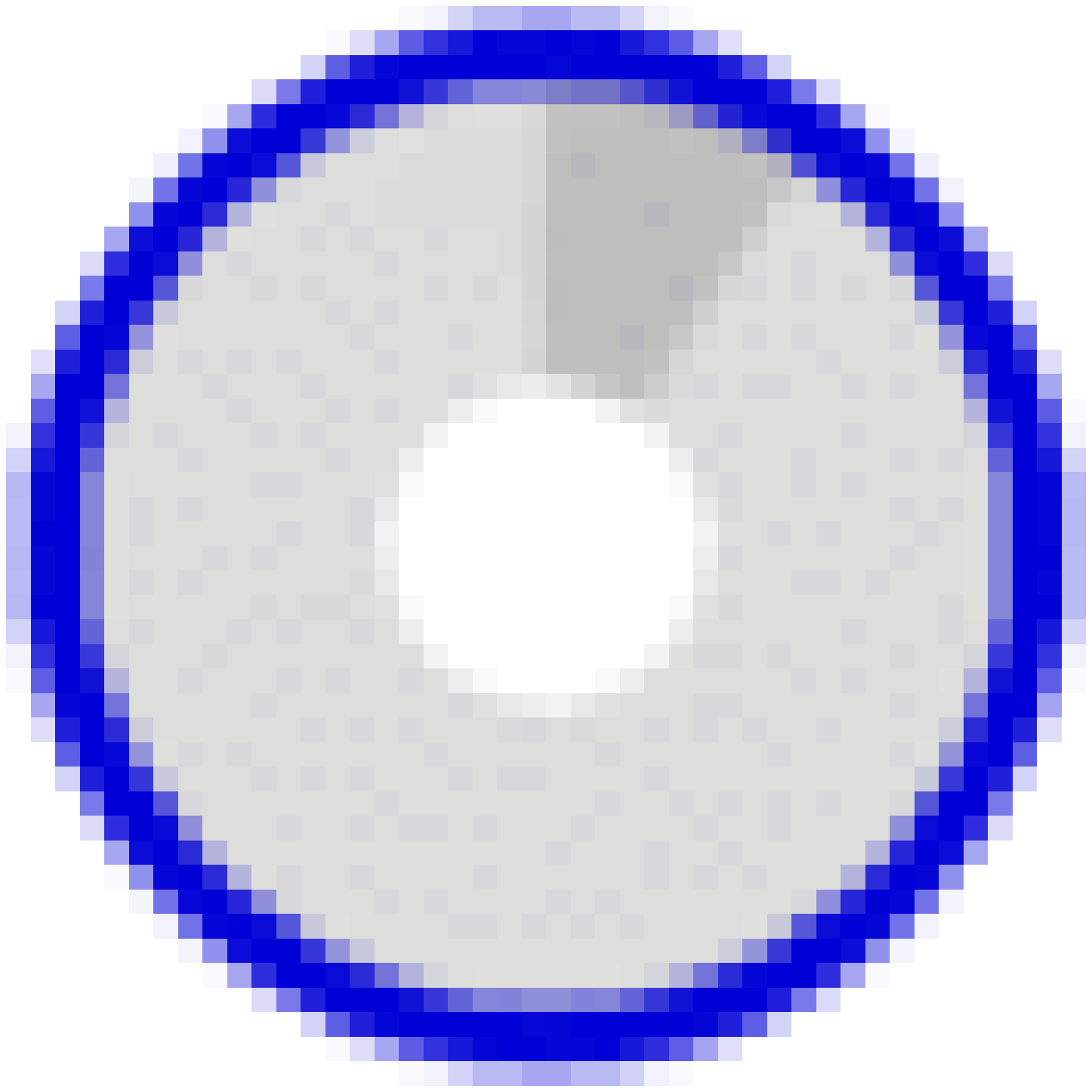}} \\
5. www.raitinvestmenttrust.com & 4 & 79/65 82\% & 24/24   100\% & 45/33   73\% & 10/8   80\% & (0.089, 0.177, 0.015) & \scalebox{0.05}{\includegraphics{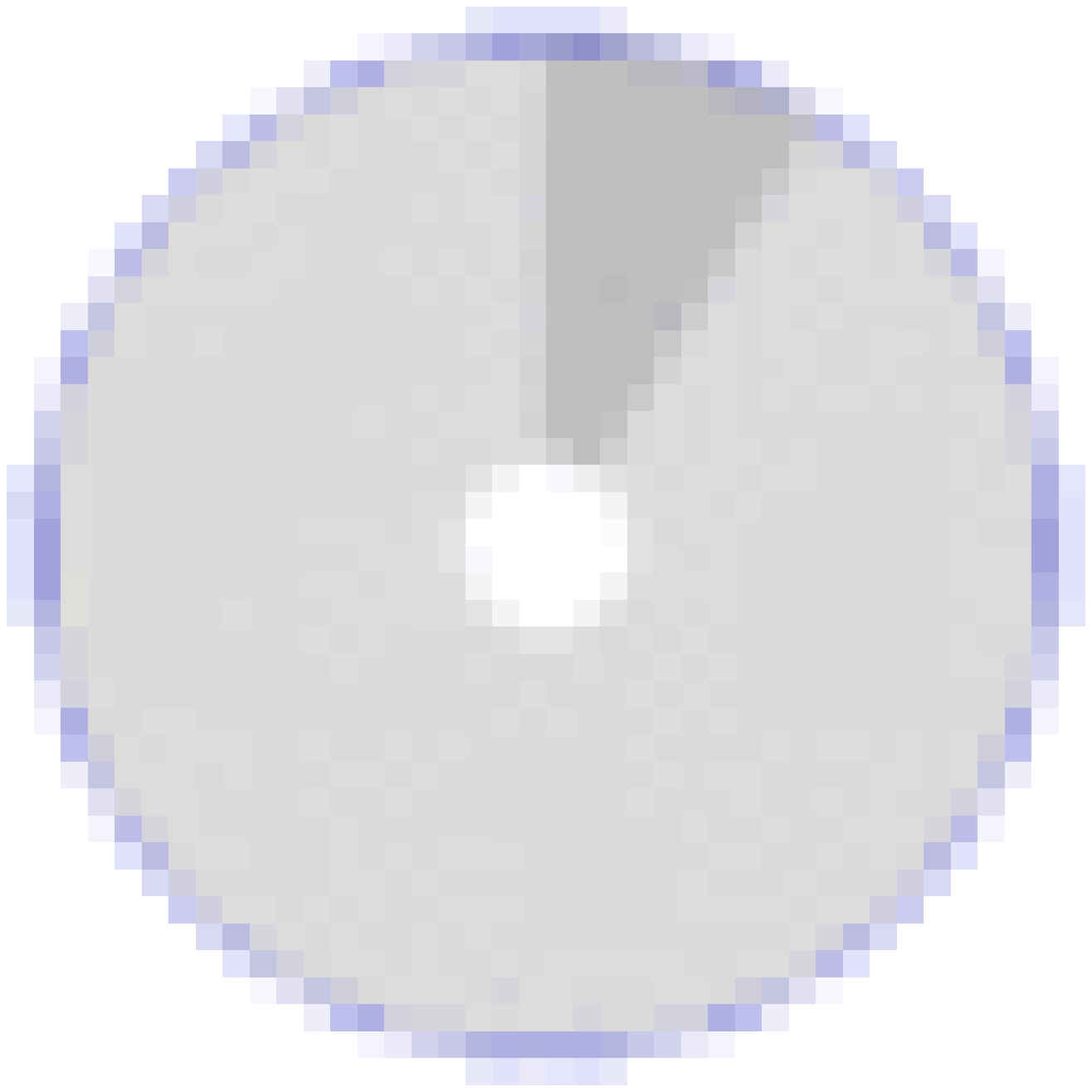}} & 33\% & \scalebox{0.05}{\includegraphics{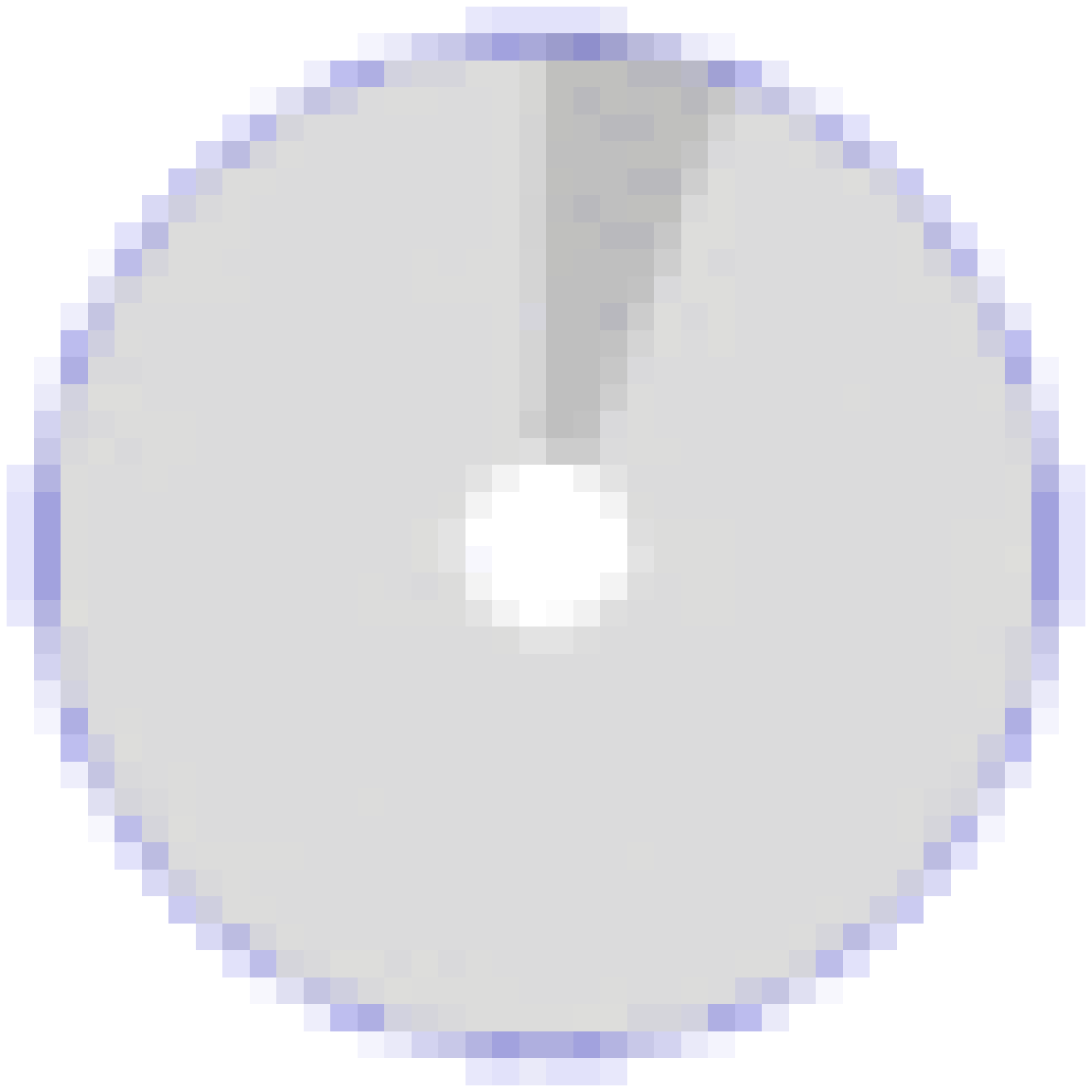}} \\
6. www.mie2005.net & 6 & 89/66 74\% & 16/15   94\% & 28/7   25\% & 45/44   98\% & (0.663, 0.258, 0.015) & \scalebox{0.05}{\includegraphics{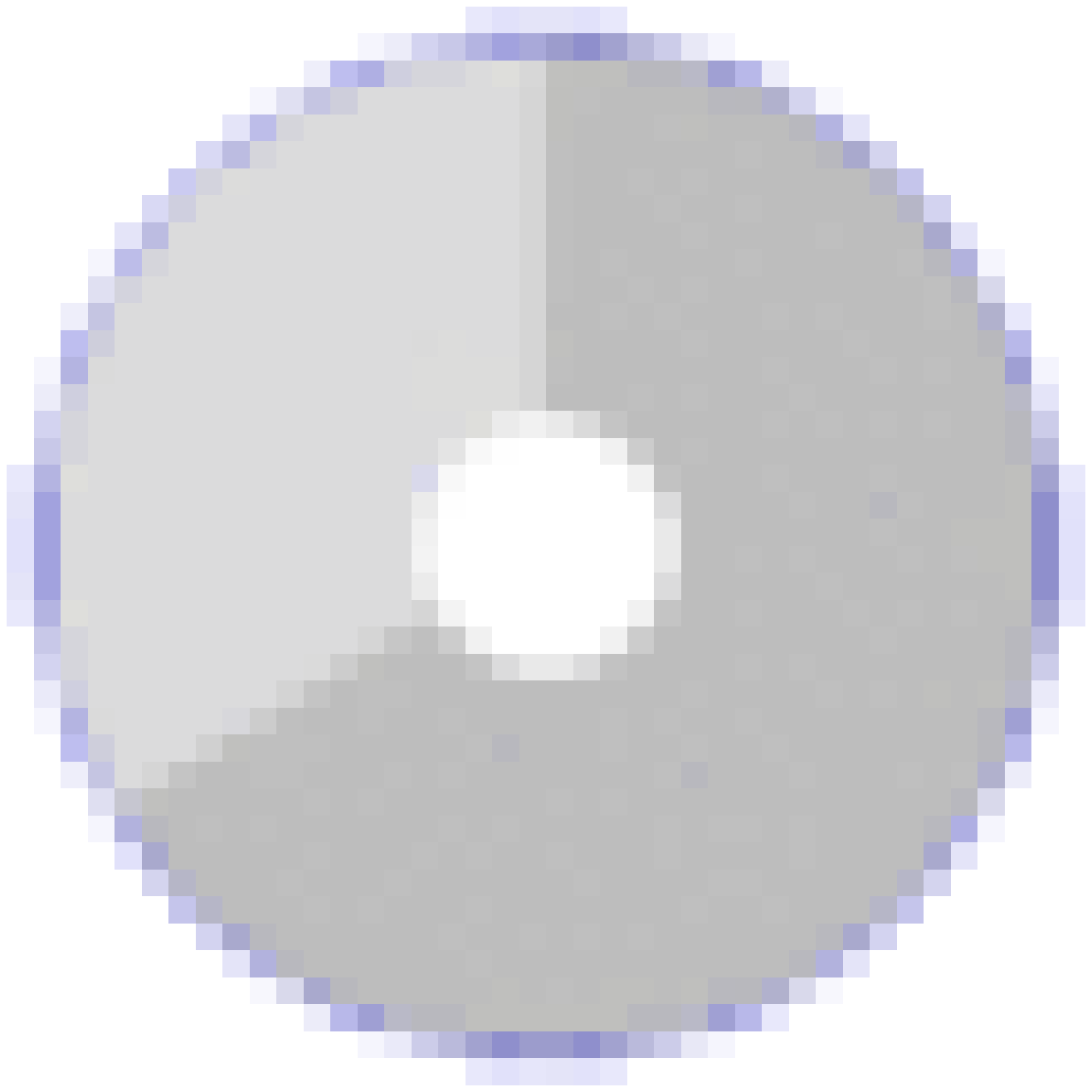}} & 89\% & \scalebox{0.05}{\includegraphics{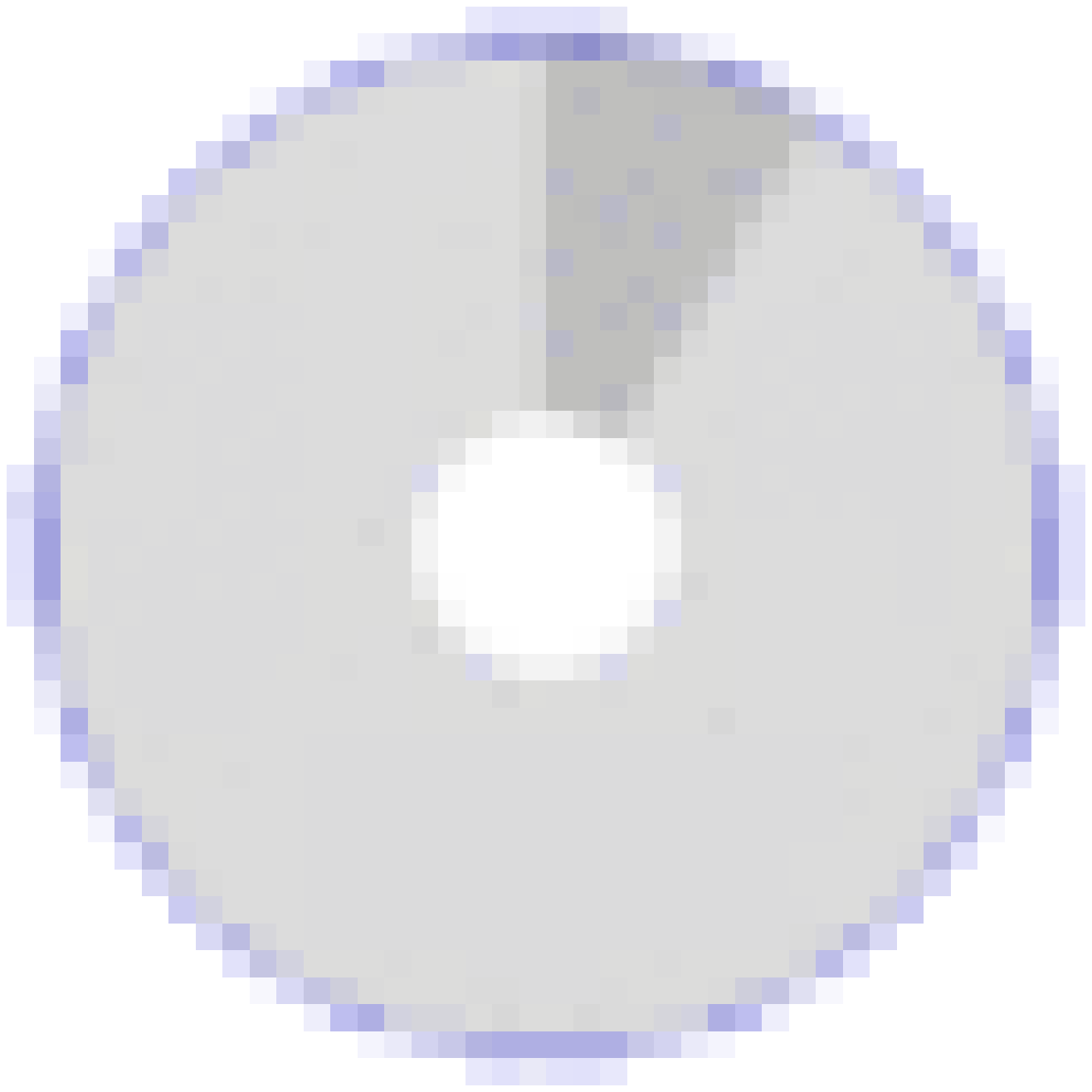}} \\
7. otago.settlers.museum & 5 & 111/48 43\% & 27/27   100\% & 82/19   23\% & 2/2   100\% & (0.171, 0.568, 0.020) & \scalebox{0.05}{\includegraphics{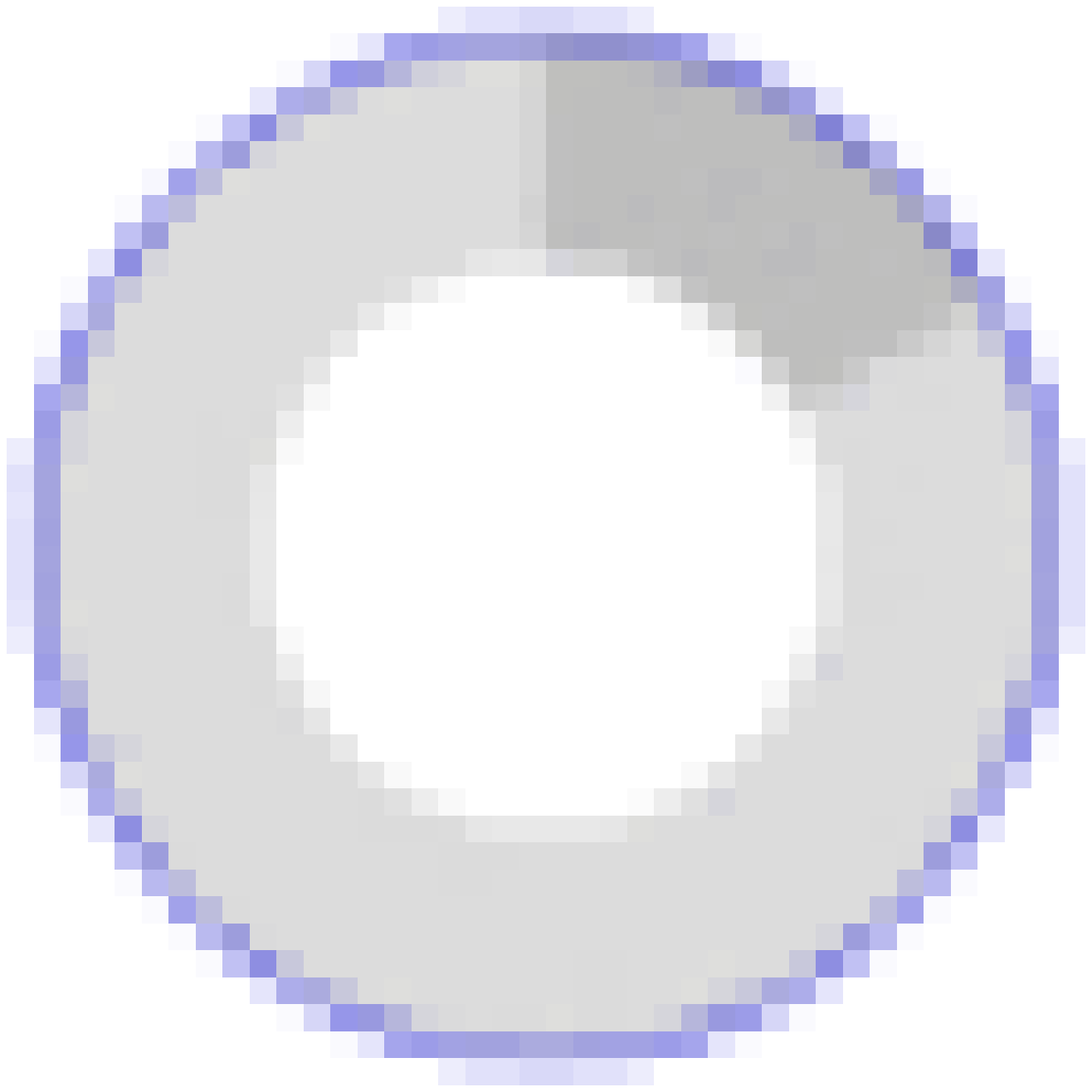}} & 40\% & \scalebox{0.05}{\includegraphics{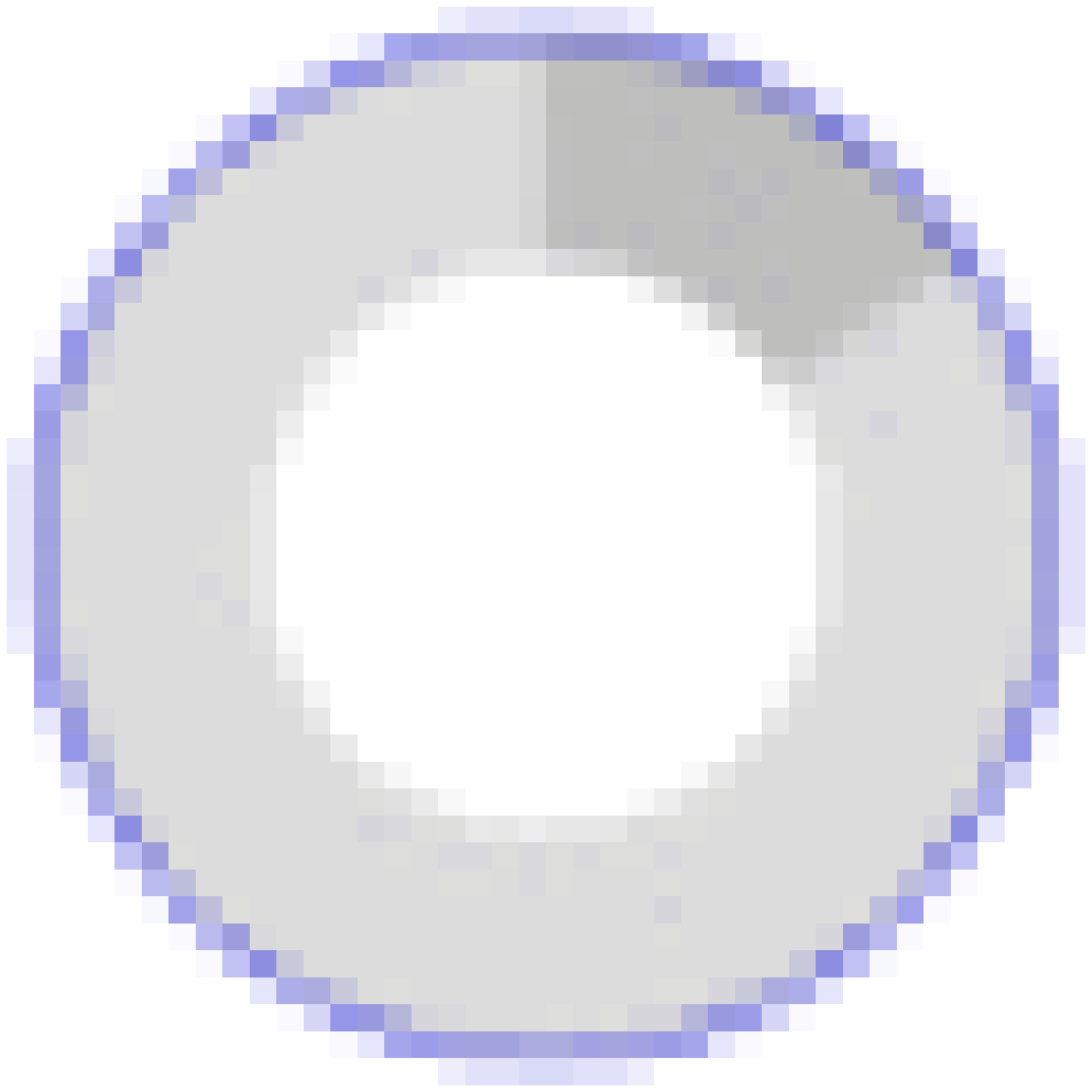}} \\
8. www.usamriid.army.mil & 7 & 142/100 70\% & 38/38   100\% & 59/19   32\% & 45/43   96\% & (0.585, 0.296, 0.000) & \scalebox{0.05}{\includegraphics{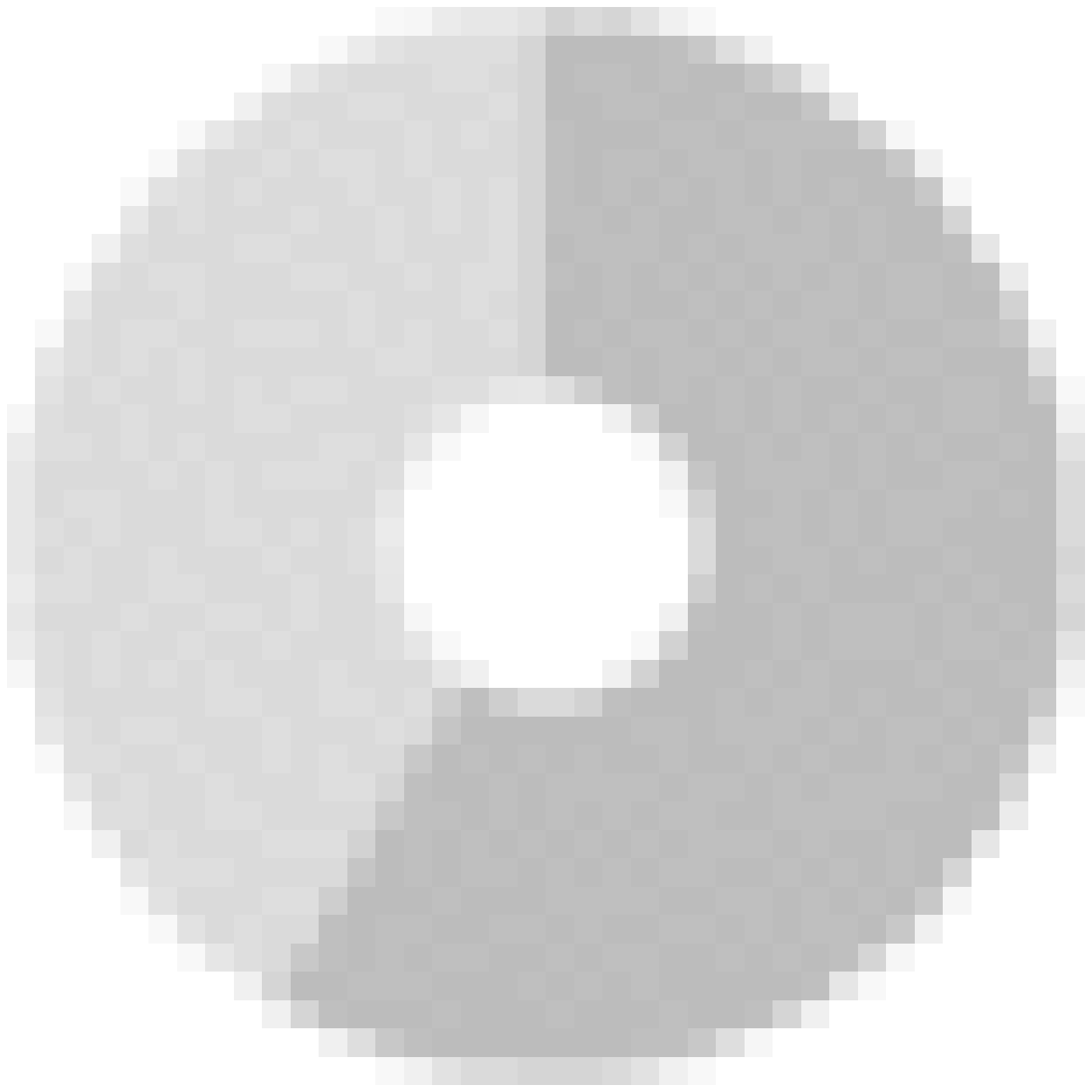}} & 50\% & \scalebox{0.05}{\includegraphics{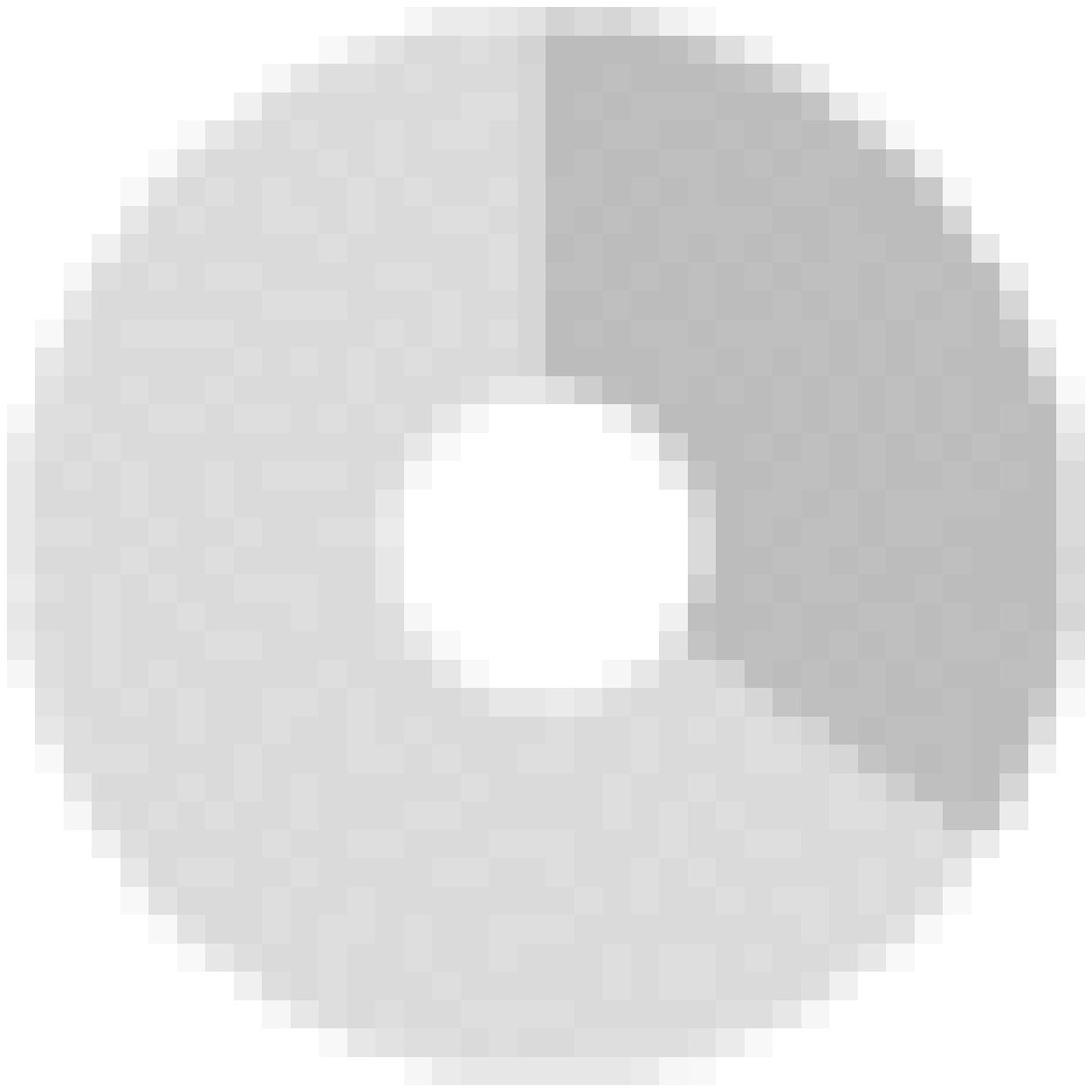}} \\
9. searcy.dina.org & 5 & 162/154 95\% & 96/95   99\% & 63/56   89\% & 3/3   100\% & (0.111, 0.049, 0.078) & \scalebox{0.05}{\includegraphics{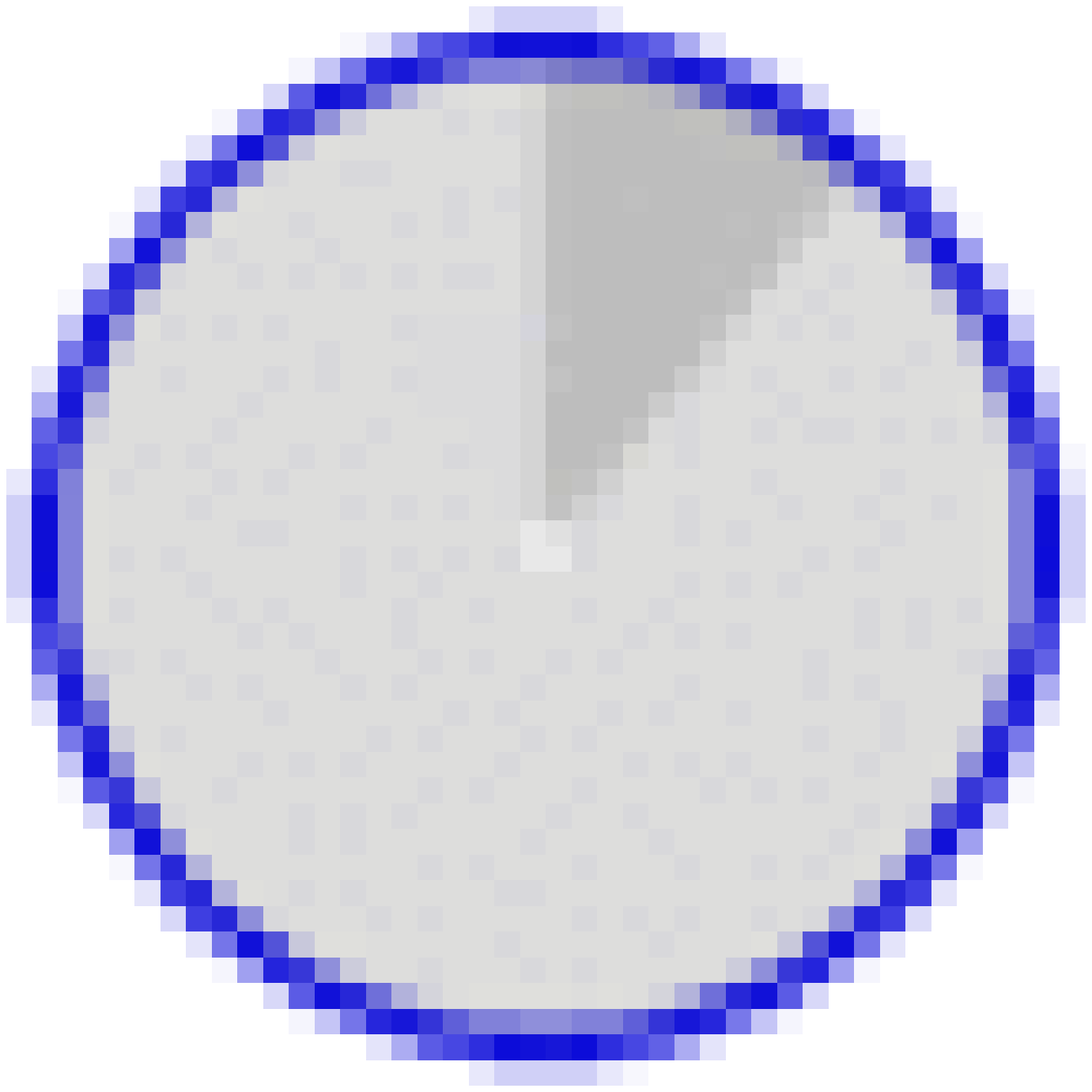}} & 43\% & \scalebox{0.05}{\includegraphics{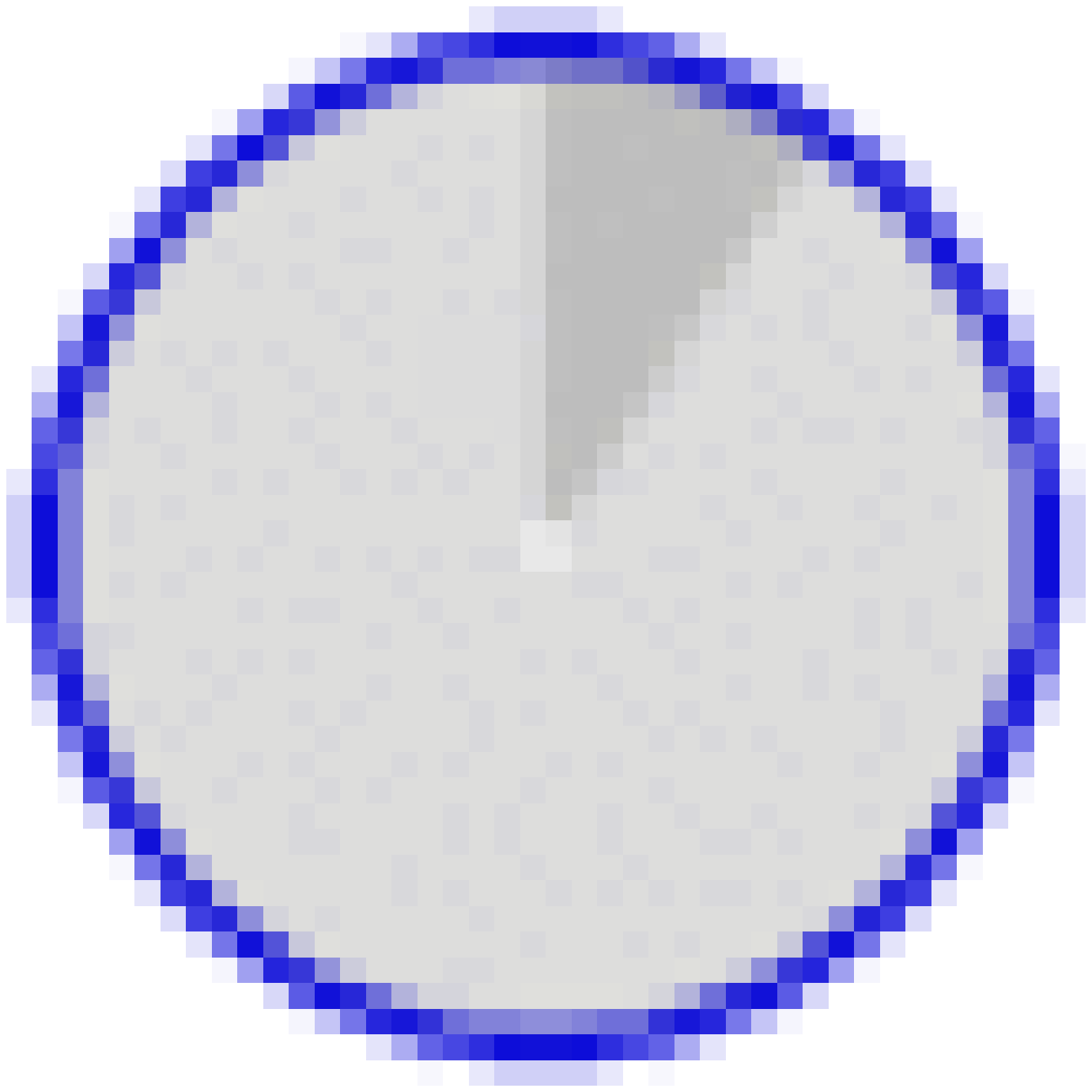}} \\
10. www.cookinclub.com & 6 & 204/187 92\% & 67/66   99\% & 136/121   89\% & 1/0 \;  0\% & (0.480, 0.083, 0.307) & \scalebox{0.05}{\includegraphics{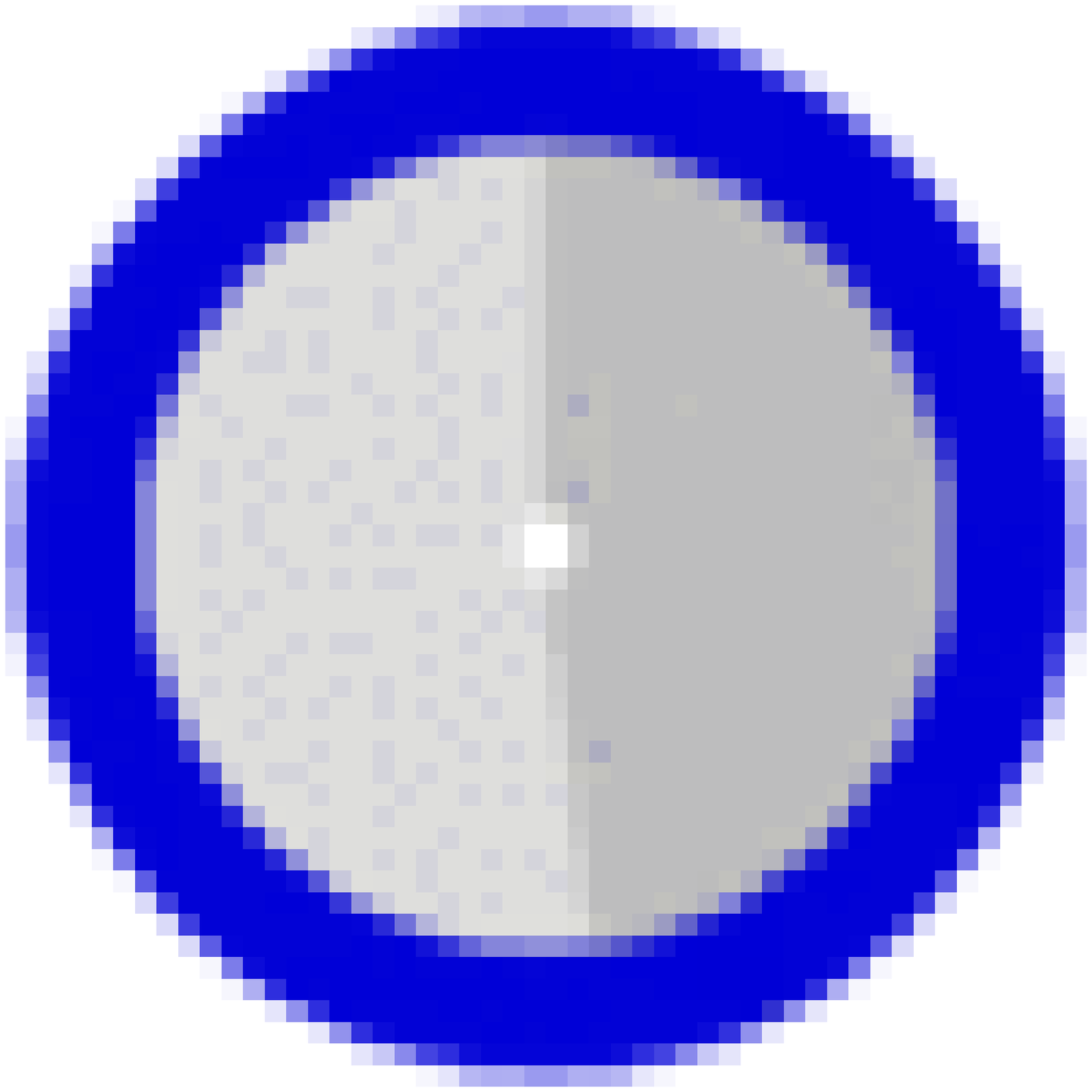}} & 100\% & \scalebox{0.05}{\includegraphics{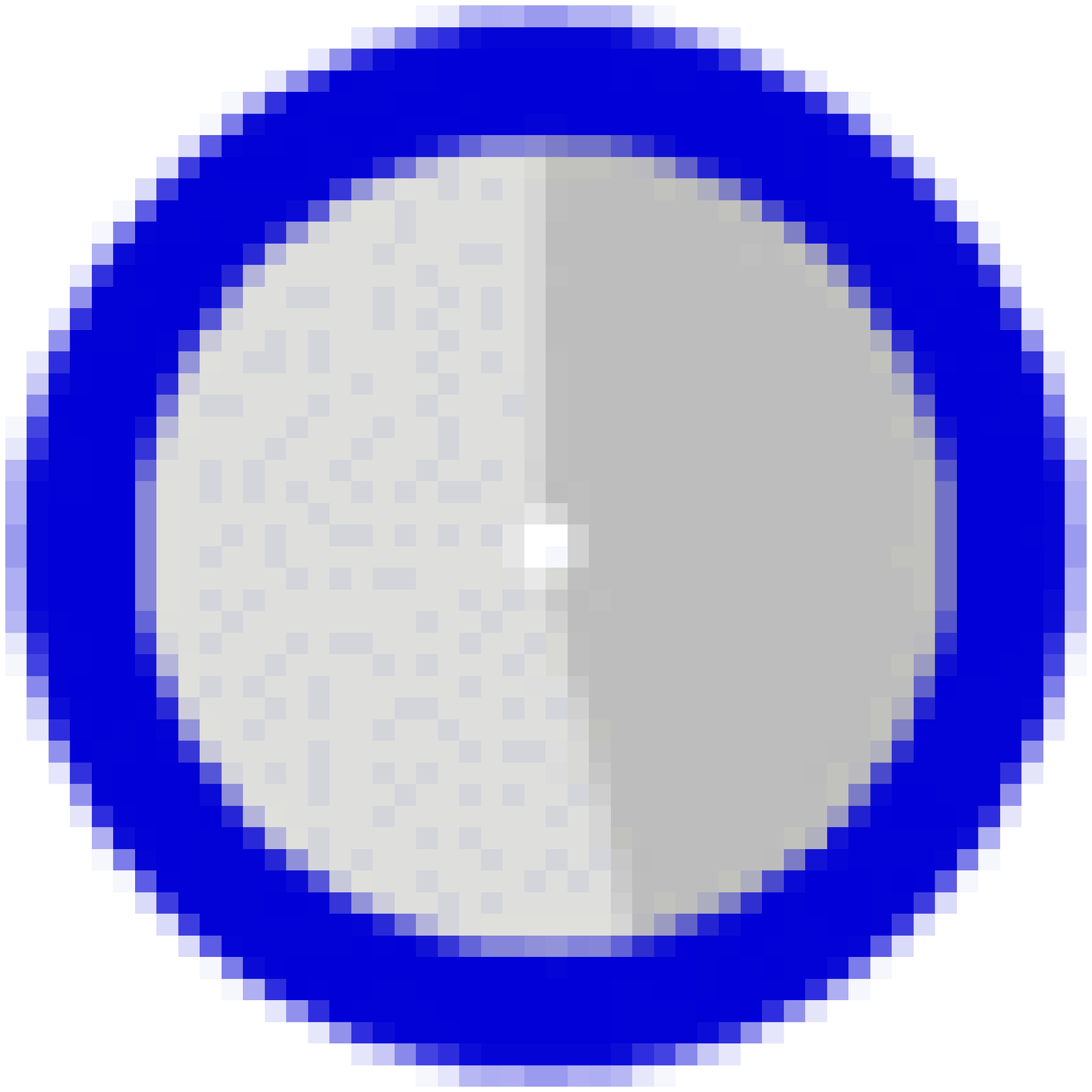}} \\
11. www.americancaribbean.com & 4 & 287/152 53\% & 60/57   95\% & 222/90   41\% & 5/5   100\% & (0.296, 0.470, 0.000) & \scalebox{0.05}{\includegraphics{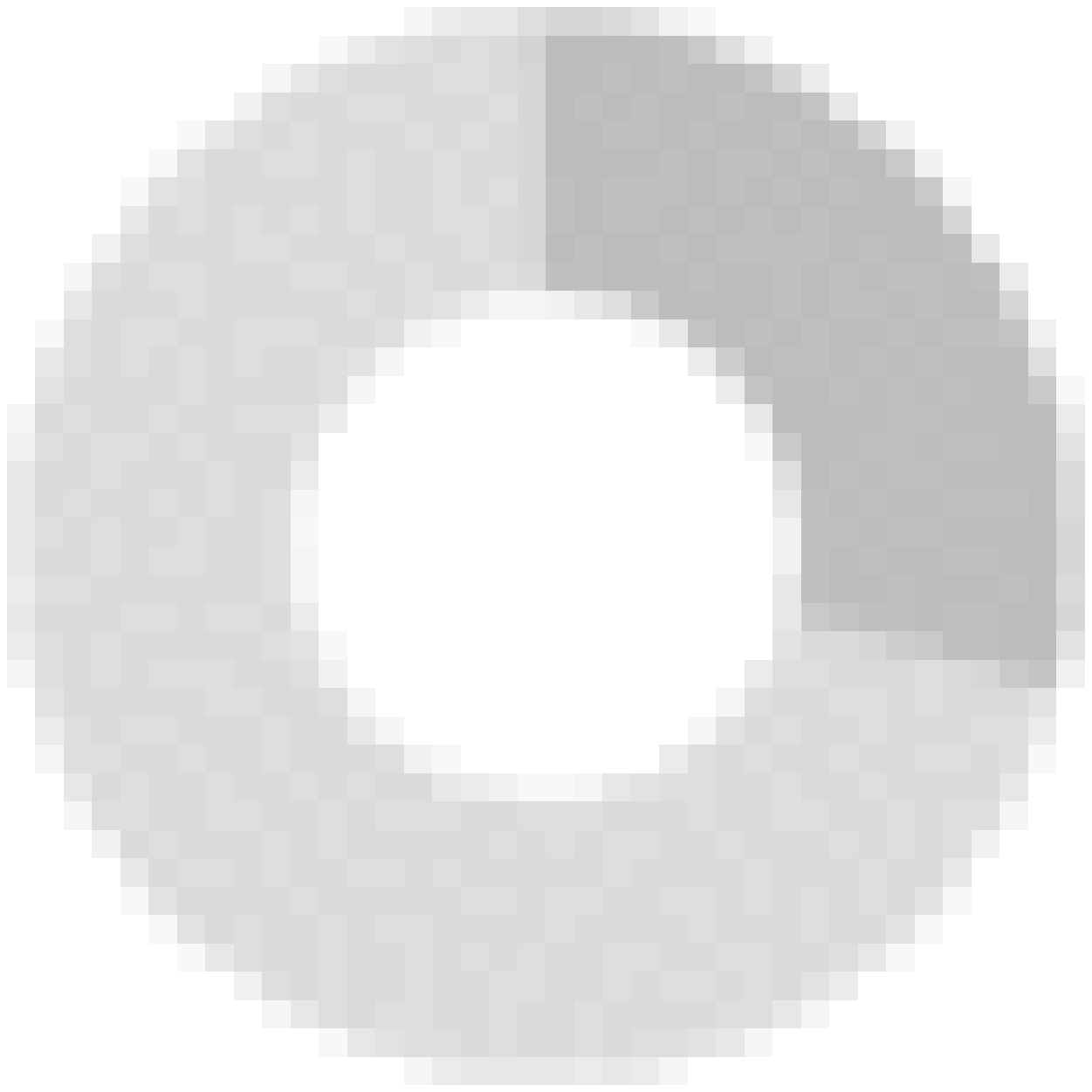}} & 100\% & \scalebox{0.05}{\includegraphics{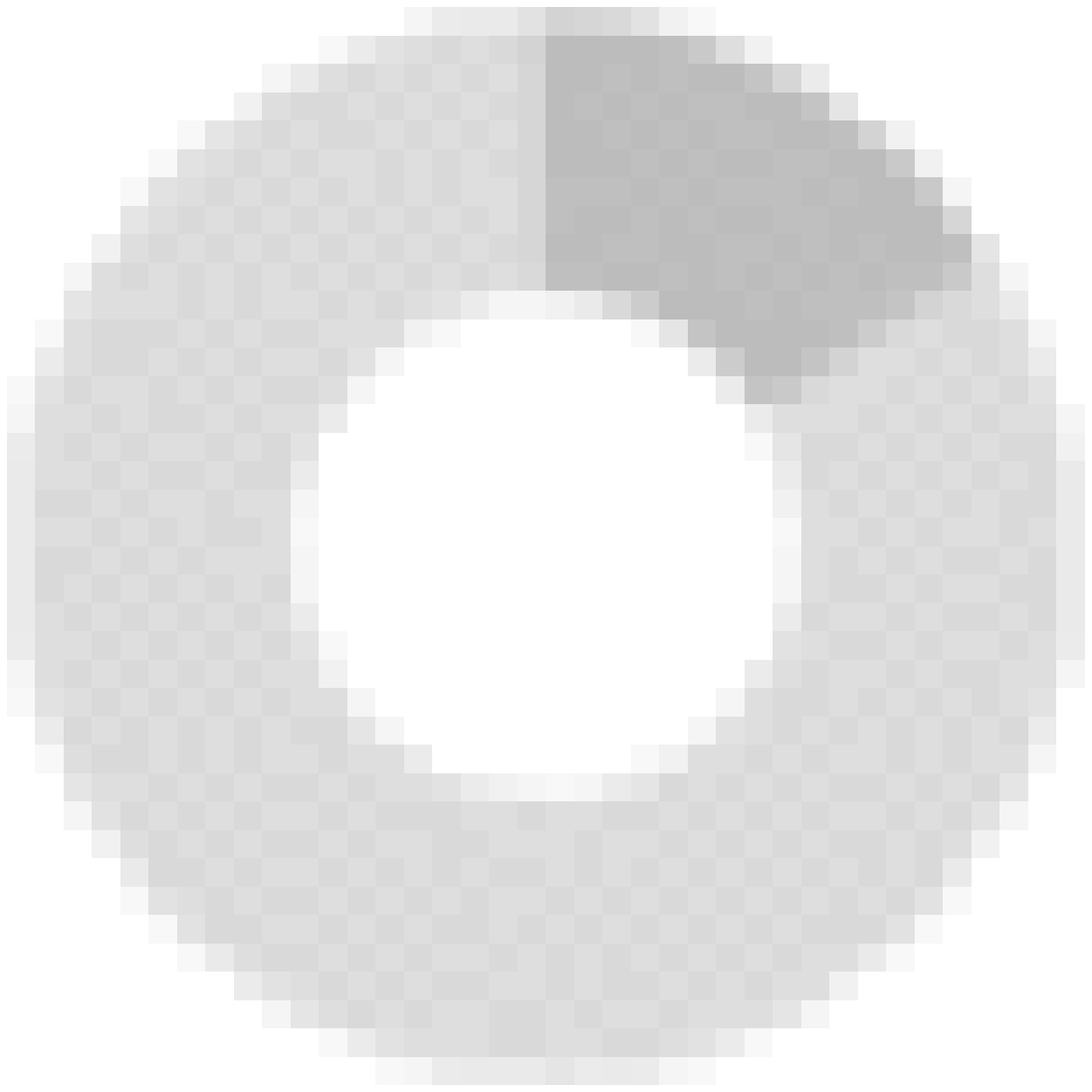}} \\
12. www.gltron.org & 6 & 294/221 75\% & 20/19   95\% & 213/190   89\% & 61/12   20\% & (0.259, 0.248, 0.005) & \scalebox{0.05}{\includegraphics{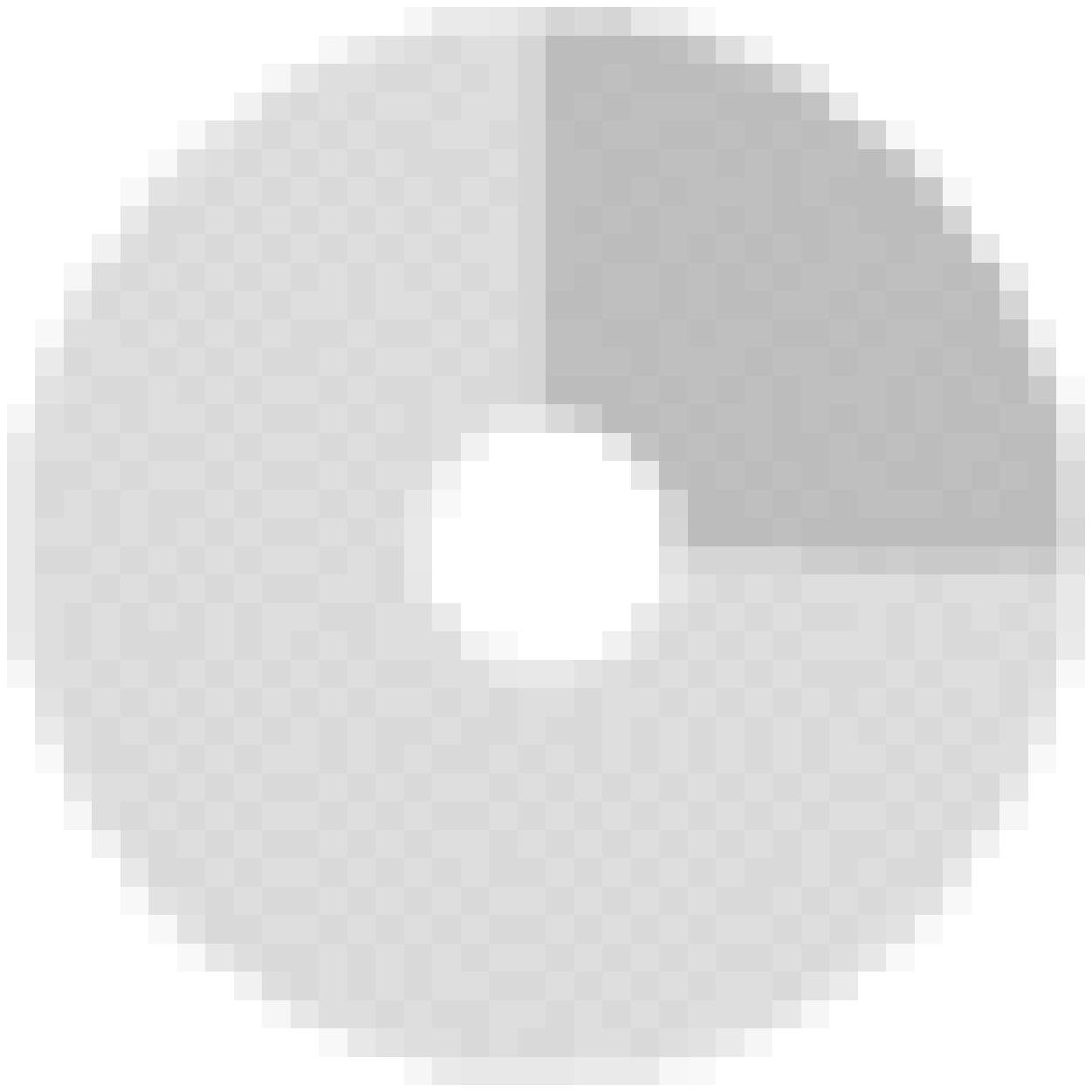}} & 90\% & \scalebox{0.05}{\includegraphics{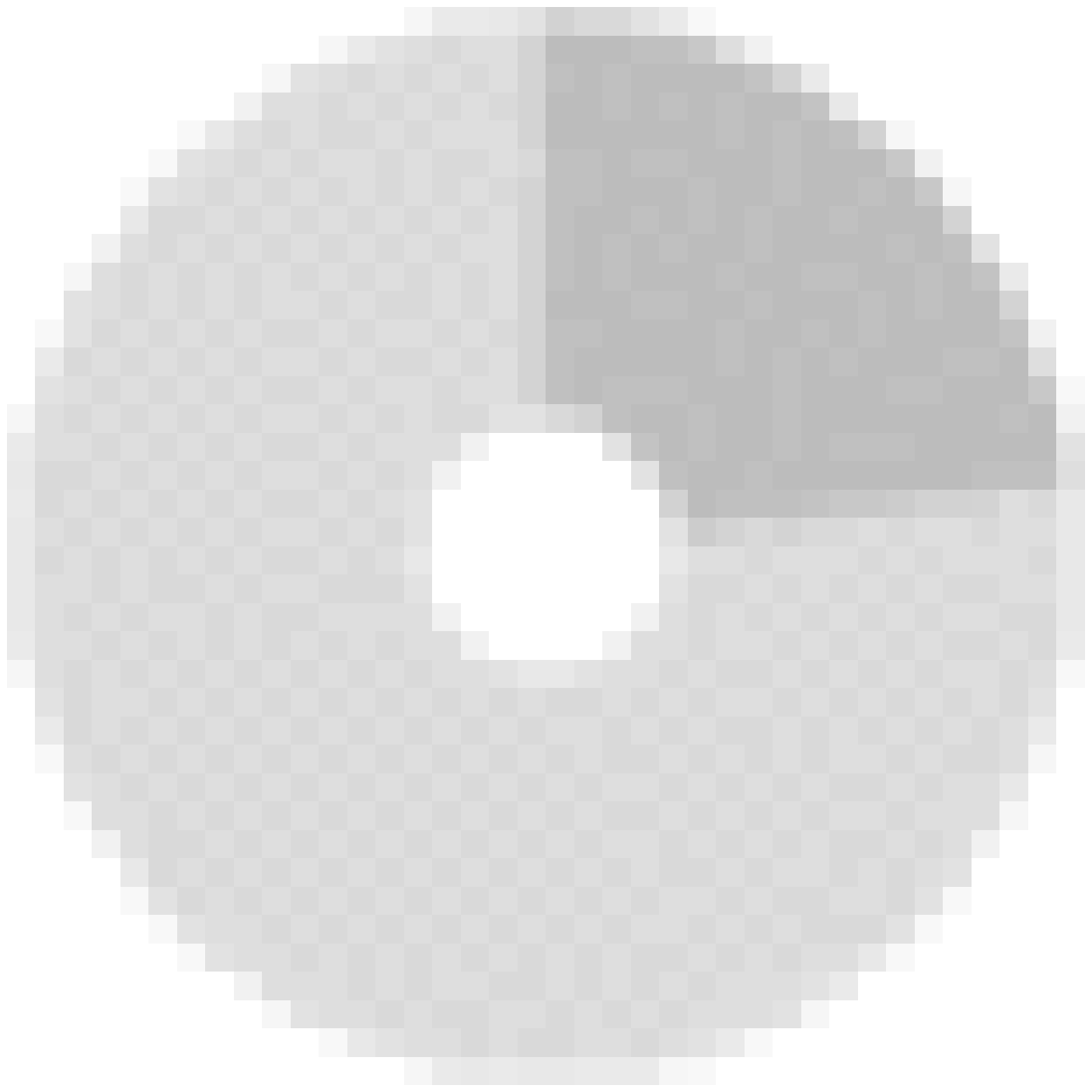}} \\
13. privacy.getnetwise.org & 8 & 305/163 53\% & 137/137   100\% & 48/25   52\% & 120/1   1\% & (0.033, 0.466, 0.201) & \scalebox{0.05}{\includegraphics{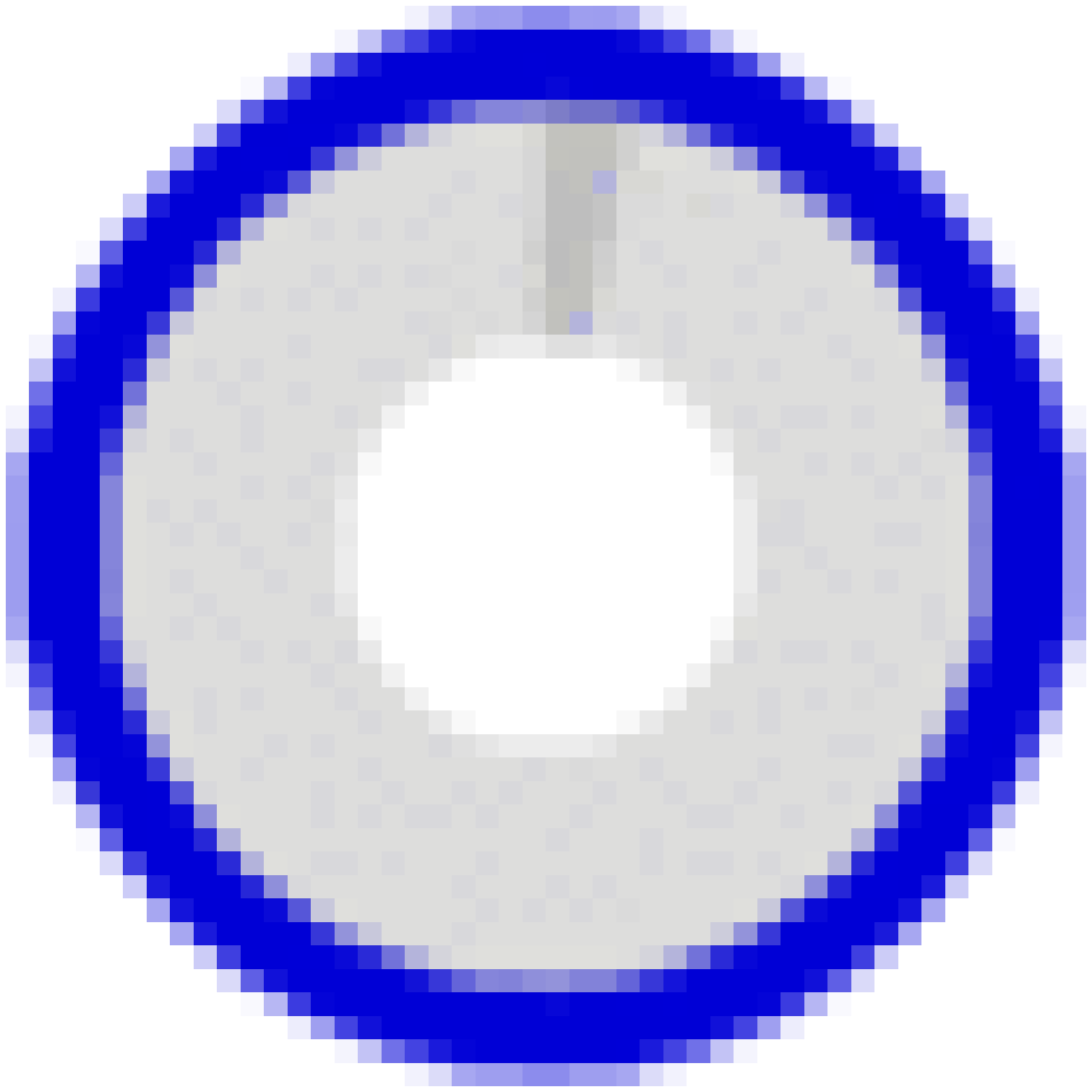}} & 70\% & \scalebox{0.05}{\includegraphics{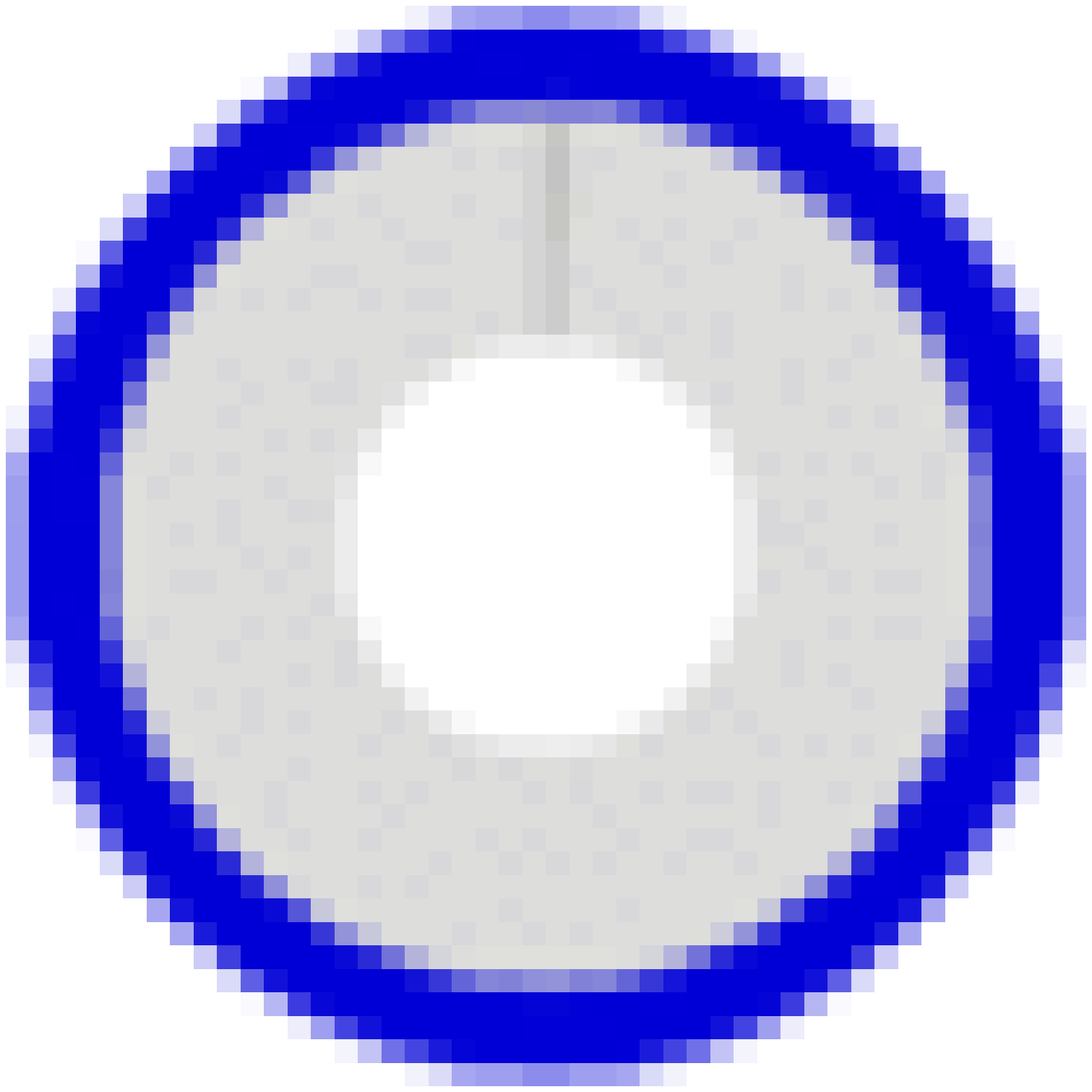}} \\
14. www.mypyramid.gov & 0 & 344/193 56\% & 158/154   97\% & 141/5   4\% & 45/34   76\% & (0.160, 0.439, 0.000) & \scalebox{0.05}{\includegraphics{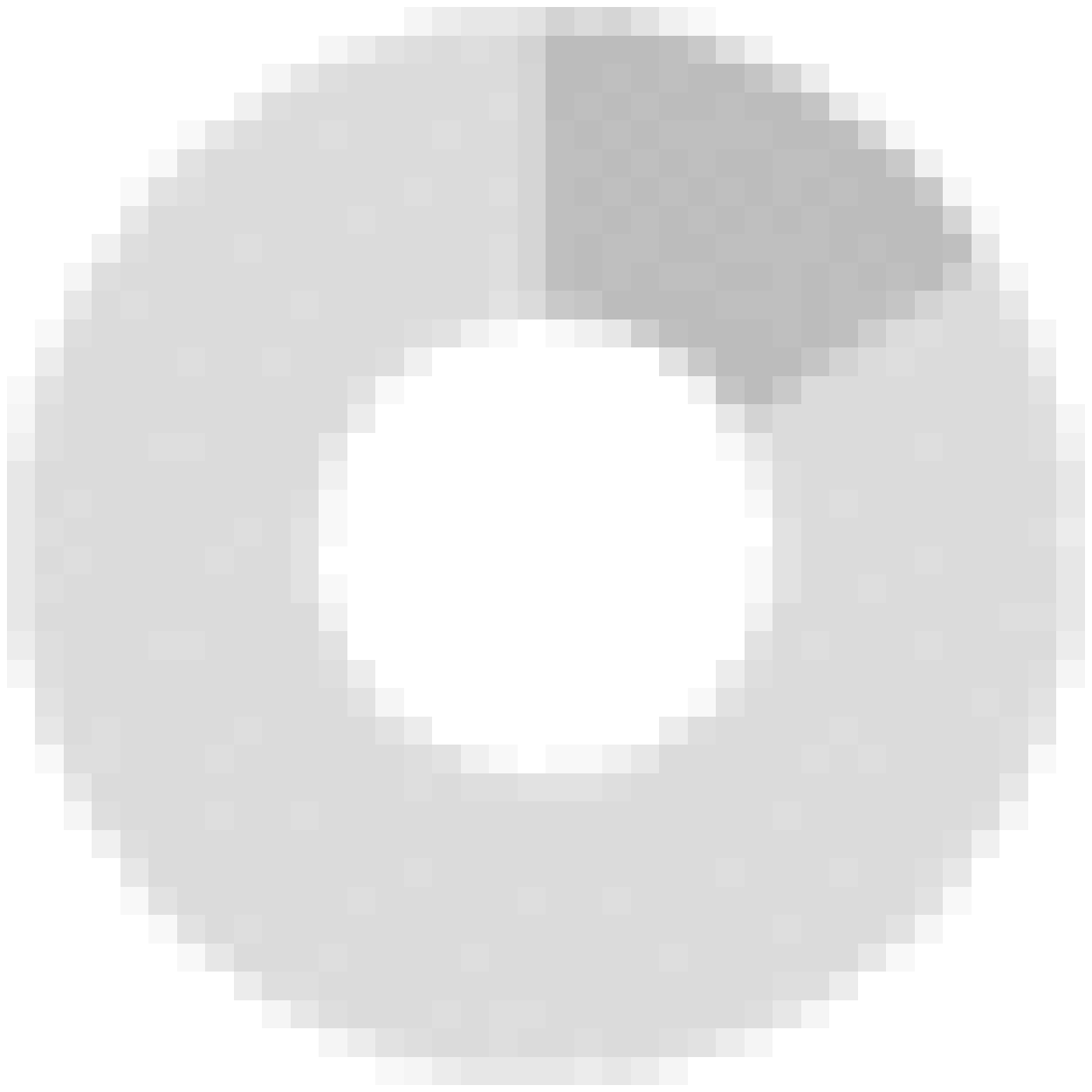}} & 32\% & \scalebox{0.05}{\includegraphics{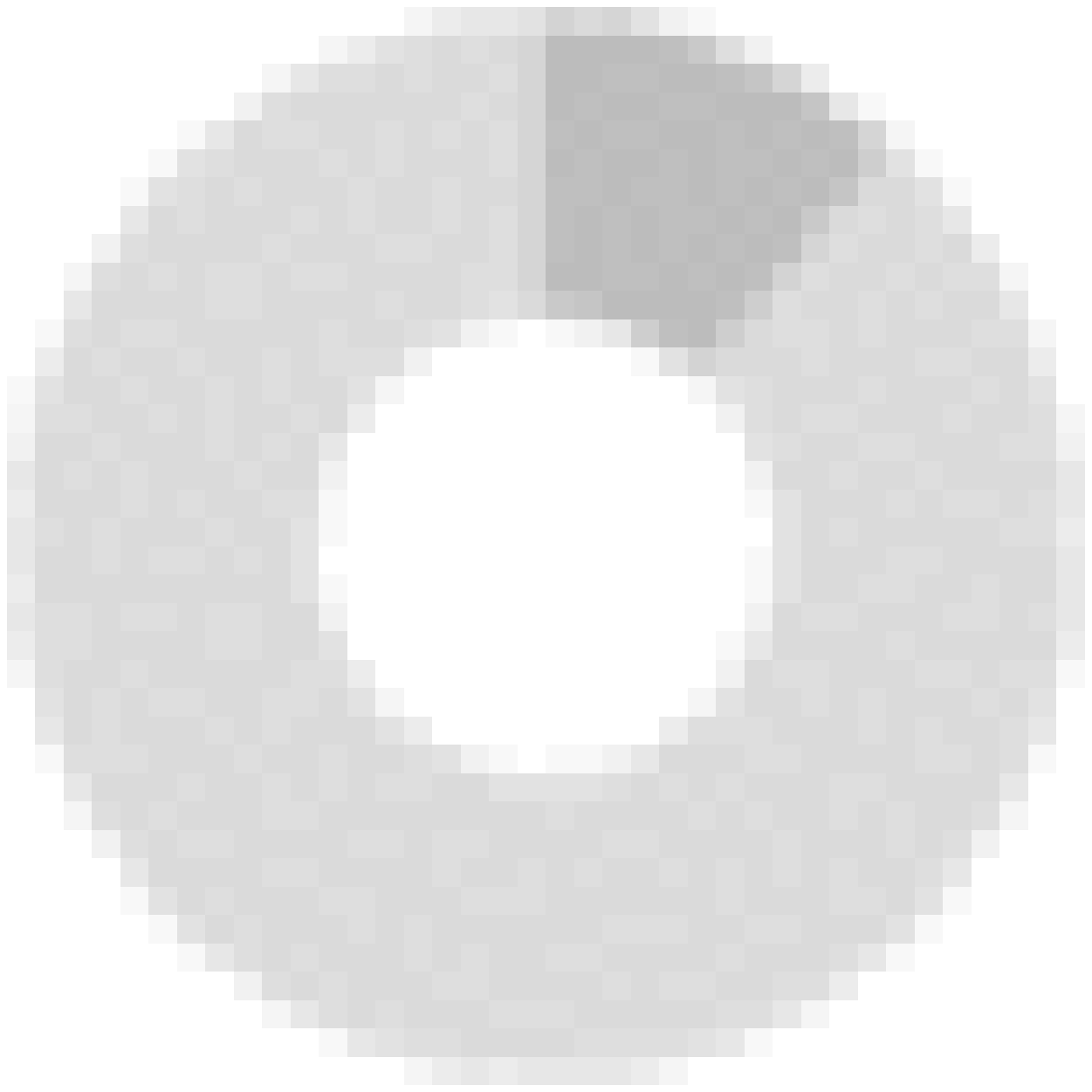}} \\
15. www.digitalpreservation.gov & 8 & 414/378 91\% & 346/329   95\% & 42/25   60\% & 26/24   92\% & (0.097, 0.087, 0.000) & \scalebox{0.05}{\includegraphics{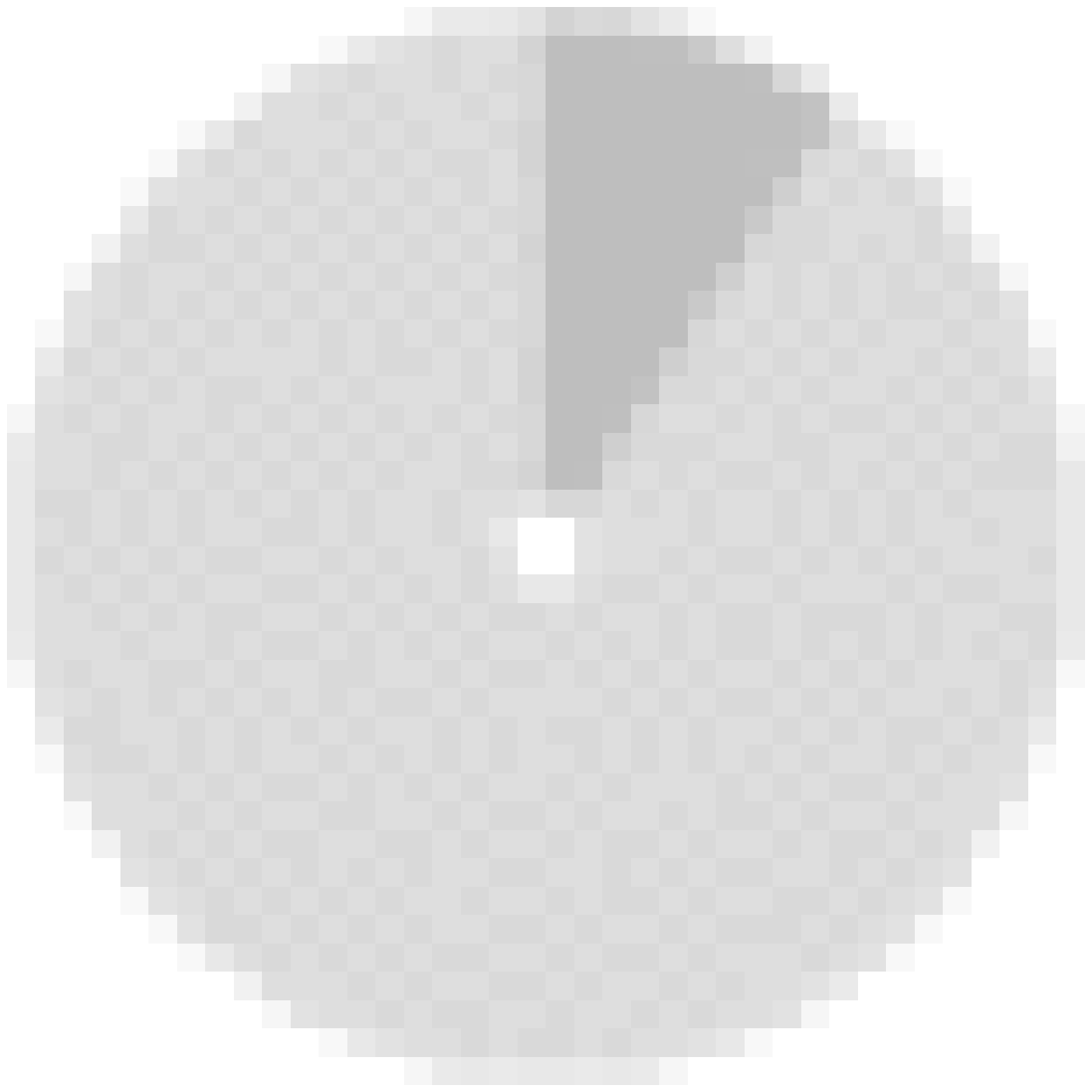}} & 44\% & \scalebox{0.05}{\includegraphics{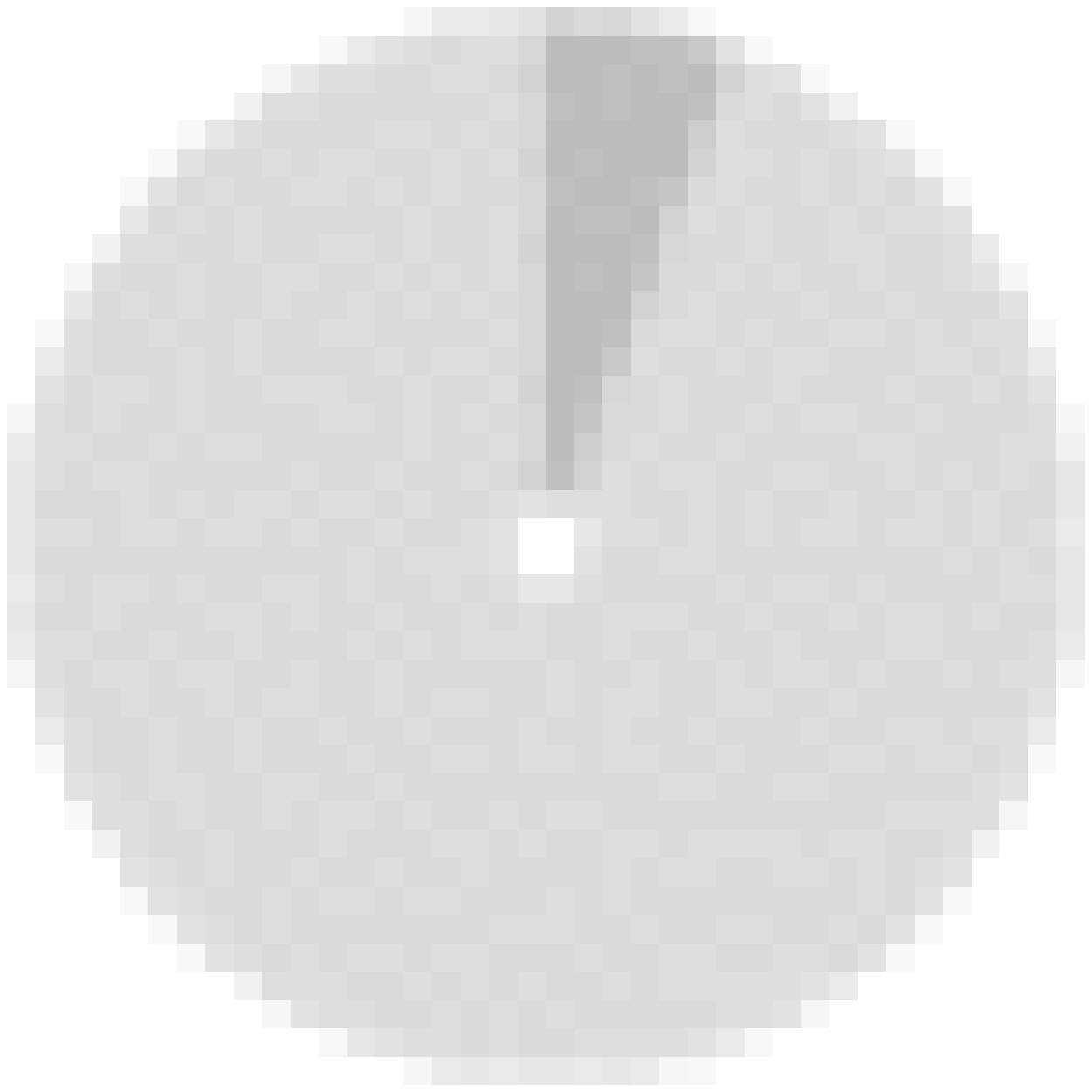}} \\
16. www.aboutfamouspeople.com & 6 & 432/430 99\% & 267/267   100\% & 165/163   99\% & 0/0    & (0.653, 0.005, 0.021) & \scalebox{0.05}{\includegraphics{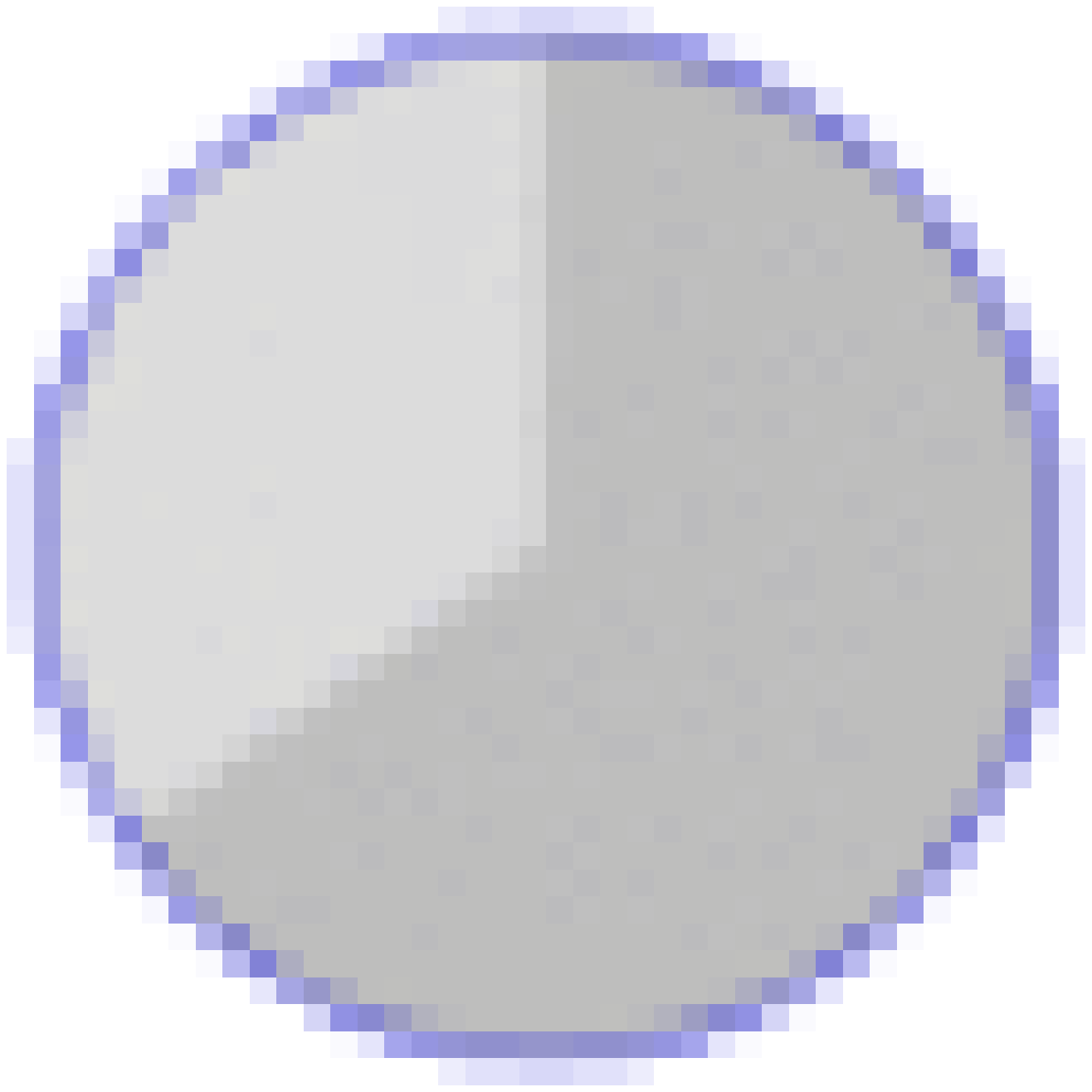}} & 100\% & \scalebox{0.05}{\includegraphics{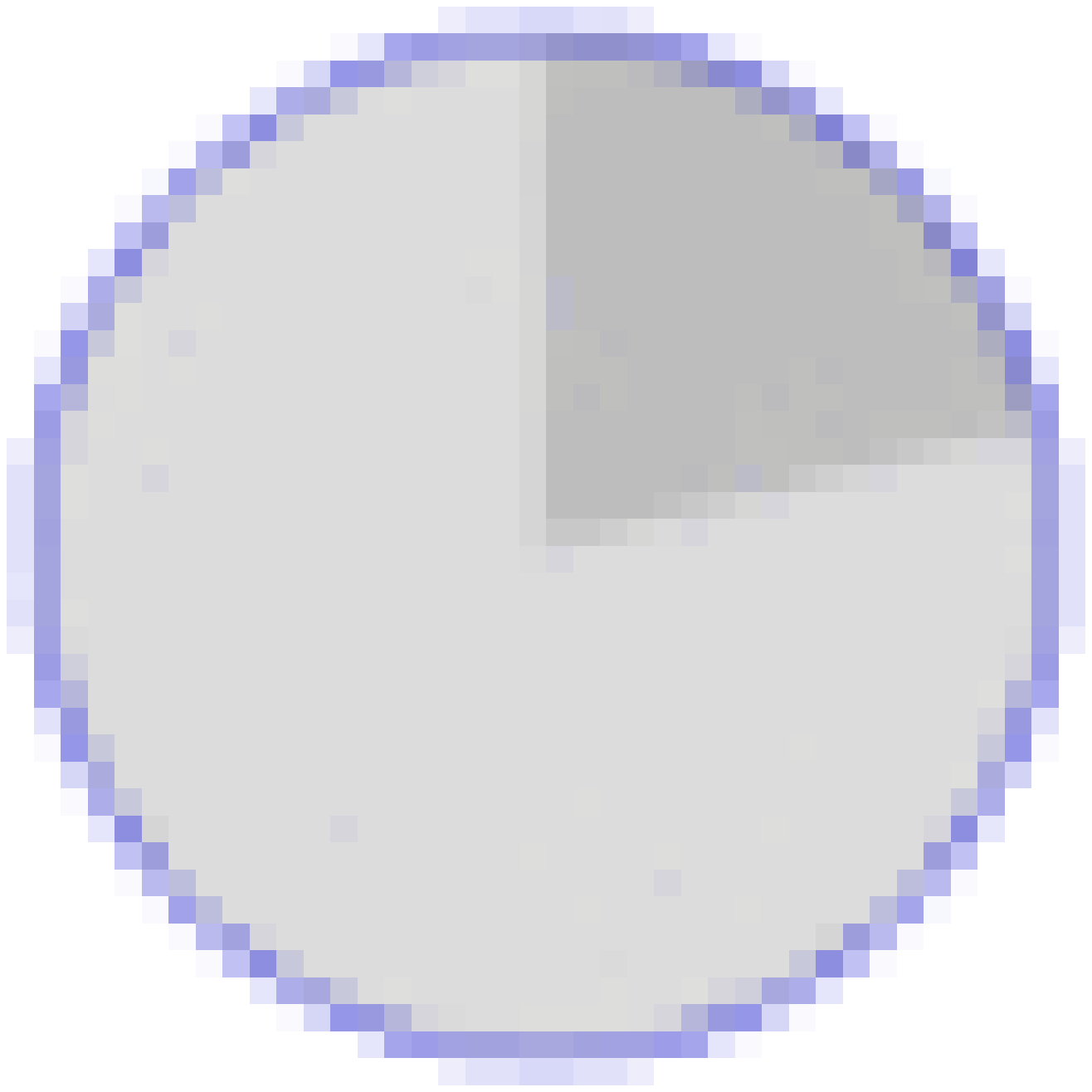}} \\
17. home.alltel.net/bsprowl & 0 & 505/112 22\% & 173/112   65\% & 332/0   0\% & 0/0    & (0.012, 0.778, 0.009) & \scalebox{0.05}{\includegraphics{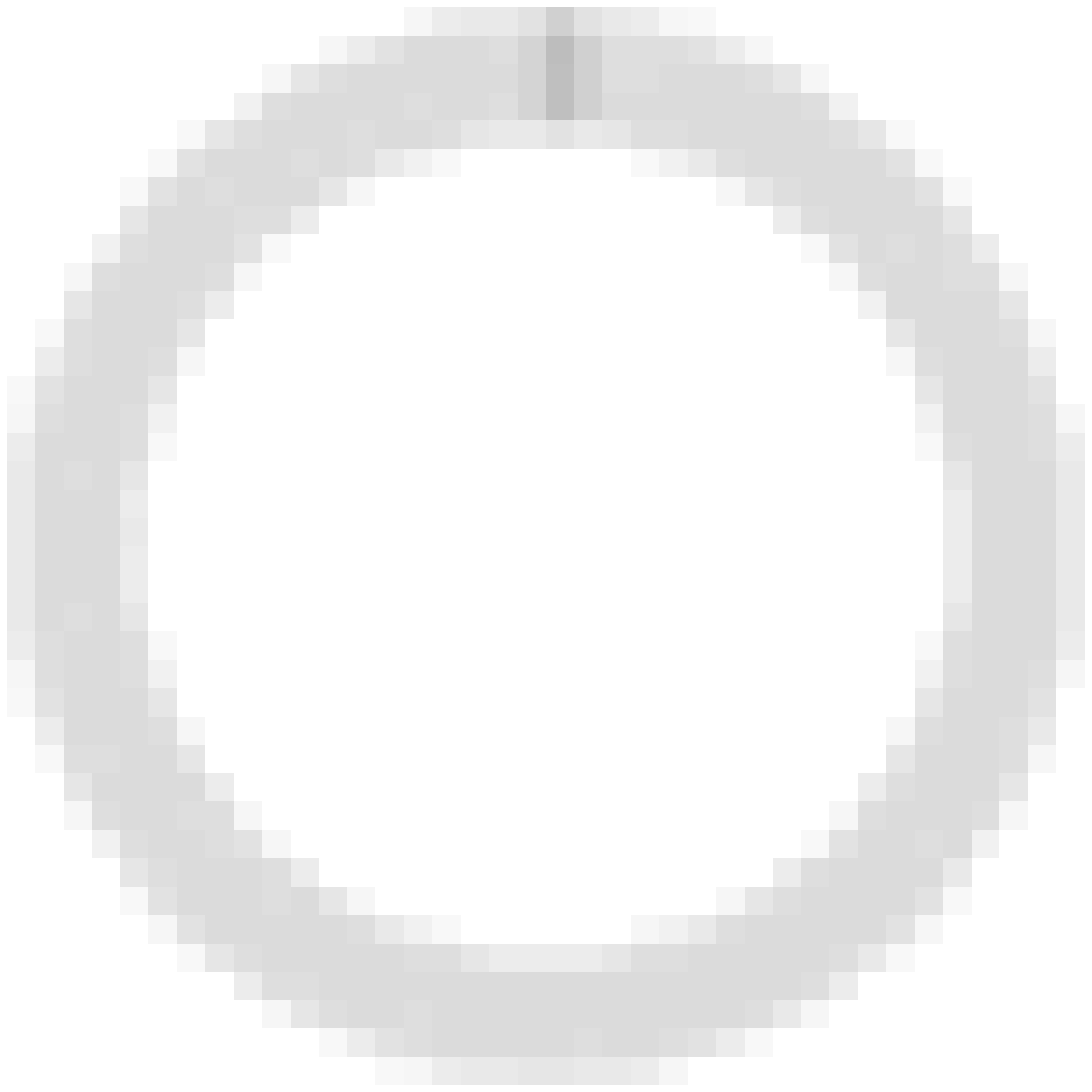}} & 100\% & \scalebox{0.05}{\includegraphics{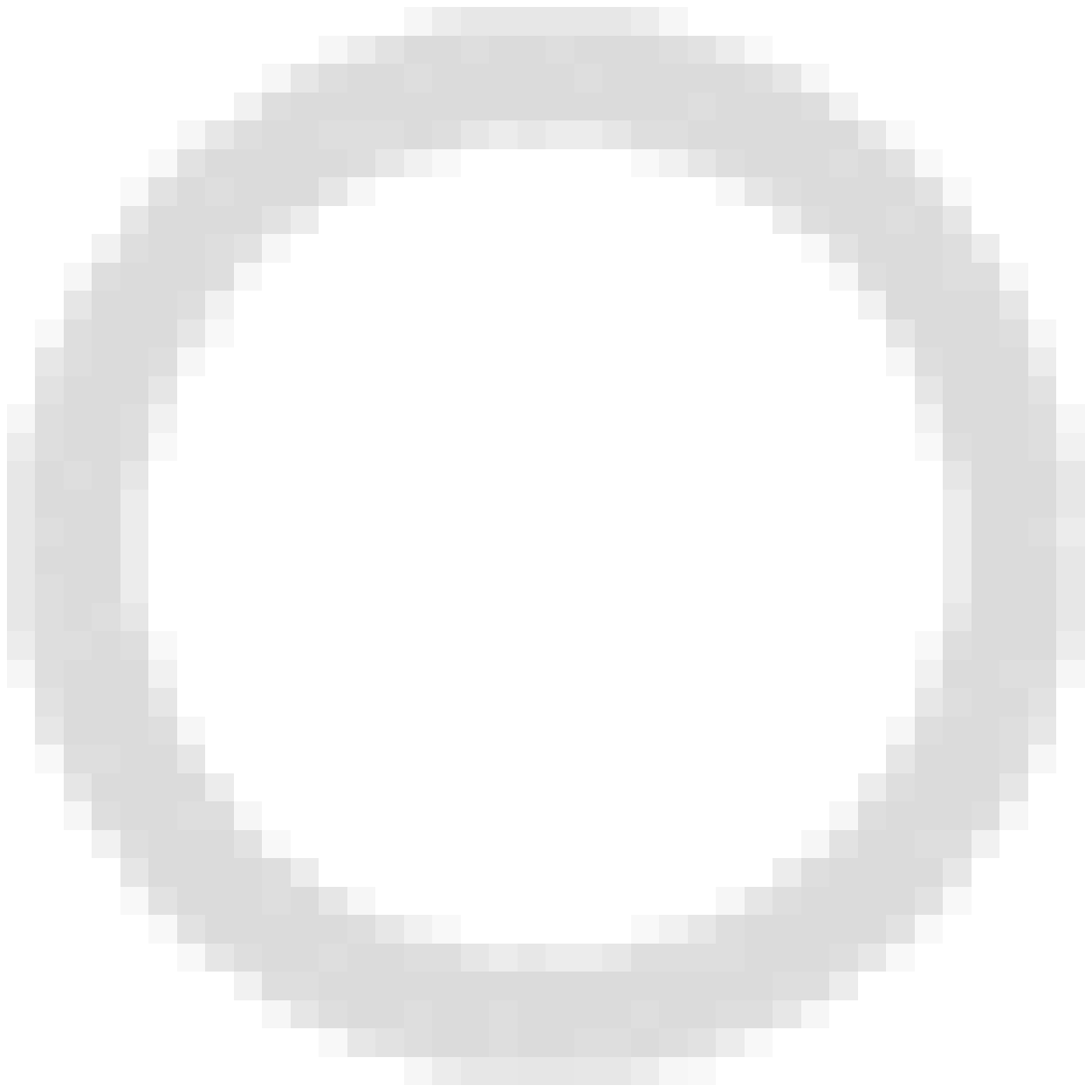}} \\
18. www.dpconline.org & 7 & 552/384 70\% & 236/227   96\% & 195/37   19\% & 121/120   99\% & (0.509, 0.304, 0.013) & \scalebox{0.05}{\includegraphics{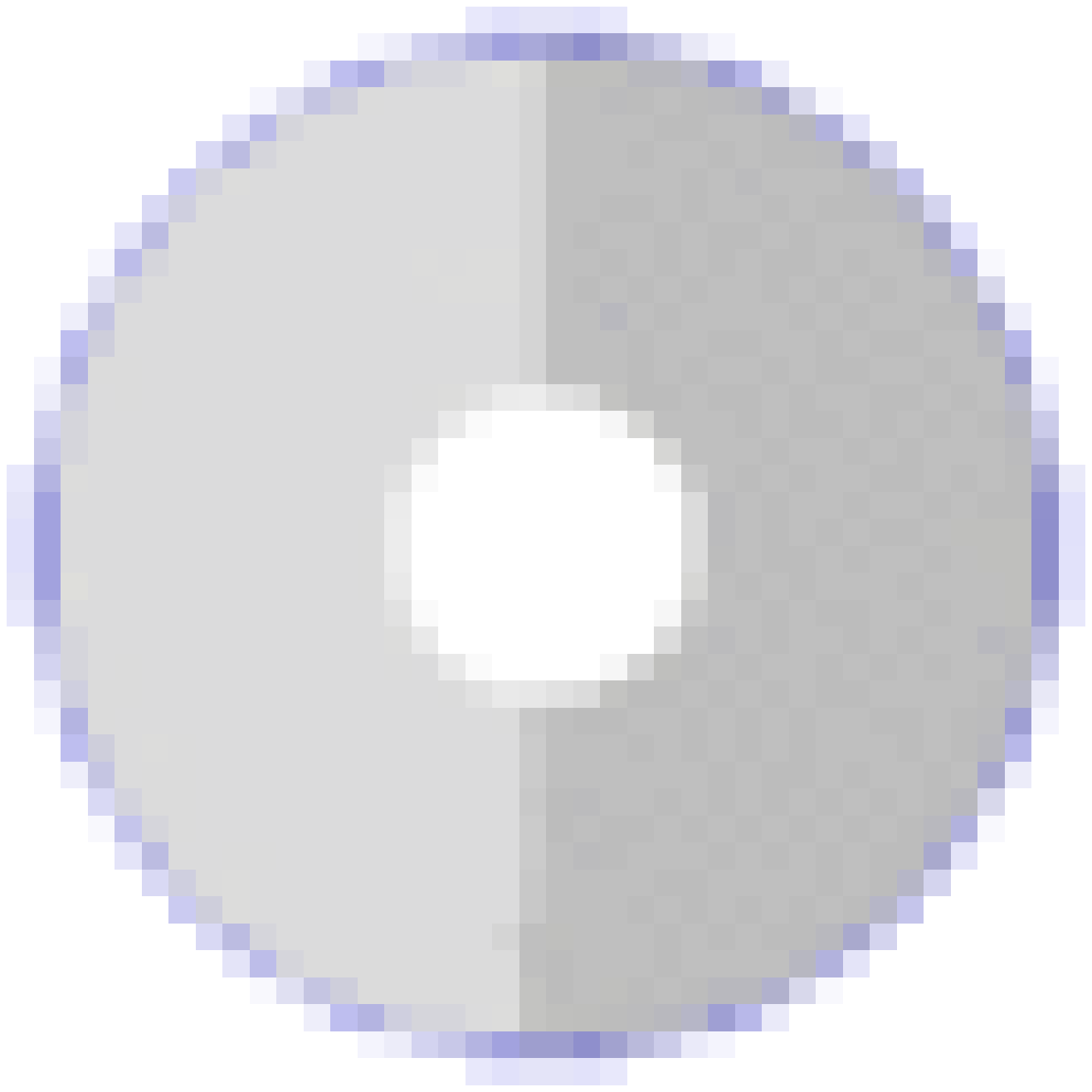}} & 66\% & \scalebox{0.05}{\includegraphics{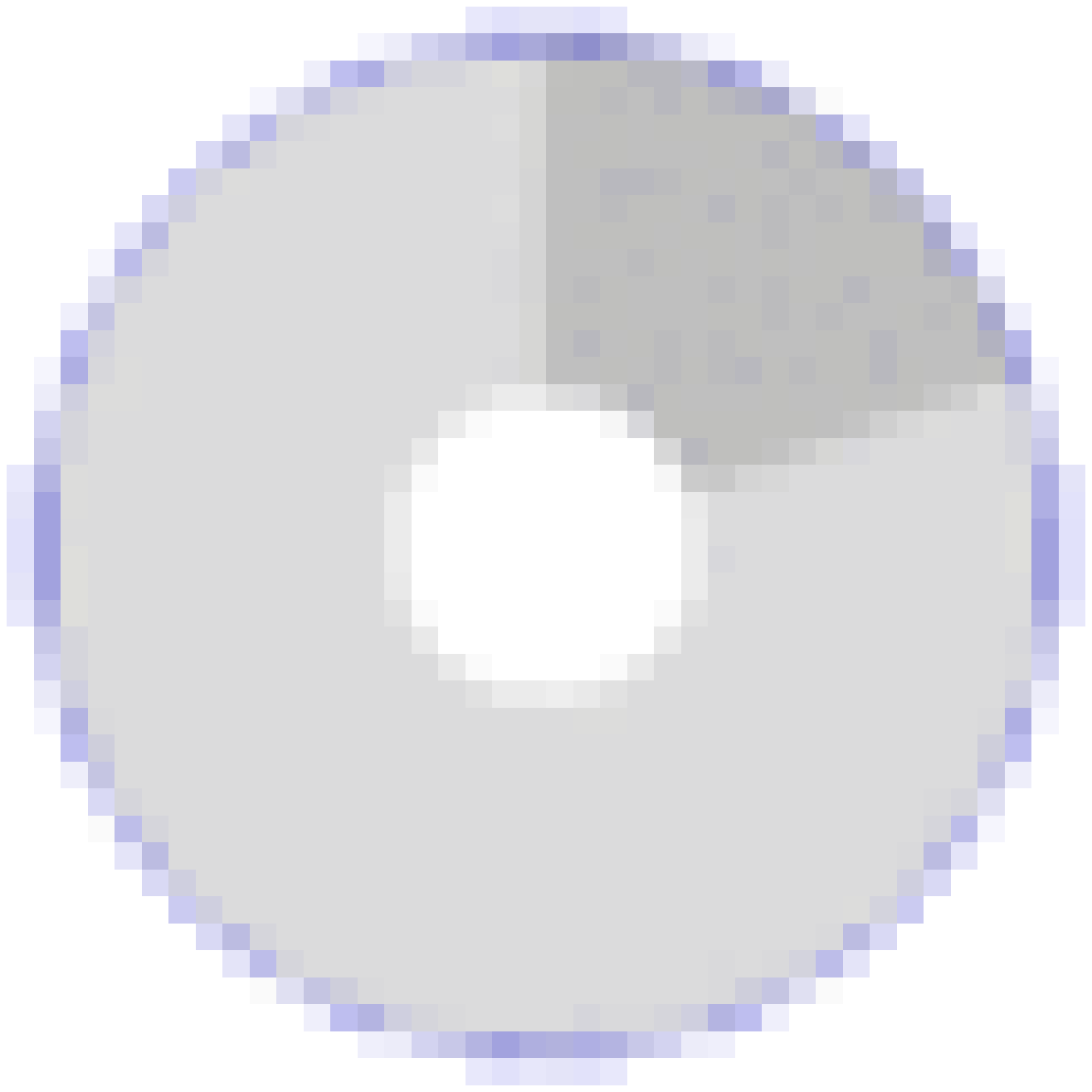}} \\
19. www.cs.odu.edu/\~{}pothen & 5 & 640/435 68\% & 160/151   94\% & 258/120   47\% & 222/164   74\% & (0.402, 0.320, 0.062) & \scalebox{0.05}{\includegraphics{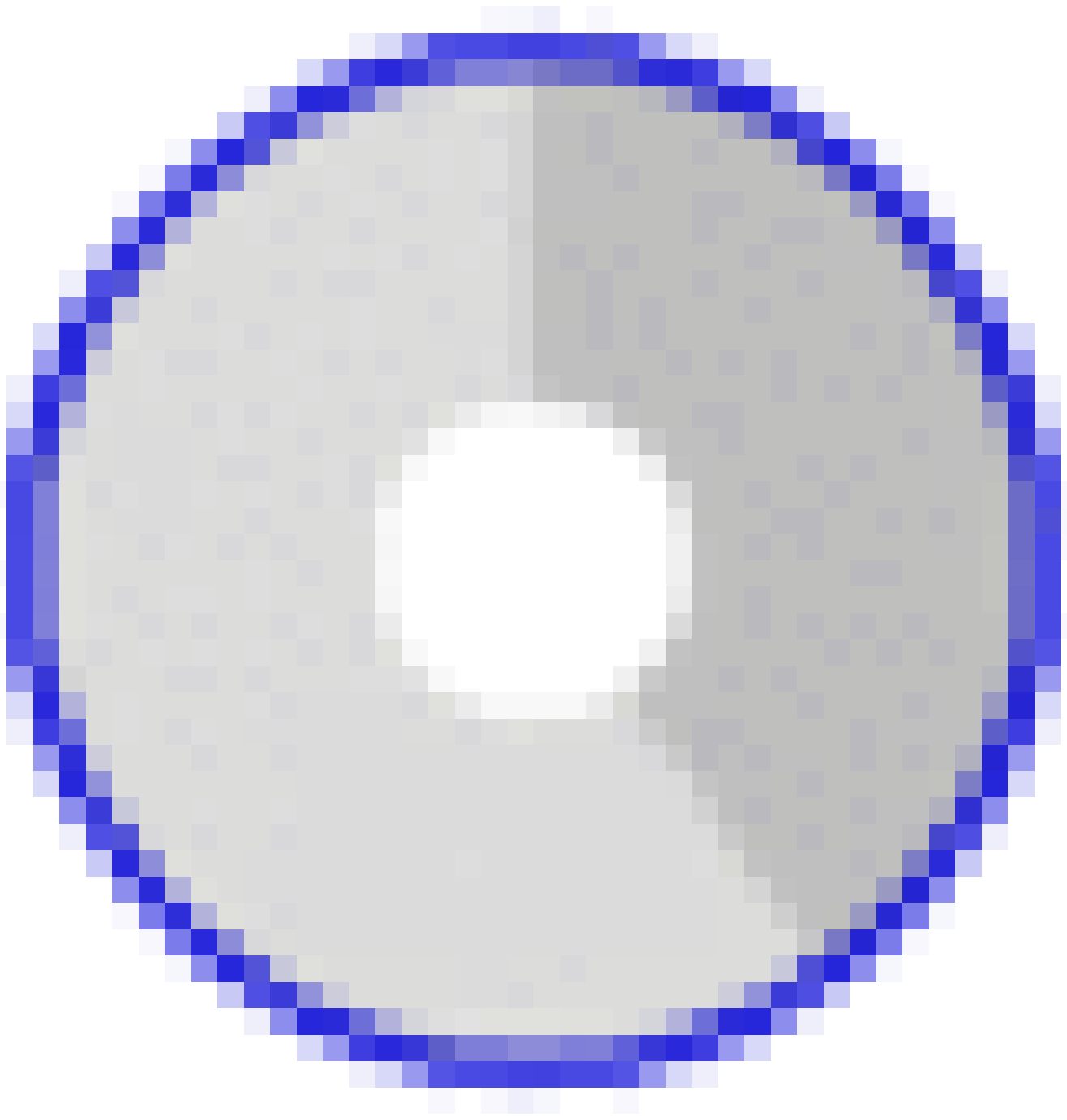}} & 28\% & \scalebox{0.05}{\includegraphics{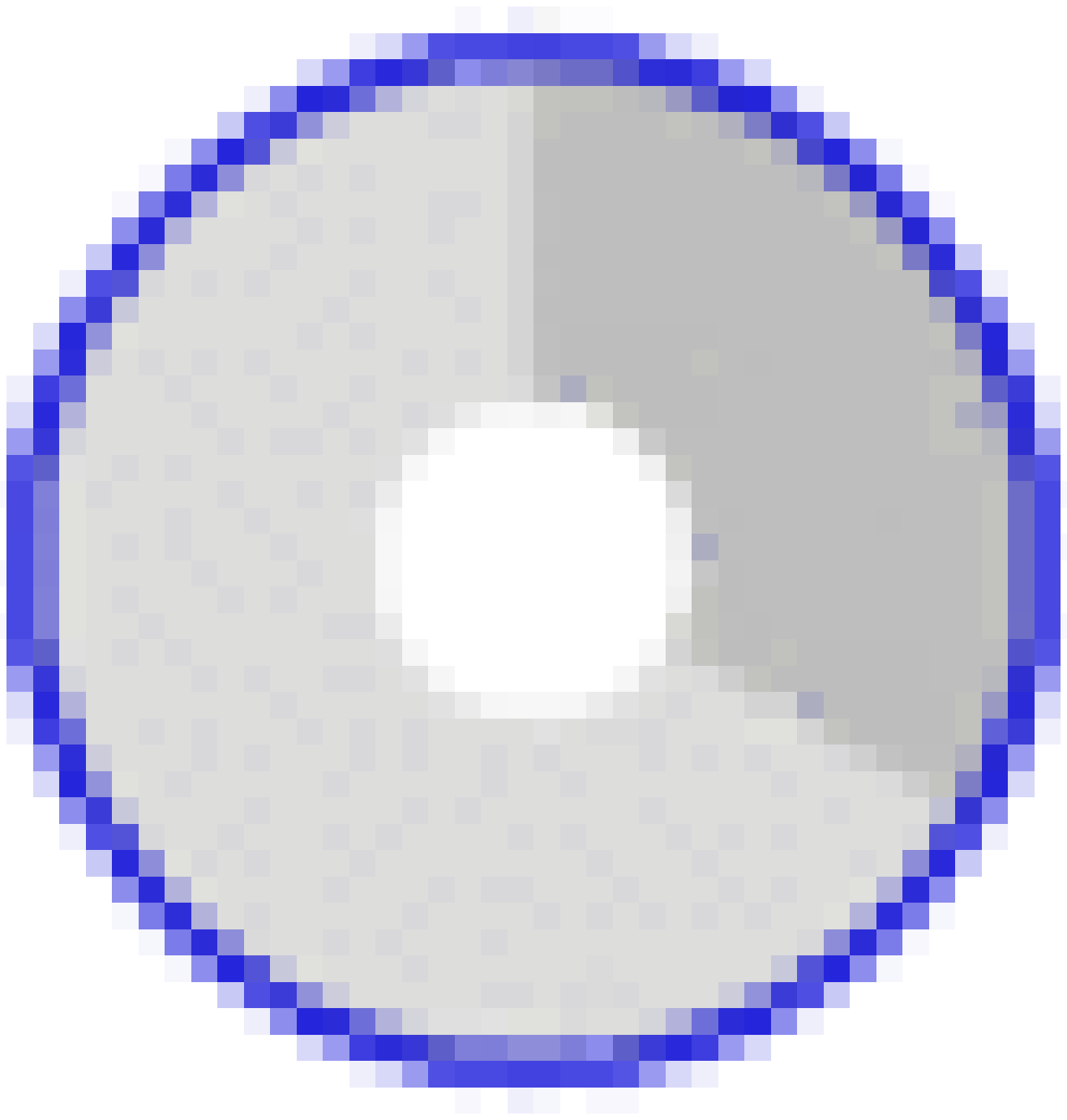}} \\
20. www.eskimo.com/\~{}scs & 6 & 719/691 96\% & 696/669   96\% & 22/21   95\% & 1/1   100\% & (0.011, 0.039, 0.001) & \scalebox{0.05}{\includegraphics{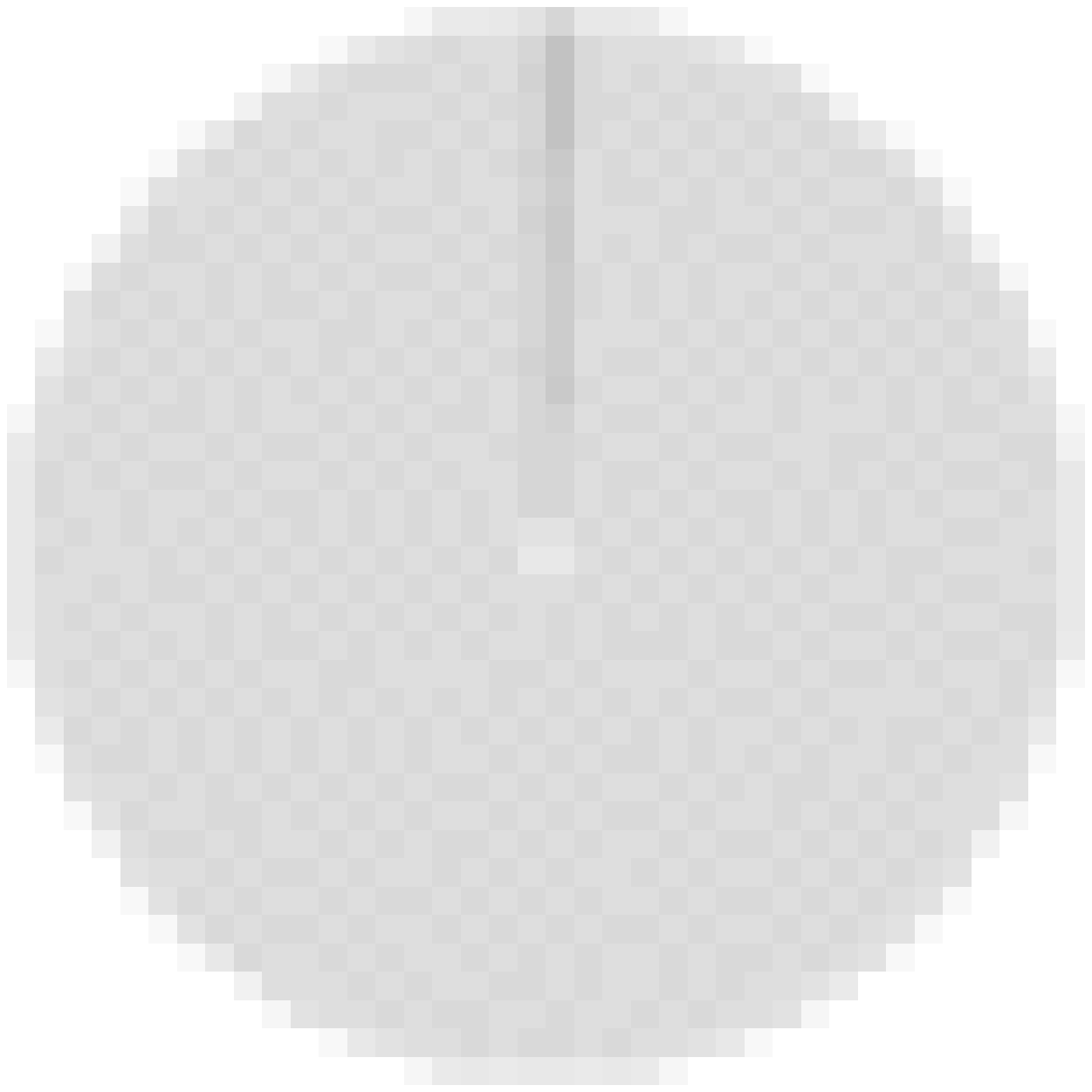}} & 50\% & \scalebox{0.05}{\includegraphics{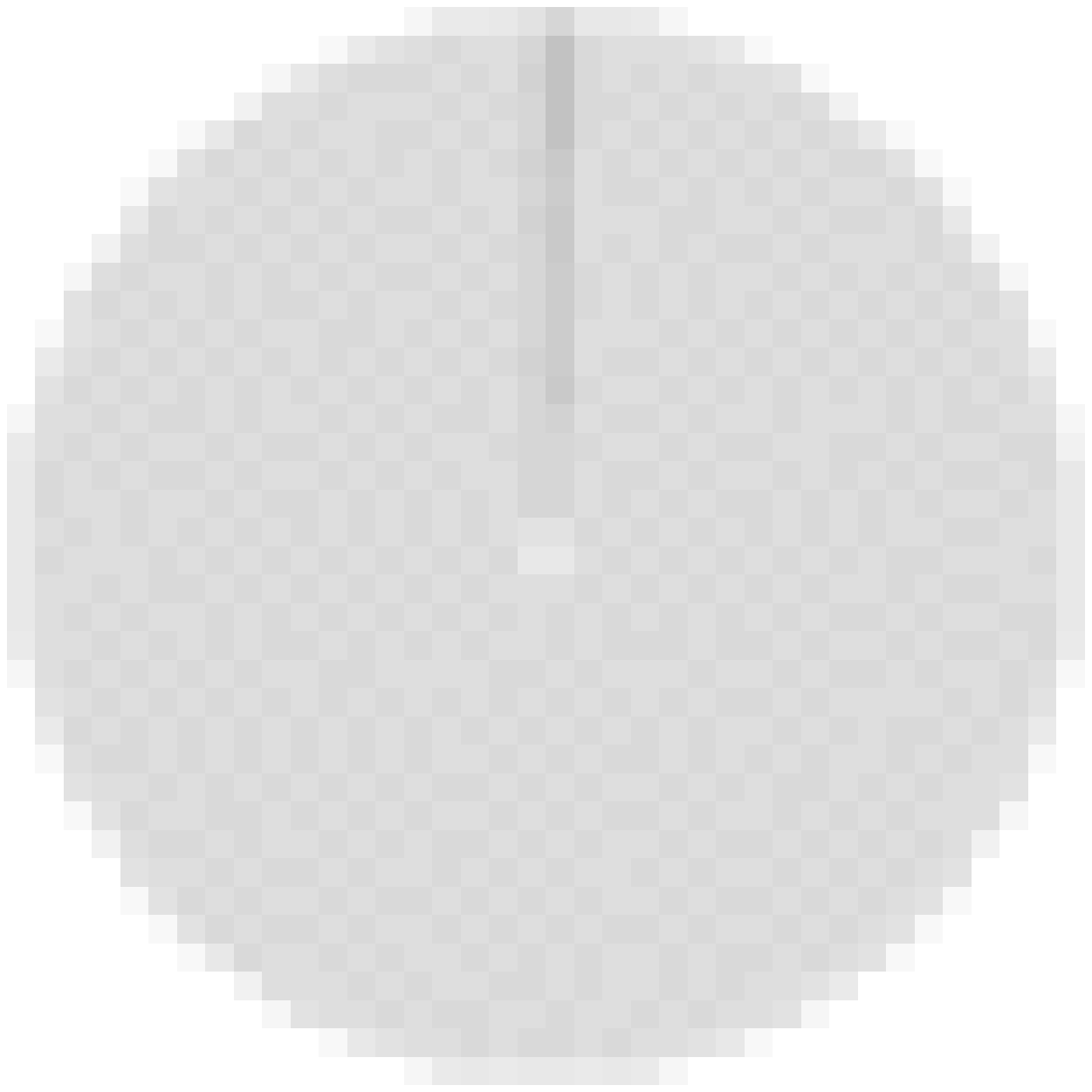}} \\
21. www.financeprofessor.com & 6 & 817/626 77\% & 455/404   89\% & 312/180   58\% & 50/42   84\% & (0.211, 0.234, 0.011) & \scalebox{0.05}{\includegraphics{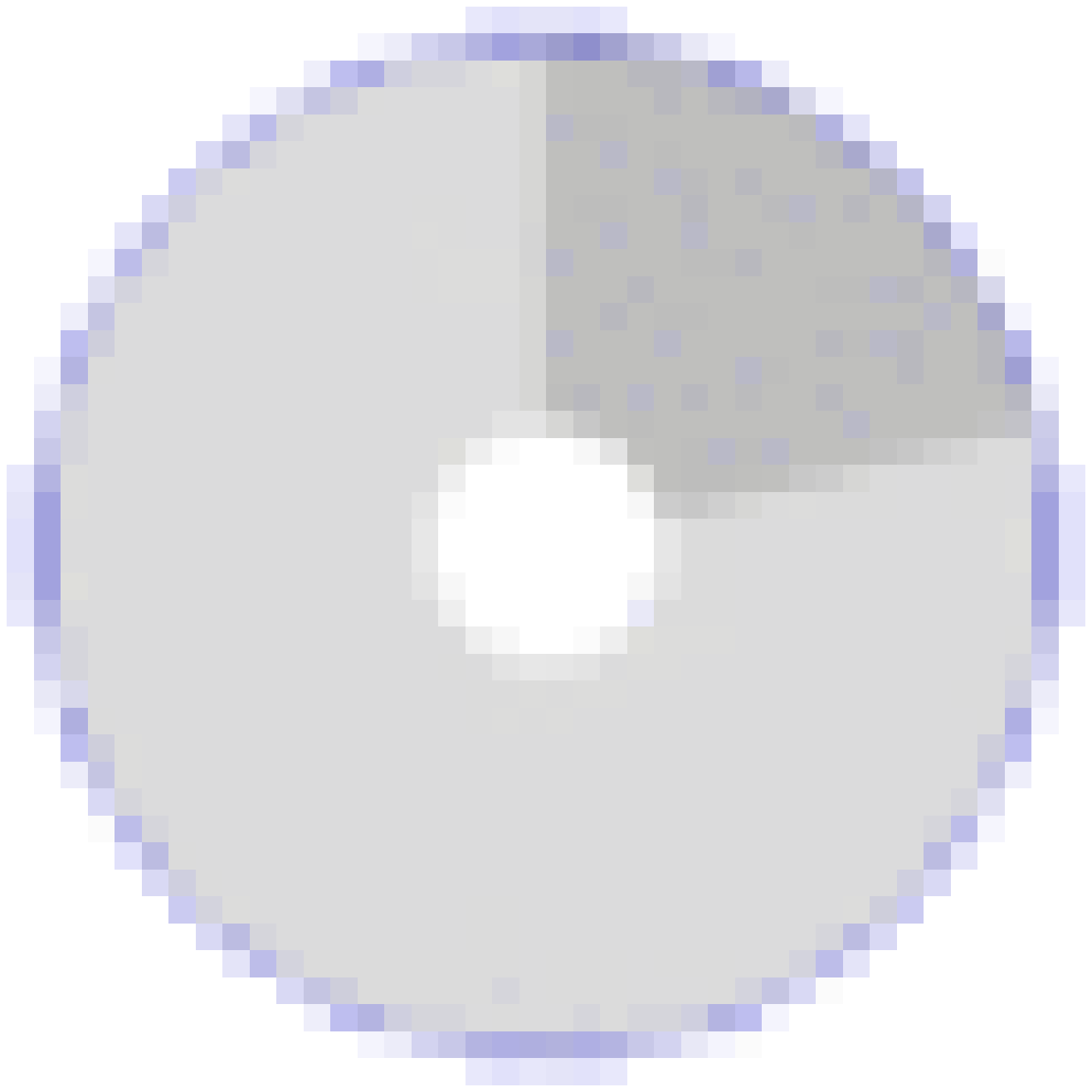}} & 72\% & \scalebox{0.05}{\includegraphics{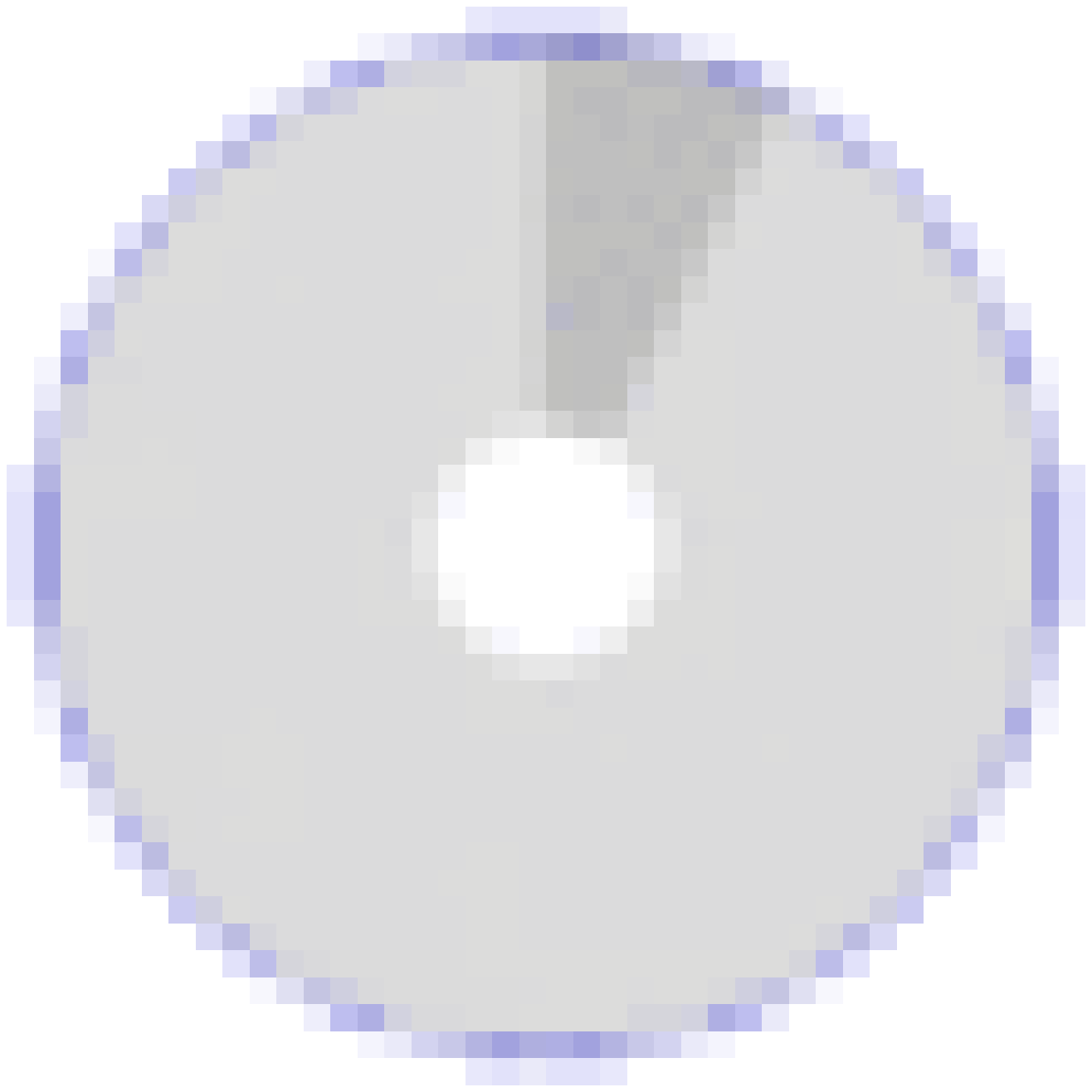}} \\
22. www.fishingcairns.com.au & 5 & 1152/1070 93\% & 259/259   100\% & 890/808   91\% & 3/3   100\% & (0.466, 0.071, 0.000) & \scalebox{0.05}{\includegraphics{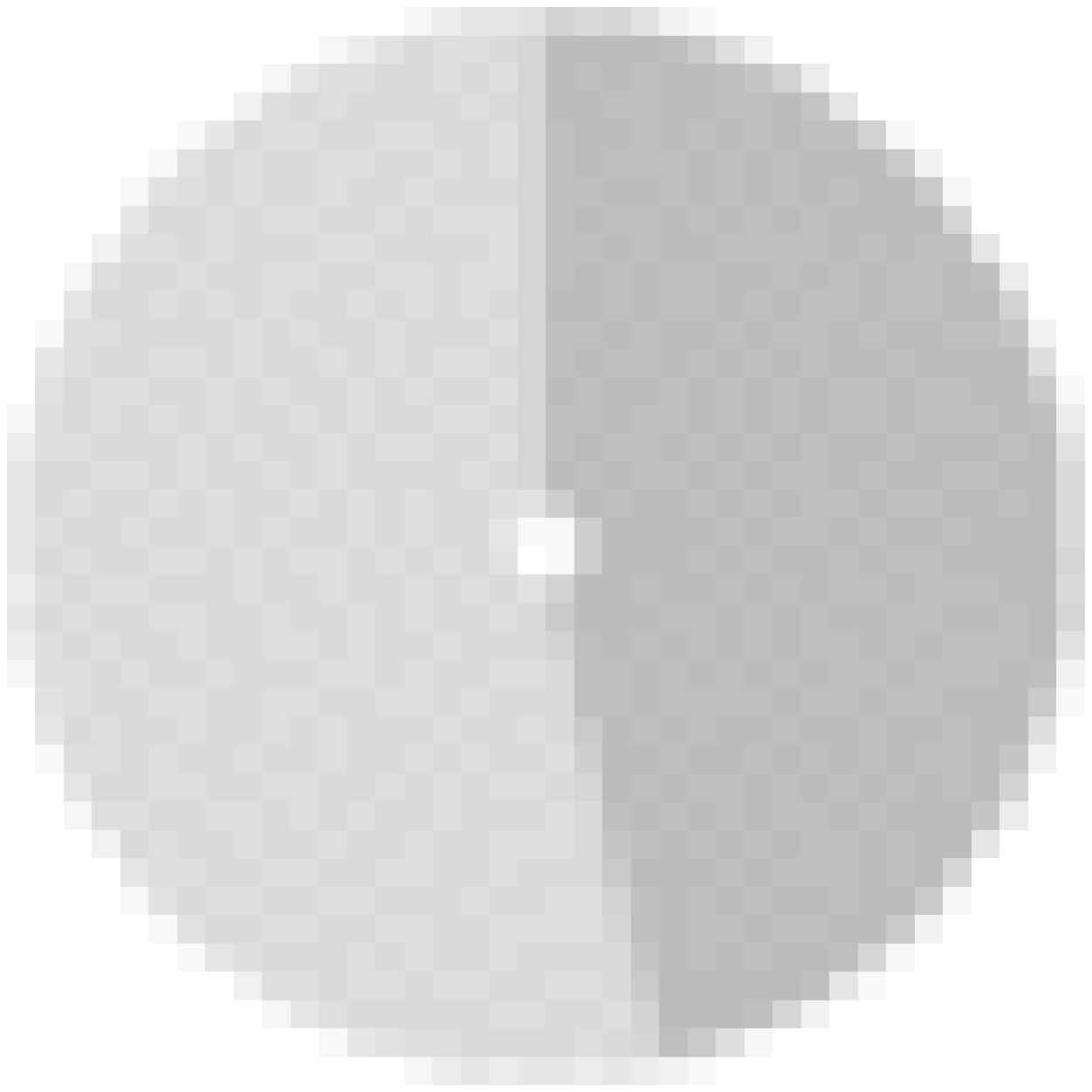}} & 95\% & \scalebox{0.05}{\includegraphics{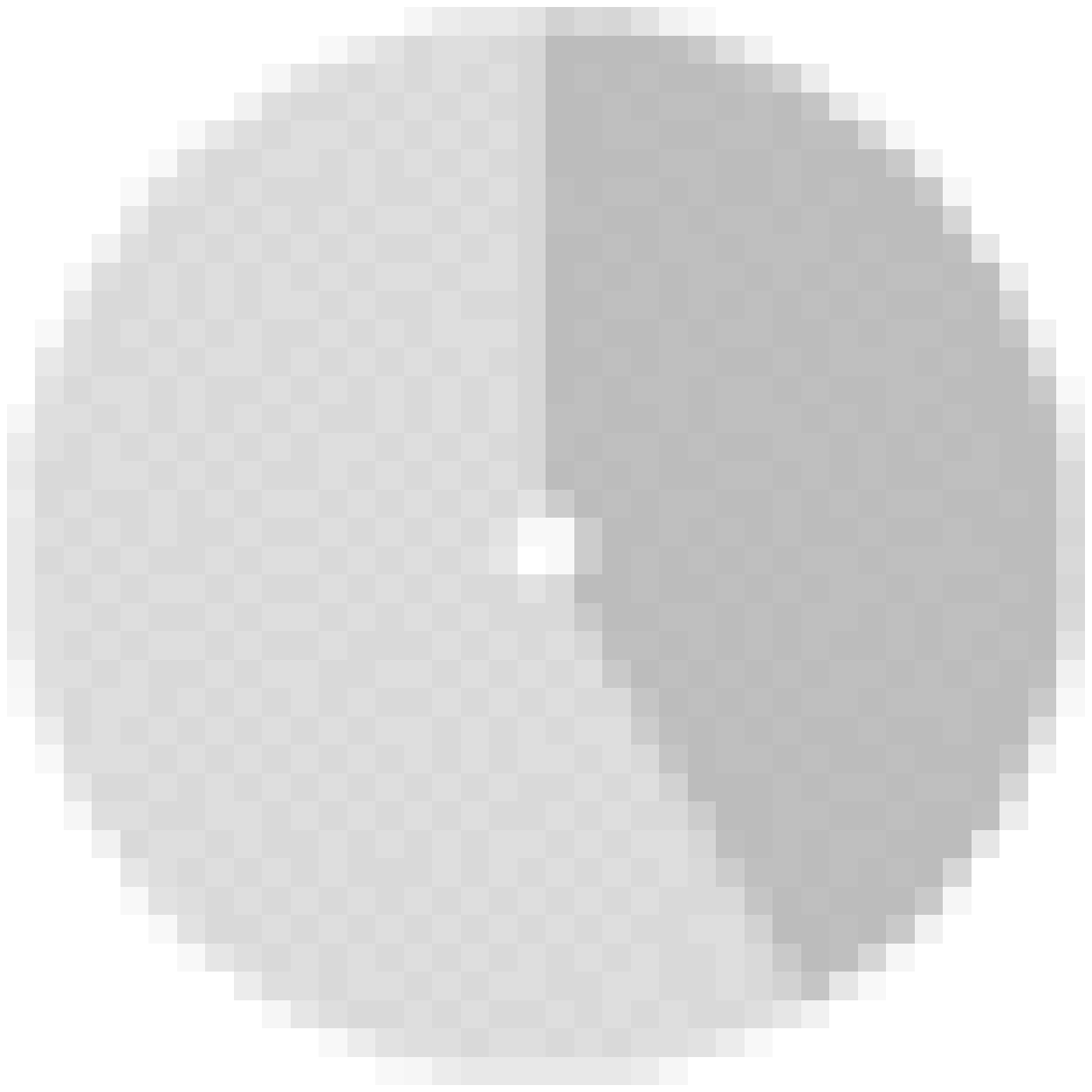}} \\
23. www.techlocker.com & 4 & 1216/406 33\% & 687/149   22\% & 529/257   49\% & 0/0    & (0.267, 0.666, 0.175) & \scalebox{0.05}{\includegraphics{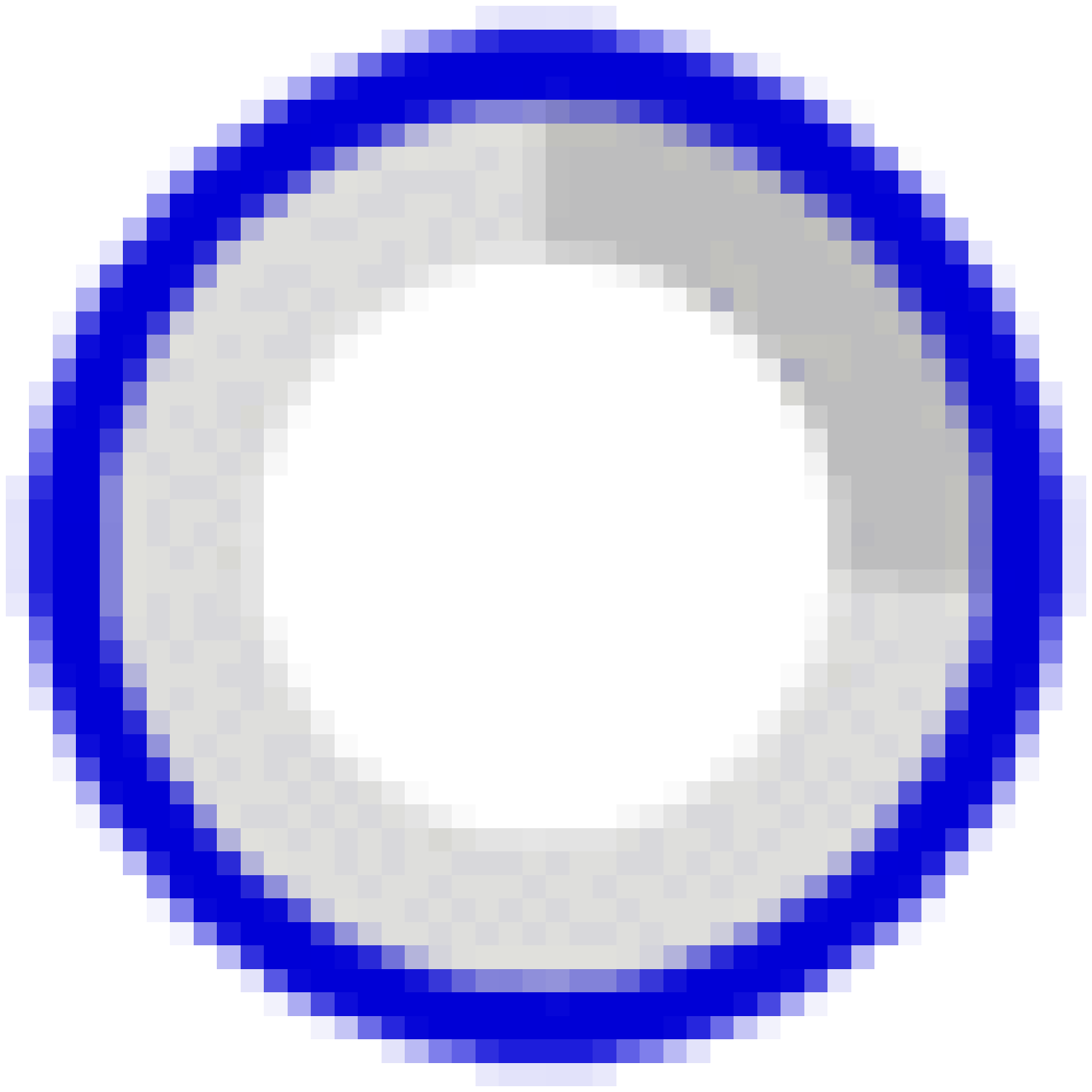}} & 99\% & \scalebox{0.05}{\includegraphics{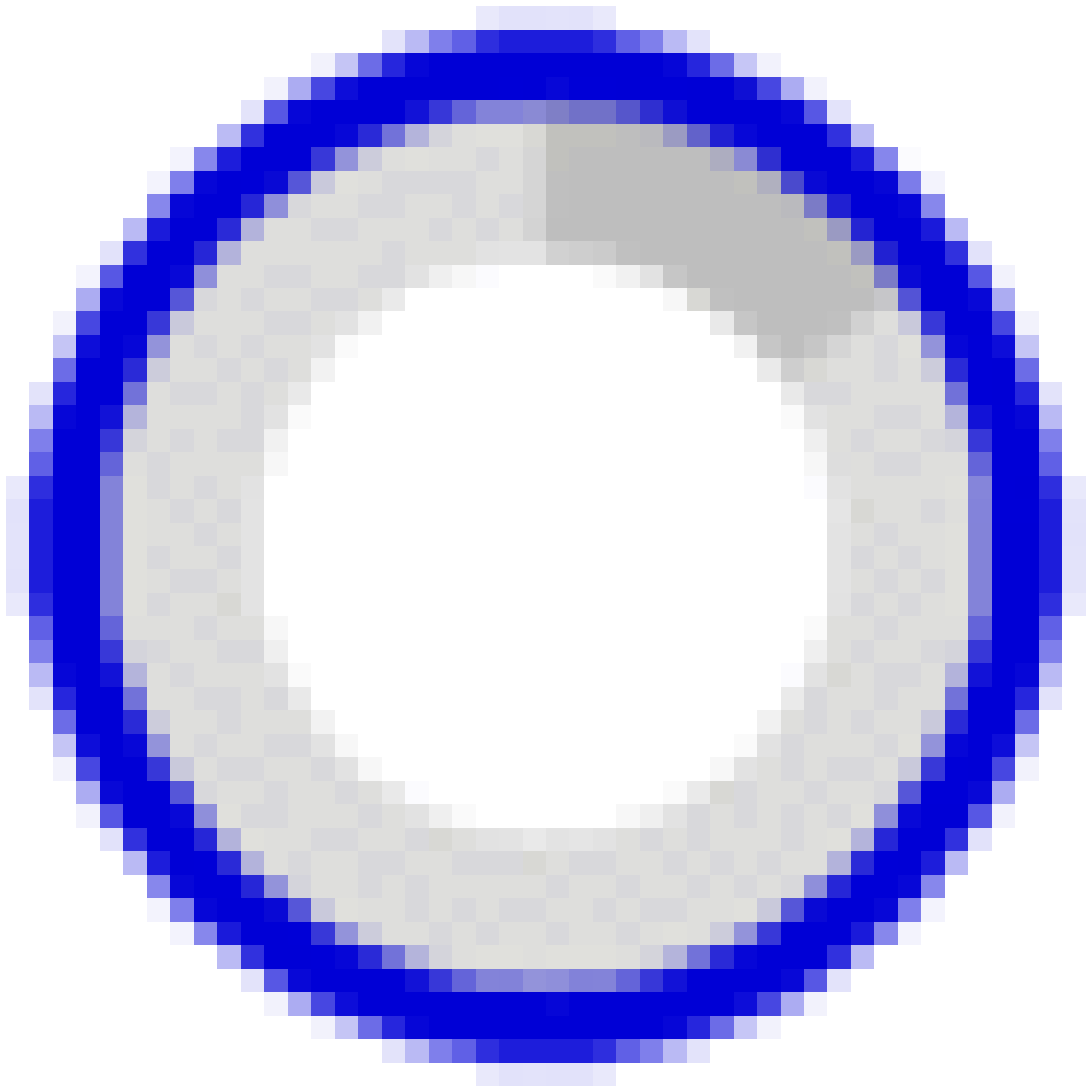}} \\
24. www.kruderdorfmeister.com & 4 & 1509/128 8\% & 1298/31   2\% & 211/97   46\% & 0/0    & (0.056, 0.915, 0.066) & \scalebox{0.05}{\includegraphics{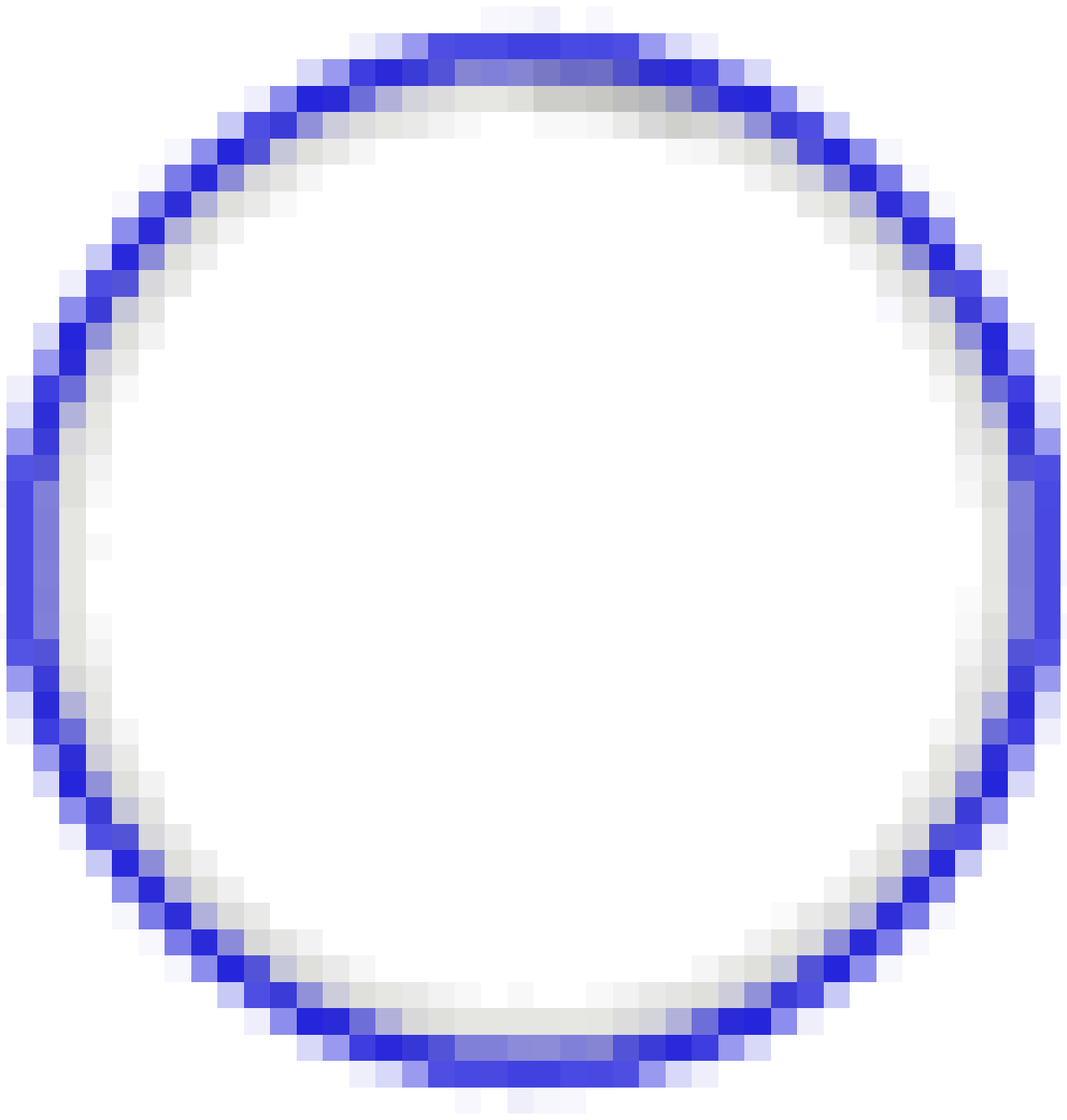}} & 50\% & \scalebox{0.05}{\includegraphics{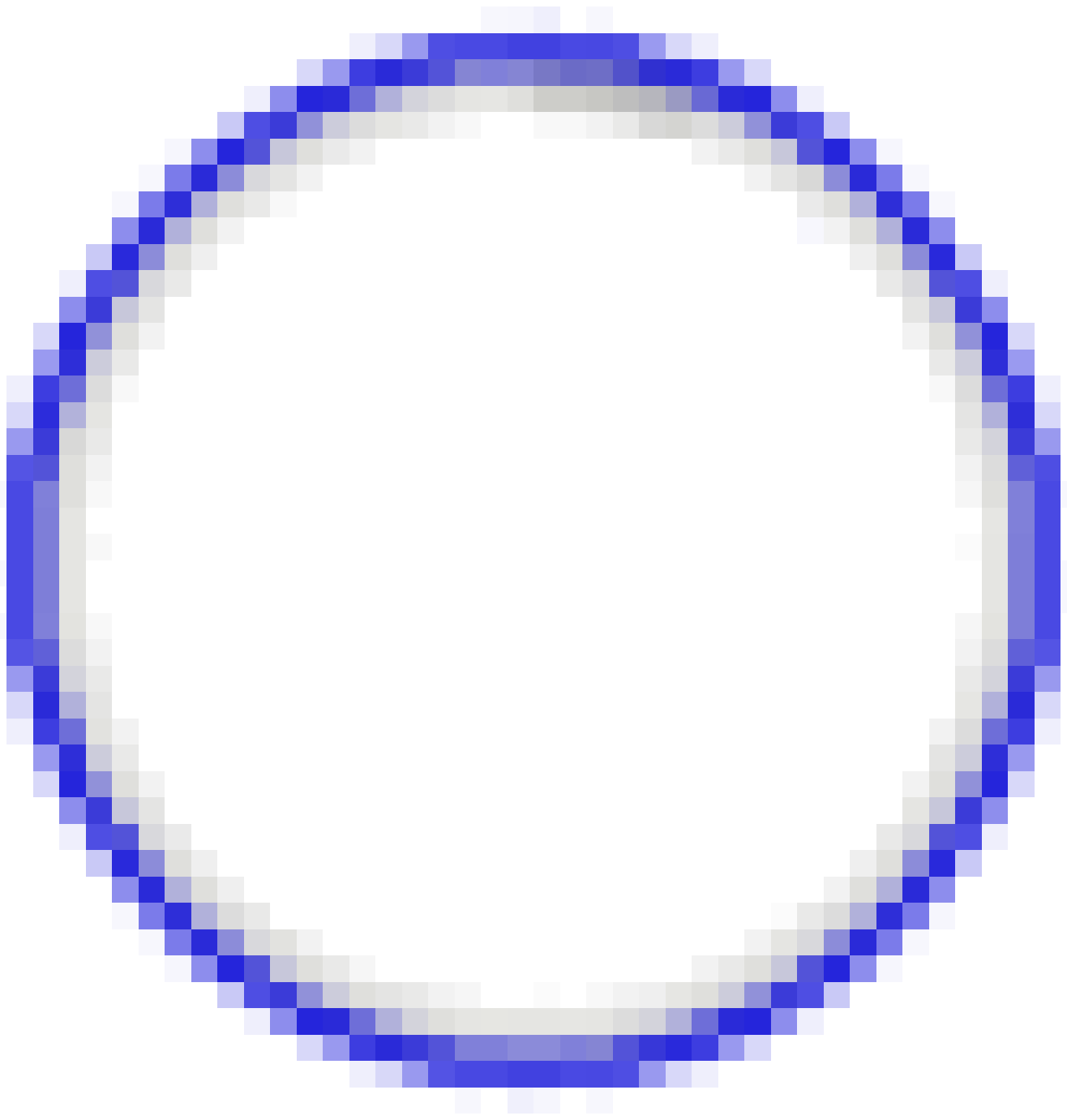}} \\
\hline
\end{tabular}
\end{small}
\label{tbl:recons}
\end{table*}

We were able to recover more than 90\% of the original files from a
quarter of the 24 websites.  For 3 quarters of the websites we
recovered more than 50\% of the files. On average we were able to
recover 68\% of the website files (median=72\%). Of those files
recovered, 30\% of them on average were not byte-for-byte
duplicates. A majority (72\%) of the `changed' text-based files were
almost identical to the originals (having 75\% of their shingles in
common). 67\% of the 24 websites had obtained additional files when
reconstructed which accounted for 7\% of the total number of files
reconstructed per website.

\begin{figure}
\begin{center}
\scalebox{0.5}{\includegraphics{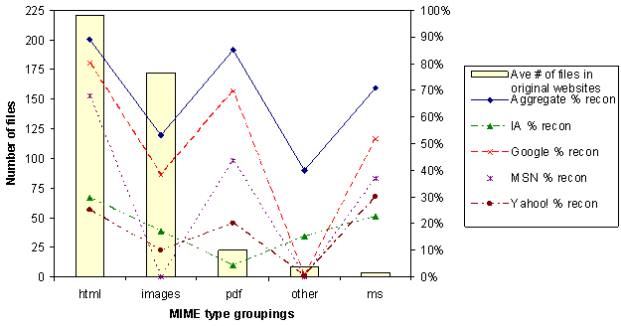}} \caption{Recovery
success by MIME type} \label{fig:graph_recon_ave}
\end{center}
\end{figure}

Figure \ref{fig:graph_recon_ave} shows how successful we were at
recovering files based on their MIME type. The percentage of
resources that were recovered from the 5 different website
reconstructions we performed (1 using all 4 web repositories, and 4
using each web repository individually) are shown along with the
average number of resources making up the original 24 websites. A
majority (92\%) of the resources making up the original websites are
HTML and images.  We were much more successful at recovering HTML
resources than images; we recovered 100\% of the HTML resources for
9 of the websites (38\%) using all 4 web repositories. In the case
of the individual web repository reconstructions, 3 of the 4 web
repositories were individually able to recover a higher percentage
of HTML resources than any other resource type.  Images and formats
with other MIME types were not as likely to be available in the web
repositories. Our experiments measuring the caching of images
verifies that SEs do not cache images as frequently as they cache
text documents.  Also MSN cannot be used to recover images; this
lowers our chance of aggregate recovery of images even further. PDF
and  Microsoft Office formats made up a small portion of the
websites.  We were more successful at recovering PDF resources
(85\%) with the aggregate reconstructions than MS Office formats
(71\%).

\begin{figure}
\begin{center}
\scalebox{0.52}{\includegraphics{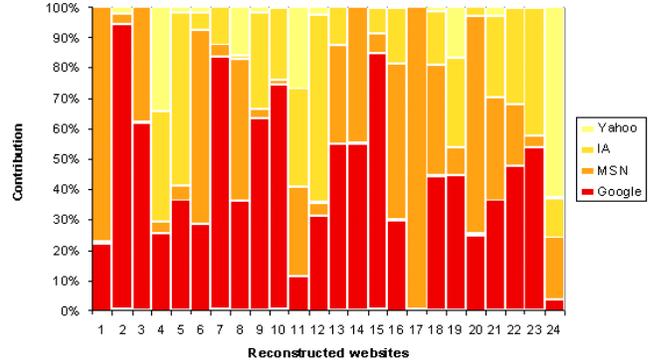}}
\caption{Web repositories contributing to each website
reconstruction} \label{fig:recon_individual_webstores}
\end{center}
\end{figure}

Figure \ref{fig:recon_individual_webstores} shows the percentage of
each web repository's contribution in the aggregate reconstructions.
The numbers of the x-axis match the numbering of Table
\ref{tbl:recons}. Google contributed the most to each website
reconstruction, providing on average 44\% of the files to each
website and failing to contribute to only one website (website 17).
MSN was second, providing on average 30\% of the files; IA was third
with 19\%, and Yahoo was last with a 7\% contribution rate. Yahoo's
poor contribution rate can be expected for a few reasons. First,
they do not consistently provide a datestamp for their resources,
and Warrick will always choose a resource with a datestamp over one
without it.  Second, Yahoo's solo recovery performance (Figure
\ref{fig:graph_recon_ave}) demonstrated they were the worst at
recovering most resources.  And finally, as we have seen in our
crawling and caching experiments, Yahoo provides very inconsistent
access to resources in their cache. MSN is not usable for recovering
images, the second most numerous type of resource in each website.
The fact that they contributed 11\% more files than IA is due to
their high success rate of recovering HTML resources as was shown in
Figure \ref{fig:graph_recon_ave}.

\begin{figure}
\begin{center}
\scalebox{0.52}{\includegraphics{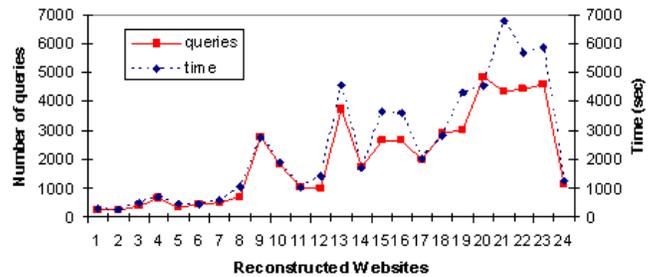}}
\caption{Number of queries performed and time taken to reconstruct
websites} \label{fig:recon_time_queries}
\end{center}
\end{figure}

Figure \ref{fig:recon_time_queries} shows the amount of time and the
number of queries required to reconstruct all 24 websites. There is
almost a 1:1 ratio of queries to seconds. Although the size of the
original websites gets larger along the x-axis, the number of files
reconstructed and the number of resources held in each web
repository determine how many queries are performed.  In none of our
reconstructions did we exceed the query limit of any of the web
repositories, and so Warrick never had to sleep for 24 hours.

\section{Future Work}
We are investigating methods that could be used to insert the
server-side components or logic into static web pages in a secure
manner so that Warrick may be able to reconstruct the server-side
logic of a website.

We are also designing a standard interface that any web repository
can implement that would like to share its holdings with Warrick.
This will allow new holdings to be added with minimal work.

In our next experiment we are designing ``tagging'' mechanisms to
allow us to track each page as they are crawled.  When we find pages
inside a cache we will be able to tell which crawler grabbed the
page and when.  This will allow us to know whether a SE is doing its
own crawling or if they are obtaining crawl data from a third party.
We are also increasing the variety of resource types in the web
collections.

\section{Conclusions}
\label{conclusions} Warrick is not a substitute for digital
preservation infrastructure and policy.  Web repositories may not
crawl orphan pages, protected pages (e.g., robots.txt, password,
IP), very large pages, pages deep in a web collection or links
influenced by JavaScript, Flash, or session IDs.  If a web
repository will not or cannot crawl it, Warrick cannot recover it.
More significantly, Warrick can only reconstruct the external view
of a website as viewed by a web crawler.  The server-side components
(CGI programs, databases, etc.) cannot be reconstructed from this
external view.

We have measured the ability of Google, MSN and Yahoo to cache 4
synthetic web collections over a period of 4 months. We measured web
resources to be vulnerable for as little as 10 days and in the worst
case, as long as our 90 day test period.  More encouragingly, many
HTML resources were recoverable for 20-90 days with TURs ranging
from 8-61 days. Google proved to be the most consistent at caching
our synthetic web collections.

We have also reconstructed a variety of actual websites from IA and
the SEs with varying success. HTML resources were the most numerous
(52\%) type of resource in our collection of 24 websites and were
the most successfully recoverable resource type (89\% recoverable).
Images were the second most numerous (40\%) resource types, but they
were less successfully recovered (53\%).   Here again, Google was
the most frequent source (44\%), but MSN was a close second (30\%),
followed by IA (19\%) and Yahoo (7\%). The probability of
reconstruction success was not correlated with PageRank or the size
of the website.

%
\bibliographystyle{abbrv}
\bibliography{lazyp-refs}  
%
%
\end{document}